\documentclass[a4paper,11pt]{article}
\usepackage{exscale}
\usepackage{epsfig}
\usepackage{multicol}
\usepackage{blindtext}
 \usepackage{fancyhdr}
 \usepackage{graphicx}
\usepackage{sidecap}
\usepackage{titlesec}
\usepackage[dvipsnames]{xcolor}
\usepackage{hyperref,graphicx,amsfonts,amssymb,amsthm,amsmath,psfrag}
\usepackage{cite}
\usepackage{physics}
\usepackage{comment}
\usepackage{enumerate}
\usepackage{mathrsfs}
\usepackage{units}
\usepackage{graphicx,amsfonts,amssymb,amsthm,amsmath,psfrag}
\usepackage[toc,page]{appendix}
\usepackage{subfigure}
\usepackage{slashed}
\usepackage{mathtools}
\usepackage{tikz}
\usepackage{booktabs}
\usepackage{siunitx}
\usepackage{bm}         
\usepackage{color}
\usepackage{amssymb}
\usepackage{amsmath}     
\usepackage{psfrag}
\usepackage{epstopdf}
\usepackage{dcolumn}
\usepackage{bm}
\usepackage[utf8]{inputenc}
\usepackage{enumitem}
\usepackage{subfigure}
\usepackage{braket}
\usepackage[normalem]{ulem}	
\usepackage{bbding}
\usepackage{amsmath}
\usepackage{multirow}
\definecolor{asparagus}{rgb}{0.53, 0.66, 0.42}

\usepackage{url}
\usepackage{hyperref,graphicx,amsfonts,amssymb,amsthm,amsmath,psfrag}
\hypersetup{colorlinks,bookmarksopen,bookmarksnumbered,citecolor=Blue,
linkcolor=Blue,pdfstartview=FitH,urlcolor=Blue}
\usepackage[color=orange]{todonotes}

\def\dd{\mathrm{d}}
 
\newcommand\redsout{\bgroup\markoverwith{\textcolor{purple}{\rule[0.5ex]{2pt}{0.8pt}}}\ULon}
\newcommand{\be}{\begin{equation}}
\newcommand{\ee}{\end{equation}}
\newcommand{\ba}{\begin{eqnarray}}
\newcommand{\ea}{\end{eqnarray}}

\newcommand{\Hkin}{H_{\rm kin}}

\newcommand{\GeV}{{\rm GeV}}

\newcommand{\mpl}{M_P}
\newcommand{\tmtextbf}[1]{{\bfseries{#1}}}

\begin{document}
\topmargin -1cm
\oddsidemargin 0cm
\evensidemargin 0cm

\begin{center}
\vspace{2cm}
{\color{Blue}{{\huge \tmtextbf{Quintessential inflation: \\  A tale of emergent and broken symmetries}} }} \\
\vspace{1.5cm}

{\large  {\bf Dario Bettoni }}
{\em  

\vspace{.4cm}
Departamento de F\'isica Fundamental and IUFFyM, \\ 
Universidad de Salamanca, \\ 
Plaza de la Merced S/N, E-37008  Salamanca, Spain 

\vspace{0.6cm}
}

{\large  {\bf Javier Rubio }}
{\em  

\vspace{.4cm}
Centro de Astrof\'{\i}sica e Gravita\c c\~ao  - CENTRA, \\
Departamento de F\'{\i}sica, Instituto Superior T\'ecnico - IST, Universidade de Lisboa - UL, \\
Av. Rovisco Pais 1, 1049-001 Lisboa, Portugal

\vspace{1.5cm}

}
\end{center}

\noindent --------------------------------------------------------------------------------------------------------------------------------

\begin{center}
{\bf \large Abstract}  
\end{center}
Quintessential inflation provides a unified description of inflation and dark energy in terms of a single scalar degree of freedom, the cosmon. We present here a comprehensive overview of this appealing paradigm, highlighting its key ingredients and keeping a reasonable and homogeneous level of details. After summarizing the cosmological evolution in a simple canonical case, we discuss how quintessential inflation can be embedded in a more general scalar-tensor formulation and its relation to variable gravity scenarios. Particular emphasis is placed on the role played by symmetries. In particular, we discuss the evolution of the cosmon field in terms of ultraviolet and infrared fixed points potentially appearing in quantum gravity formulations and leading to the emergence of scale invariance in the early and late Universe. The second part of the review is devoted to the exploration of the phenomenological consequences of the paradigm. First, we discuss how direct couplings of the cosmon field to matter may affect neutrinos masses and primordial structure formation. Second, we describe how Ricci-mediated couplings to spectator fields can trigger the spontaneous symmetry breaking of internal symmetries such as, but not limited to, global U(1) or Z$_{\bf 2}$ symmetries, and affect a large variety of physical processes in the early Universe.
  \vspace{0.9cm}

\noindent --------------------------------------------------------------------------------------------------------------------------------

\vspace{2cm}

\noindent bettoni@usal.es \\ javier.rubio@tecnico.ulisboa.pt

\newpage
\tableofcontents
\newpage

\section{Introduction}\label{sec:intro}

\textit{``Once upon a time the Universe sped up, out of emptiness. Then, there was something all over the place. Lucky we were that it dawdled to hasten again, so we can be here to observe".} \\

The current best-fitting picture of the statistical properties of the Universe, from horizon scales ($15000$ Mpc) to typical galactic separations ($1$ Mpc), is the so-called $\Lambda$ Cold Dark Matter scenario ($\Lambda$CDM). This $6$-parameter model assumes the Universe to be composed of photons, neutrinos, ordinary matter and a non-relativistic or cold dark matter (CDM) component with gravitational interactions only. On top of that, it presumes the existence of two accelerated expansion eras taking place in the very early and the very late Universe: Inflation and Dark Energy.
\begin{enumerate}
 \item Inflation: This epoch provides a natural solution to the shortcomings of the hot Big Bang (hBB) model, being able to generate, without finetunings, a flat, homogeneous and isotropic Universe while removing unwanted relics such as monopoles or other topological defects. More importantly, it provides a causal mechanism for the generation of the primordial density perturbations seeding structure formation. These fluctuations are unavoidably created out of the quantum vacuum and stretched out to superhorizon scales, where they become classical. In the simplest realizations, they are also adiabatic, Gaussian and display a nearly scale-invariant spectrum at all observable scales. On top of that, they remain frozen outside the horizon until they reenter it again long after the end of inflation.  When doing so, they generate the \textit{coherent} set of oscillations we observe in the  Cosmic Microwave Background (CMB) and seed the large-scale distribution of matter and galaxies in the Universe. Interestingly enough, the results of three decades of observations are remarkably consistent with the simplest incarnation of the inflationary paradigm: a canonical scalar field slowly rolling in a featureless potential and eventually approaching a global minimum, where oscillations are effectively damped by the quanta production of the fields coupled to it. 

 Beyond temperature and density fluctuations, inflation predicts a so-far unobserved background of tensor perturbations displaying also a quasi scale-invariant spectrum. The main signature of this stochastic gravitational wave background is a curl-like pattern in the CMB polarization. These $B$-modes are the target of several ongoing and forthcoming initiatives, ranging from ground-based experiments to stratospheric balloons and space missions \cite{Kamionkowski:2015yta}. Its detection would constitute the first test of the quantum nature of spacetime, while providing an indirect window to fundamental physics all the way up to the unification scale.
 
\item Dark energy: The discovery in the late 90's of the current accelerated expansion of the Universe revolutionized modern cosmology \cite{SupernovaCosmologyProject:1997zqe,SupernovaSearchTeam:1998fmf}. Since then, this result has been indirectly confirmed, among others, by CMB observations \cite{Planck:2018jri,BICEP:2021xfz} and large scale structure surveys \cite{2dFGRS:2001csf,2dFGRS:2001ybp,SDSS:2005xqv}. In the context of General Relativity, the simplest mathematical explanation is a non-vanishing cosmological constant $\Lambda$ leading to a time-independent energy density of order $\rho_\Lambda\simeq (1\, {\rm meV})^4$.  The inclusion of this everlasting component in a Universe where all other matter species redshift with expansion raises immediately two naturalness issues: i) \textit{why is the cosmological constant so small as compared to all particle physics scales?} and ii) \textit{why did it start dominating precisely now?} The lack of satisfactory answers to these questions has motivated a plethora of alternative explanations to the present accelerated expansion of the Universe, usually coined under the generic name of dark energy models. Many of these proposals involve a time-dependent (non-homogeneous) scalar field with an evolving equation of state different from that of baryons, neutrinos, dark matter, or radiation. This fifth or quintessential entity is generically designed to mimic, with sufficient accuracy, a de Sitter-like behaviour at the present cosmological epoch, in full analogy with the inflationary stage but at an energy scale many orders of magnitude lower. Interestingly enough, some quintessence scenarios involve appealing scaling solutions in which the dark energy density remains comparable to the dominant radiation and matter components for an extended period of time, eventually jumping out of this behaviour to drive the late-time acceleration. Provided that the event triggering this transition does not involve any additional tuning of times or scales, this alleviates the aforementioned \textit{``why now"} problem.
\end{enumerate}

Since inflation and quintessence  typically require a beyond the Standard Model field to drive an almost de Sitter stage,  it seems natural to seek for the unification of these two stages into a common framework, potentially based on an underlying symmetry principle.  This is the main idea behind the so-called  \textit{Quintessential inflation} paradigm, where the inflaton and dark energy fields are identified as a single scalar component customarily dubbed cosmon \cite{Peccei:1987mm}. This approach reduces the required number of degrees of freedom by one, with the further advantage of eliminating the problem of setting the initial conditions for the quintessence field, which are now determined by the standard inflationary attractor. When written in terms of a canonical scalar field, these scenarios involve a potential of runaway-type with two flat or slow-roll regions connected by a steep transition spanning more than hundred orders of magnitude. This large hierarchy of scales can be imposed by hand or follow naturally from some underlying principle, as happen for instance in quantum chromodynamics, where the existence of an ultraviolet (UV) fixed point makes the infrared (IR) confinement scale exponentially smaller than the UV cutoff of the theory.  At the phenomenological level, the steep transition leads generically to an intermediate cosmological epoch in which the energy density of the Universe is dominated by the kinetic energy density of the cosmon field. During this period, the global equation of state of the Universe approaches unity and the total energy density decreases as the sixth  power of the scale factor. This peculiar scaling  might allow to discriminate quintessential inflation scenarios from standard oscillatory models. In fact, as we will discuss in detail in this review, there are several interesting phenomenological consequences associated to it.  First, the fast dilution of the cosmon energy density implies that any other form of matter will eventually dominate the background evolution. This amplification effect has advantages and drawbacks. On the one hand, it allows to recover the standard hBB picture even if the cosmon field is only mildly depleted after inflation. On the other hand, it induces a substantial growth of any stochastic background of primordial gravitational waves at the corresponding scales, leading to potential violations of Big Bang Nucleosynthesis (BBN) constraints in scenarios involving very extended periods of kination.

\vspace{5mm}
\noindent\textbf{Aim and spirit of this review:} To make quantitative predictions within the quintessential inflation paradigm one can follow two major strategies: perform case-by-case studies involving specific choices of the cosmon potential  or follow a first-principles observational-oriented approach. The first point of view has been widely explored in the literature \cite{WaliHossain:2014usl,deHaro:2021swo}, from the early works of Refs.~\cite{Spokoiny:1993kt,Peebles:1998qn,Peloso:1999dm,Dimopoulos:2001ix,Giovannini:2003jw,Brax:2005uf,BuenoSanchez:2006fhh} to the most recent developments \cite{Hossain:2014xha,Agarwal:2017wxo,Ahmad:2017itq,Geng:2017mic,Dimopoulos:2018eam,Dimopoulos:2019ogl,Benisty:2020xqm,Benisty:2020qta,Karciauskas:2021fdu}. In this review, we will follow the second  approach, trying to connect the behavior of the cosmon field to a minimum of quantities and principles. In particular, we aim to provide an overview of quintessential inflation for a wide audience, including both cosmologists and particle physicists. To cope with this goal, we have focused on the generic ingredients of the paradigm, keeping a reasonable and homogeneous level of details throughout the whole manuscript and intentionally avoided the exhaustive description of the myriad of models in the literature. Overall, this approach brings clarity at the expense of disguising the technical complexity of the issues we discuss, for which we refer the reader to Refs.~\cite{WaliHossain:2014usl,deHaro:2021swo}. For the sake of concreteness, we also exclude from this review a number of theoretical topics, such as warm quintessential inflation \cite{Dimopoulos:2019gpz}, curvaton-inspired scenarios \cite{Feng:2002nb}, higher-dimensional settings \cite{Kamali:2019xnt} and non-Riemannian approaches related to scale symmetry \cite{Guendelman:2014bva,Guendelman:2015liz}.

This article is organized as follows. In Section \ref{sec:standard_picture}, we overview the cosmological history of a canonically normalized cosmon field minimally coupled to gravity. After discussing the generalities of the inflationary stage in Section \ref{subsec:inflation}, we concentrate in Section \ref{subsec:kination} on the unusual kinetic dominated era, presenting the constraints on its duration imposed by BBN. Special attention is then given to the unconventional heating stage, which we describe in detail in Section \ref{subsec:std_heating}, leaving for Section \ref{subsec:hbb_DE} the discussion of the standard hBB evolution and the onset of scalar domination at late times.  The embedding of quintessential inflation in a more general variable gravity framework is presented in Section \ref{sec:fieldrel}, where we summarize the alternative representations and equivalence classes in the literature. Section \ref{sec:scalingframe} discusses the potential association of quintessential inflation with the emergence of scale symmetry in the vicinity UV and IR fixed points, presenting the conditions for the absence of fifth-forces and the temporal non-variation of the Standard Model coupling constants. The last part of the review, Section \ref{sec:pheno}, presents an extensive discussion of the expected phenomenology in the presence of interactions with a potential beyond the Standard Model sector. In particular, Section \ref{subsec:CQ} describes the coupled quintessence paradigm and its applications in the context of growing neutrino quintessence and primordial structure formation, while Section \ref{subsec:spectator} discusses the impact of non-minimal gravitational interactions and their effects on (re)heating, gravitational waves' production and baryogenesis. Finally, in Section \ref{sec:conclusions}, we comment on the challenges and open avenues in this field of research.
\newline

\noindent
\textbf{Notation and conventions.} We will work in natural units ($c=1$ and $\hbar=1$) and use a mostly plus signature for the metric (-,+,+,+). 

\section{Cosmological evolution} \label{sec:standard_picture}

In its simplest form, quintessential inflation can be formulated in terms of a canonically-normalized scalar field minimally coupled to gravity. The corresponding action takes the form
\begin{equation}\label{actioncosmon}
S= \int d^4x \sqrt{-g}\left[\frac{M_P^2}{2}R-\frac12 (\partial\phi)^2-V(\phi)\right]+S_{\rm I}(g_{\mu\nu},\phi,\chi,A_\mu,\psi)\,,
\end{equation}
with $M_P= (8\pi G)^{-1/2}=2.44\times 10^{18}$ GeV the reduced Planck mass, $R$ the Ricci scalar and $(\partial \phi)^2$ a short hand notation for the kinetic term $g^{\mu \nu}  \partial_{\mu}\phi \,\partial_{\nu}\phi$, with $g_{\mu\nu}$ the metric and $g={\rm det} \,(g_{\mu\nu})$ the metric determinant. The cosmon potential 
$V(\phi)$ is assumed to be of a runaway form and, as in any other dark energy-related models, it cannot innately explain  why the observed cosmological constant is so negligibly small, making necessary the implementation of a symmetry principle or cancellation mechanism in order to explain its value. For the purposes of this review, we will simply assume that the cosmon potential vanishes asymptotically at large, positive field values, $V(\phi\rightarrow \infty)\rightarrow 0$, albeit presenting in Section \ref{sec:scalingframe} several arguments supporting this assumption. Finally, the action $S_{\rm I}$ encodes the interactions of the cosmon field with scalar ($\chi$), fermion ($\psi$) and vector ($A_\mu$) degrees of freedom in the Standard Model sector and beyond. 

\begin{figure}
    \centering
    \includegraphics[scale=0.8]{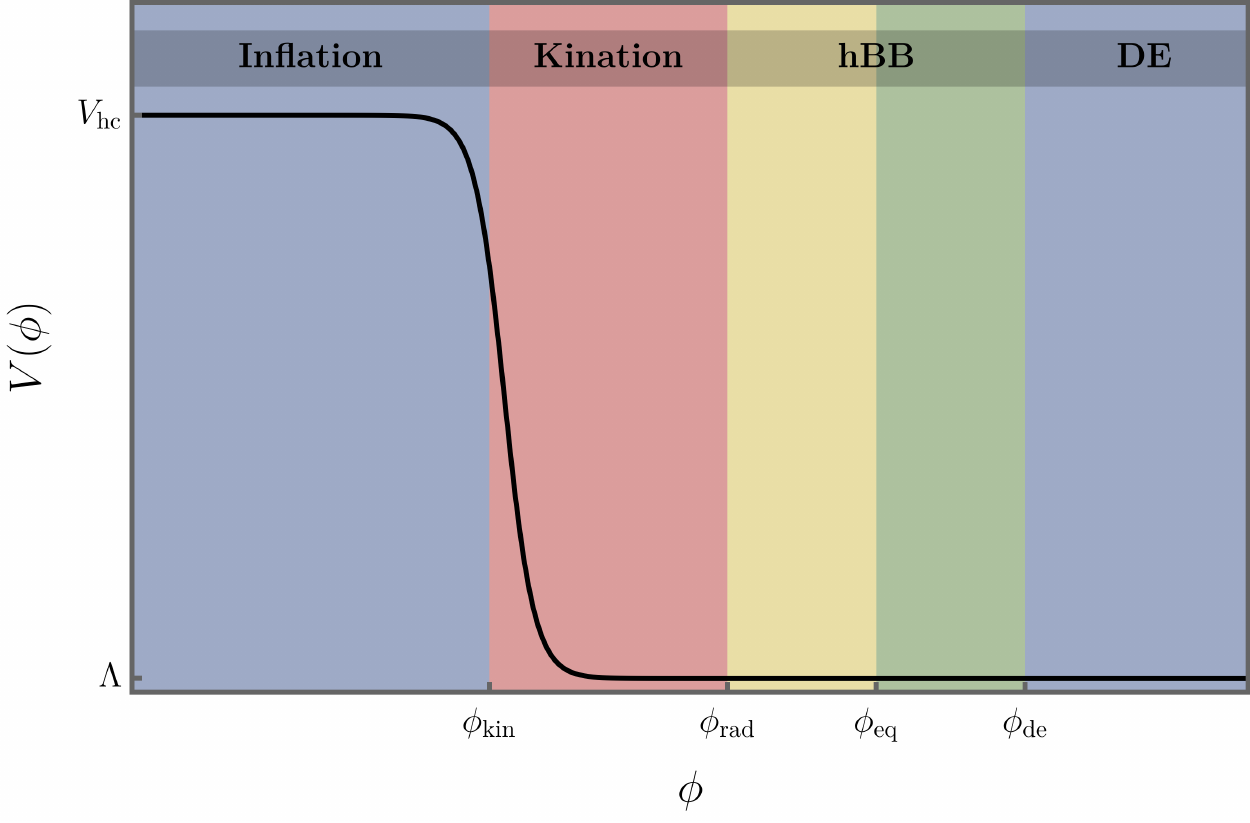}
    \caption{Schematic illustration of the quintessential inflation potential in canonically-normalized scenarios. The presence of two flat or slow-roll regions connected by a steep transition is a priori sufficient to generate two periods of exponential expansion separated by a stage of kinetic domination. Provided a sufficiently efficient creation of particles at the end of inflation, this setting allows also for the standard hot big bang evolution (radiation and matter domination).}
    \label{fig:potential}
\end{figure}

\begin{figure}
    \centering
    \includegraphics[scale=0.19]{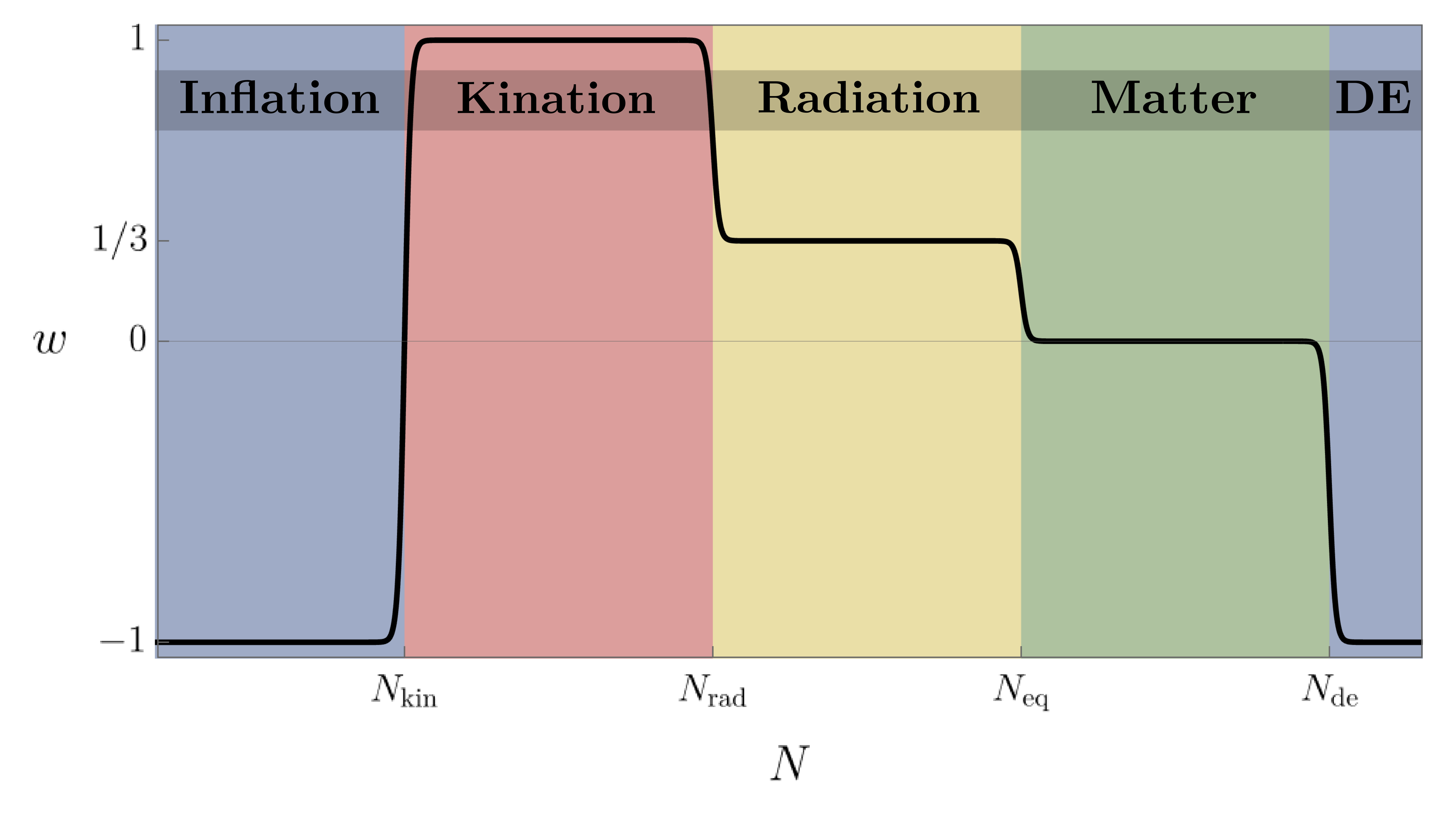}
    \caption{Cosmological evolution of the background equation of state in quintessential inflation scenarios. The lack of a potential minimum translates generically into the appearance of a kinetic dominated era immediately after the end of inflation ($w\simeq-1$), during which the global equation of state parameter is stiff ($w\simeq 1$). This unusual expansion history extends till the onset of radiation ($w=1/3$) and matter domination ($w=0$),  where the cosmon field remains energetically subdominant. Eventually, it starts dominating again, giving rise to the current accelerated expansion of the Universe ($w\rightarrow -1$).
    }
    \label{fig:w_evol}
\end{figure}

\begin{figure}
    \centering
    \includegraphics[scale=0.19]{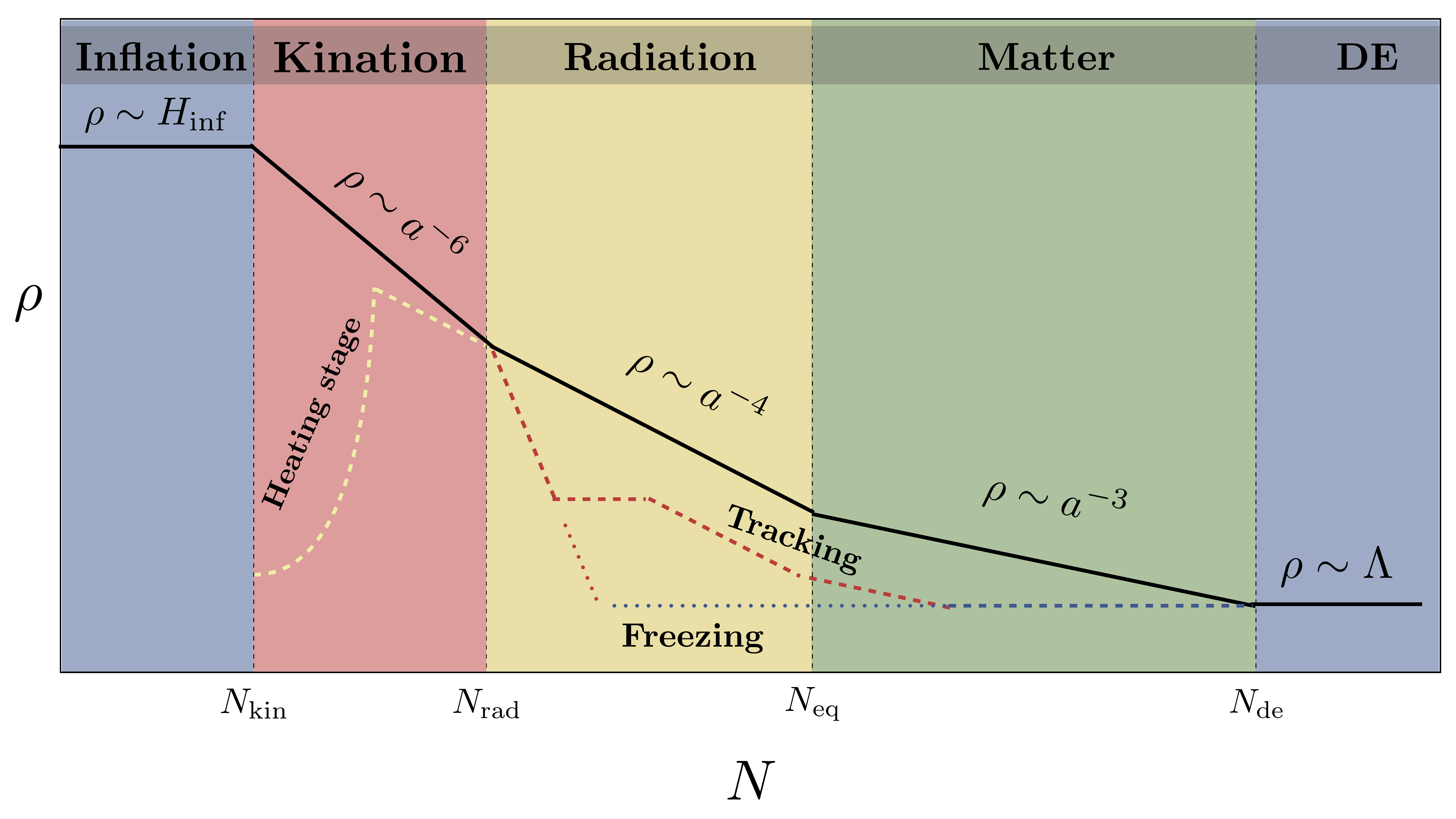}
    \caption{Schematic description of the cosmological evolution of the energy densities in quintessential inflation scenarios. The black, solid line corresponds to the dominant component density. The dashed lines correspond to subdominant radiation (yellow) and cosmon field (purple).}
    \label{fig:rho_evol}
\end{figure}

On a flat Friedman-Lema\^itre-Robertson-Walker (FLRW) Universe with diagonal metric $g_{\mu\nu}={\rm diag}(-1,a^2(t)\delta_{ij})$ and scale factor $a(t)$, the Friedman and Klein-Gordon equations following from the variation of the action \eqref{actioncosmon} with respect to $g_{\mu\nu}$ and $\phi$ can be written as 
\begin{equation}\label{eq:FriedmanandKG} 
H^2=\frac{1}{3M_P^2}\sum_{i} \rho_i  \,, \hspace{6mm}  \dot H=-\frac{1}{2 M_P^2} \sum_i (\rho_i +p_i)\,,\hspace{6mm}   \ddot \phi +3H\dot \phi-a^{-2}\nabla^2 \phi  +V_{,\phi}=Q(\phi,\chi,A_\mu,\psi)\,,
\end{equation}
with the sum running over the cosmon field and all matter species in $S_{\rm I}$ and the $Q$ term accounting for the potential interactions among these components.
Here $H\equiv \dot a/a$ is the Hubble expansion rate, the dots denote derivatives with respect to the coordinate time $t$, $\rho_i$ and $p_i$ stand for the energy densities and pressures of each component in $S$ and the subscript $\phi$ indicates a derivative with respect to the cosmon field.
Assuming the contribution of all matter fields in $S_{\rm I}$ to be or become subdominant at very early and very late times, the dynamics of the system  of equations in \eqref{eq:FriedmanandKG} becomes controlled by the energy density and pressure of the cosmon field at those epochs,
\begin{equation}\label{rho&p}
\rho_\phi=\frac12 \dot{\phi}^2+\frac{1}{2a^2}\vert \nabla \phi \vert^2 +V\,,\hspace{10mm} \textrm{and} \hspace{10mm} p_\phi=\frac12 \dot{\phi}^2-\frac{1}{6 a^2}\vert \nabla \phi \vert^2  -V\,,
\end{equation}
or equivalently by its energy density and the equation-of-state parameter
\begin{equation}
w_{\phi}\equiv \frac{p_\phi}{\rho_\phi}\,.
\end{equation}
For a homogeneous scalar field, the latest quantity interpolates between a de Sitter value $w_{\phi}\simeq -1$ for potential domination, $V\gg \dot \phi^2$, and a stiff value $w_{\phi}\simeq +1$ for kinetic domination, $\dot \phi^2 \gg V$. Thus, as illustrated in Fig.~\ref{fig:potential}, the presence in the cosmon potential of two flat or slow-roll regions connected by a steep transition is a priori sufficient to generate two periods of exponential expansion at very different energy scales separated by a stage of kinetic domination. Of course, the wealth of available observational data ranging from CMB observations \cite{Planck:2018jri,Planck:2018vyg} to BBN constraints \cite{ParticleDataGroup:2020ssz} places several requirements in this model building process \cite{Chiba:2012cb,Geng:2015fla,Geng:2017mic,Durrive:2018quo}. For instance, a phenomenologically successful scenario should be able to generate the correct amplitude and tilt for the primordial power spectrum of density fluctuations, as well as to incorporate a stage of relativistic particle production giving rise to the onset of the hBB era. On top of that, this ambitious program requires the cosmon field to remain quiet and energetically subdominant for an extended period of time after particle creation, such that the standard radiation and matter dominated epochs have the sufficient duration. Eventually, the cosmon field must also exit its quiescence and start behaving as a quintessence component able to support the present accelerated expansion of the Universe, as depicted in Figs.~\ref{fig:w_evol} and \ref{fig:rho_evol}.  In what follows, we will provide a quick overview of the full cosmological evolution in this type of scenarios, going through the various stages with a minimum amount of details and referring to the extensive literature for more comprehensive analyses.


\subsection{Inflation}\label{subsec:inflation}

The inflationary stage is assumed to start in a region of the Universe in which the cosmon evolution is sufficiently slow and the energy density stored in the potential term dominates the total energy budget, $V\gg 1/2\, \dot \phi^2,\, 1/2 \vert \nabla\phi\vert^2$. As any other single-field scenario, these models benefits from a late-time attractor solution that makes the dynamics in phase space almost insensitive to the specific initial conditions \cite{Salopek:1990jq}. In order to sustain inflation for an extended period of time, it is also necessary that the acceleration experienced by the cosmon  is sufficiently small as compared with its velocity per Hubble time, $\vert \ddot{\phi}\vert \ll H\vert \dot\phi\vert$. Under these conditions, the first Friedman and Klein-Gordon equations in \eqref{eq:FriedmanandKG} become respectively
\begin{equation}\label{eq:FriedmannSR}
3 M_P ^2 H^2 \simeq V\,, \hspace{20mm}
-V_{,\phi} \simeq 3H\dot{\phi}\,.
\end{equation}
This slow-roll regime can be supported by a variety of potential shapes around the region of interest, including power-law forms \cite{Peebles:1998qn,Peloso:1999dm,Rubio:2017gty}, exponential profiles \cite{Geng:2017mic} and plateau-like behaviours \cite{Dimopoulos:2017zvq,Akrami:2017cir}. The main requirement is that the so-called slow-roll parameters
\begin{equation}\label{eq:SRcond}
    \epsilon \equiv \frac{M_P^2}{2}\left(\frac{V_{,\phi}}{V}\right)^2 \,,\hspace{20mm} \eta\equiv M^2_P\frac{V_{,\phi\phi}}{V}\,,
\end{equation}
encoding the local slope and curvature of the potential remain small for an extended period of time.  More precisely, the condition $\epsilon\ll 1$ is required for the accelerated expansion to take place in the first place, while the requirement $\vert \eta\vert \ll 1$ ensures that it lasts long enough to solve the flatness and horizon problems. In the limit $\epsilon\approx 0$, the rate of expansion is approximately constant and the scale factor grows exponentially with time, $a\propto e^{Ht}$, leading to an almost de Sitter expansion. 
During the inflationary stage, scalar (s) and tensor (t) perturbations are created out of quantum vacuum fluctuations and spatially stretched to  superhorizon scales \cite{Mukhanov:1981xt, Guth:1982ec, Starobinsky:1982ee, Hawking:1982cz, Bardeen:1983qw}, where they become classical and inherit almost scale-invariant spectra \cite{Mukhanov:1981xt, Guth:1982ec, Starobinsky:1982ee, Hawking:1982cz, Bardeen:1983qw,Lesgourgues:1996jc, Polarski:1995jg,Kiefer:1998qe,Kiefer:1998pb,Kiefer:1998jk}. At the leading order in the slow-roll parameters, these spectra take the parametric form
\begin{equation}\label{PsPt}
{\cal P}_s = {\cal A}_s \left(\frac{k}{k_{\rm hc}}\right)^{n_s-1}\,,  \hspace{10mm}    
{\cal P}_t = {\cal A}_t \left(\frac{k}{k_{\rm hc}}\right)^{n_t}\,,
\end{equation}
with the corresponding amplitudes and tilts,
\begin{eqnarray}
 \mathcal A_{\,s} &=& \left. \frac{1}{2 M_P^2 \,\epsilon} \left(\frac{H}{2\pi}\right)^2 \right\vert_{\rm hc}\,, \hspace{20mm}  n_s=1-6\epsilon_{\rm hc} +2\eta_{\rm hc}\,, \label{eq:As} \\ 
  \mathcal A_{\,t} &=&\left. \frac{8}{M_P^2}\left(\frac{H}{2\pi}\right)^2\right\vert_{\rm hc}\,,
\hspace{25mm}   n_t=-2\epsilon_{\rm hc}\,, 
\end{eqnarray}
depending only on the potential term and its derivatives through the relation $H^2\simeq V/(3 M_P ^2)$, and the subscript hc indicating the evaluation of the corresponding quantities at the horizon exit of a given reference scale $k^{-1}_{\rm hc}=(a_{\rm hc} H_{\rm hc})^{-1}$. Note that both ${\cal P}_s$ and ${\cal P}_t$ are time-independent as long as the associated fluctuations are outside the horizon, meaning that their amplitudes at horizon re-entry, and therefore the initial conditions for the subsequent causal evolution, are the same as those at the first horizon crossing. The ratio of the spectral amplitudes defines the so-called tensor-to-scalar ratio 
\begin{equation}\label{eq:r}
 r \equiv \left.  \frac{\mathcal A_{\,t}}{{\mathcal A}_s}\right\vert_{\rm hc} =16\epsilon_{\rm hc}=- 8 n_t\,.
\end{equation}
The last equality can be understood as a \textit{consistency relation} for single field inflationary models. In particular, since we are working in Einstein gravity and the cosmon field satisfies the null energy condition, the Hubble rate decreases with time, leading to a red-tilted GW spectrum with negative tensor spectral tilt.

The horizon crossing of a given momentum scale can be related to the number of $e$-folds of inflation obtained by integrating the second equation in \eqref{eq:FriedmannSR}, namely
\begin{equation}
    N \equiv \ln\frac{a_{\rm end}}{a}\simeq \frac{1}{M_P}\int_{\phi_{\rm end}}^\phi \frac{d\phi}{\sqrt{2\epsilon}}\,,
\end{equation}
with $\phi_{\rm end}$ the value of the cosmon field following from the violation of the slow-roll conditions, ${\rm max}[\epsilon(\phi_{\rm end}), \vert \eta(\phi_{\rm end} ) \vert] = 1$. In particular, for a scale $k^{-1}_{\rm hc}=(a_{\rm hc} H_{\rm hc})^{-1}$ reentering the horizon at the present cosmological epoch ($a_{\rm hc} H_{\rm hc}=a_0 H_0$), we have 
\begin{equation}\label{eq:Nneed}
N_{\rm hc}=-\ln  \left(\frac{H_{\rm end}}{H_{\rm hc}}\right)+\ln \left( \frac{a_{\rm end} H_{\rm end}}{a_0 H_0}\right)    \,.
\end{equation}
The first term in this expression depends implicitly on the specific cosmon potential and therefore on the overall energy scale of inflation. The second one accounts for the complete post-inflationary evolution and, in particular, for any nonstandard expansion period, such as kinetic domination. 

Recent observational constraints on the inflationary observables at a pivot scale
$k_{\rm hc} = 0.05\, {\rm Mpc}^{-1}$ are provided by the Planck and  BICEP/Keck collaborations \cite{Planck:2018jri,BICEP:2021xfz},
\begin{eqnarray}
{\cal A}_s &=(2.099 \pm 0.014)\times 10^{-9}  \hspace{10mm} &\textrm{at 68\% CL}\,, \label{Asmeasured} 
\\
n_s&=0.9649 \pm 0.0042  \hspace{17mm} &\textrm{at 68\% CL}\,, 
\\
r &<  0.036 \hspace{34mm} &\textrm{at 95\% CL}\,. \label{eq_rlimit}
\end{eqnarray}
Note that while the deviation of the scalar spectral tilt from unity is statistically significant, the precise value of the tensor-to-scalar ratio remains unknown, and with it the Hubble rate and the energy scale of inflation at the pivot scale. In particular, combining Eqs.~\eqref{eq:As} and \eqref{eq:r} with the observed  amplitude \eqref{Asmeasured}, we get 
\begin{eqnarray} \label{eq:max_inf_scale}
H_{\rm hc}&=&  7.9\times 10^{12}\,{\rm GeV} \left(\frac{r}{0.001}\right)^{1/2}\lesssim  4.7 \times 10^{13}~\rm GeV\,,  \\
V_{\rm hc}^{1/4} &=&5.8 \times 10^{15}\,{\rm GeV} \left(\frac{r}{0.001}\right)^{1/4}  \lesssim  1.4 \times 10^{16}~\rm GeV\,,
\end{eqnarray}
with the final bounds following directly from Eq.~\eqref{eq_rlimit}. The combination of this upper limit with the consistency relation \eqref{eq:r} implies also that the tensor spectrum at CMB scales is almost exactly scale invariant, $-n_t\leq 0.0045 \ll 1$, a condition that we will assume in what follows. 

\subsection{Kination}\label{subsec:kination}

After the end of inflation, the cosmon field acquires a large kinetic energy density at the expenses of its potential counterpart, which rapidly becomes subdominant and irrelevant for the subsequent dynamics. This gives rise to a kinetic dominated era dubbed \textit{deflation} \cite{Spokoiny:1993kt} or \textit{kination} \cite{Joyce:1996cp} where the global equation of state of the Universe approaches the maximum value allowed by causality, $w=1$. During this unusual cosmological epoch, the Ricci scalar is negative [cf.~Eq.~\eqref{eq:Rw}] and the first Friedmann and Klein-Gordon equations in \eqref{eq:FriedmanandKG} take the approximate form 
\begin{equation}\label{eq:FRkin}
3 M_P^2H^2 \simeq \frac{\dot \phi^2}{2}\,,   \hspace{20mm}  \ddot\phi +3H\dot\phi\simeq 0\,.
\end{equation}
This system of equations admits a solution 
\begin{equation}\label{eq:Hofa_kin}
\phi-\phi_{\rm kin}=\sqrt{6}\, M_P \, \Delta N_{\rm kin}\,, \hspace{20mm}     H^2 = H^2_{\rm kin}\left(\frac{a_{\rm kin}}{a}\right)^6\,,
\end{equation}
with 
\begin{equation}
\Delta N_{\rm kin} \equiv \log \left(\frac{a}{a_{\rm kin}}\right)
\end{equation}
the number of $e$-folds of kination. The quantitative differences among the times, field values and energy scales in this expression (``kin'') and those at the end of inflation (``end'') depend on the specific model under consideration. However, for most practical purposes, they can be considered small and safely neglected. This allows us to identify the end of inflation with the onset of kination, i.e., $a_{\rm kin}=a_{\rm end}$,  $\phi_{\rm kin}=\phi_{\rm end}$, $H_{\rm kin}=H_{\rm end}$. 

The scaling of the Hubble rate in \eqref{eq:Hofa_kin} affects a plethora of particle physics processes in the very early Universe, including, for instance, the generation of the matter-antimatter asymmetry \cite{Joyce:1996cp,Joyce:1997fc,Bettoni:2018utf} or the abundance of dark matter relics \cite{Kamionkowski:1990ni,Salati:2002md, Profumo:2003hq, Chung:2007vz, Visinelli:2009kt, Redmond:2017tja, DEramo:2017gpl, DEramo:2017ecx, Visinelli:2017qga,Bernal:2020bfj}. Additionally, it enhances any stochastic background of gravitational waves of primordial origin, such as those generated by inflation \cite{Giovannini:1998bp,Giovannini:1999bh,Tashiro:2003qp,Rubio:2017gty, Caprini:2018mtu,Figueroa:2018twl, Figueroa:2019paj, Bernal:2019lpc}, preheating \cite{Khlebnikov:1997di,Easther:2006gt,Easther:2006vd,Dufaux:2007pt,Garcia-Bellido:2007nns,Garcia-Bellido:2007fiu}, cosmic strings \cite{Cui:2017ufi,Cui:2018rwi,Bettoni:2018pbl, Chang:2019mza,Gouttenoire:2019rtn,Chang:2021afa} or phase transitions \cite{Chung:2010cb}; see Ref.~\cite{Allahverdi:2020bys} for a review. For the specific case of inflationary perturbations, a global stiff equation of state translates into an explicit breaking of scale invariance for modes reentering the horizon during kinetic domination. To see this, let us consider the present density parameter for GWs, customarily defined as the current GW energy-density  $\rho_{\rm GW}(\tau_0,k)$ per logarithmic interval of comoving momentum $k$ normalized to the critical density $\rho_{\rm crit}=3M_P^2H^2$,  \cite{Giovannini:1998bp,Giovannini:1999bh,Rubio:2017gty,Caprini:2018mtu}, i.e.
\begin{equation}
 \Omega_{\rm GW}^{(0)} (k)\equiv\dfrac{1}{\rho_{\text{crit}}}\dfrac{d\rho_{\rm GW}(\tau_0,k)}{d\ln k} = \frac{1}{12}\left(\frac{k}{a_0 H_0}\right)^2 \Delta_h^2(\tau_0,k)\,,
\end{equation}
with 
\begin{equation}
\Delta_h^2(\tau_0,k)\equiv \frac{2k^3}{\pi^2} \left<|h_k(\tau_0)|^2\right>
\end{equation}
the ensemble-averaged spectrum of tensor perturbations $h_{k}$ at present time.
These modes follow themselves from a Fourier decomposition
\begin{equation}
h_{ij}(\tau,\mathbf{x})=\sum_{\lambda}\int \frac{d^3\textbf{k}}{(2\pi)^3} e^{i\mathbf{k}.\mathbf{x}} \epsilon_{ij}^\lambda(\mathbf{k})h^\lambda_{\mathbf{k}}(\tau)\,,
\end{equation}
with $h_{ij}$ a transverse-traceless metric perturbation  ($\partial_j h_{ij} = h^i_i = 0$) entering the linearized Einstein equations
\begin{equation}
h_{ij}^{\prime\prime}+2\frac{a^{\prime}}{a}h_{ij}^{\prime}-\nabla^2 h_{ij} = 0\,.
\end{equation}
Here  $\lambda = +, \times$ stand for the two physical graviton polarization states and $\epsilon_{ij}^\lambda(\mathbf{k})$ denotes a basis of polarization tensors, with 
$\epsilon_{ij}^\lambda(\mathbf{k})=\epsilon_{ji}^\lambda(\mathbf{k})$, 
$\epsilon_{ii}^\lambda(\mathbf{k})=0$,
$k_i\epsilon_{ij}^\lambda(\mathbf{k})=0$,  $\epsilon_{ij}^\lambda(\mathbf{k})=\epsilon_{ij}^{\lambda^*}(-\mathbf{k})$, 
$\epsilon_{ij}^\lambda(\mathbf{k})\epsilon_{ij}^{\sigma^*}(\mathbf{k})=2\delta^{\lambda\sigma}$. Assuming the tensor spectrum to be completely unpolarized and neglecting its small tilt $n_t\ll 1$ \cite{BICEP:2021xfz}, the present GW energy density parameter in a Universe undergoing an epoch of kination between inflation and radiation domination takes the form
\cite{Figueroa:2019paj}
\begin{eqnarray}\label{GWkination}
\Omega_{\text{GW}}^{(0)}(k) \simeq \Omega_{\rm GW}^{(0)}{\Big |}_{\rm no-kin}
\times\left\lbrace
\begin{array}{crl}
1 & \,, & \hspace{5mm} k \ll k_{\rm rad}\,, \vspace*{0.4cm}\\
\frac{4}{\pi}\,\left(\frac{k}{k_{\rm rad}}\right)
& \,, & \hspace{5mm} k \gg k_{\rm rad}\,, \\
\end{array}
\right.
\end{eqnarray}
with
\begin{equation}
 \Omega_{\rm GW}^{(0)}{\Big |}_{\rm no-kin} \simeq ~5 \cdot10^{-22} \left( \frac{H_{\rm kin}}{ 10^{11} \, {\rm GeV}} \right)^2
\end{equation}
the corresponding scale-invariant amplitude in the absence of kination and $k_{\rm rad}$ the momentum associated with the horizon scale at the onset of radiation domination.

The linear raising of \eqref{GWkination} at high-momenta serves as an observational signature of quintessential inflation. At the same time, it limits the corresponding parameter space since the energy density of this relativistic component affects directly the expansion rate of the Universe and with it the abundance of light elements. In particular, BBN observations impose a limit on the integrated GW contribution to the total energy density \cite{Maggiore:1999vm,Caprini:2018mtu}, namely
\begin{equation}\label{GWbounds}
h^2 \int_{k_{\text{BBN}}}^{k_{\rm kin}}\Omega^{(0)}_{\text{GW}}(k)d\ln k  \leq 5.6 \times 10^{-6} \, 
\Delta N_{\nu} \simeq 1.12\times 10^{-6}\,, 
\end{equation}
with $h=0.678$ the reduced Hubble constant, $k_{\rm kin}$ and $k_{\rm BBN}$ the momenta associated to the horizon scale at the onset of kination and BBN, and $\Delta N_{\nu}\lesssim 0.2$ an effective parameter encoding any extra amount of radiation beyond the Standard Model content~\cite{Mangano:2005cc,Cyburt:2015mya}. This bound applies to all sources of primordial gravitational waves and in particular to the specific inflationary background in Eq.~\eqref{GWkination}. Taking into account the power law behaviour of this quantity, the integral in Eq.~\eqref{GWbounds} can be approximated by its upper end, getting a simpler expression $h^2\Omega^{(0)}_{\text{GW}}(k\gg k_{\rm rad}) \lesssim 1.12\times 10^{-6}$ that tends to exclude models with large inflationary scales and extended periods of kination.

\subsection{Heating} \label{subsec:std_heating}

The measured abundance of light elements indicates that our Universe was radiation dominated during BBN \cite{ParticleDataGroup:2020ssz}, thus setting a limit on the duration of the kinetic dominated era discussed in the previous section \cite{Gouttenoire:2021jhk},
\begin{equation}
\Delta N_{\rm kin}\lesssim 27 +\frac13\ln\left(\frac{H_{\rm kin}}{10^{11} {\rm GeV}}\right) \,.
\end{equation}
The end of this peculiar cosmological epoch will take place when the Universe becomes dominated by the energy density of a relativistic component, created potentially through the so-far irrelevant interaction terms in Eq.~\eqref{actioncosmon}. Note, however, that the generation of a thermal bath in quintessential inflation cannot proceed as in the usual oscillatory models of inflation (for a review, see e.g. Refs.~\cite{Bassett:2005xm,Allahverdi:2010xz}). 
First, the runaway character of the potential prevents the standard amplification of bosonic fields via parametric resonance \cite{Kofman:1997yn}. Second, the cosmon field cannot directly interact with the Standard Model particles since that would induce the decay and destabilization of the scalar potential tail. Hence, a successful heating mechanism must proceed either through an indirect production of particles or involve a set of sufficiently efficient couplings to an auxiliary matter sector quickly shutting down the interaction and preventing the complete decay of the inflaton condensate. In either case, the cosmon energy density  must dilute faster than the produced species or their decay products, in order to allow for the onset of a radiation dominated epoch. 

The background energy density during kination is initially dominated by the cosmon field, which decays as $\rho_\phi \sim a^{-6}$.  This rapid scaling translates into a relative amplification of any radiation component $\rho_r \sim a^{2} \rho_\phi$, making it eventually dominate the background evolution \cite{Ford:1986sy}. 
A convenient way of quantifying the duration of this stage while being agnostic on the particular model of heating is to introduce the concept of \textit{heating efficiency} \cite{Rubio:2017gty}. The simplest definition of this quantity involves the ratio between the energy density of the relativistic particles produced at the onset of kination and the energy density of the cosmon field at that time,
\begin{equation}
\label{eq:heateff}
    \Theta \equiv \frac{\rho_r^{\rm kin}}{\rho_\phi^{\rm kin}} = \left(\frac{a_{\rm kin}}{a_{\rm rad}}\right)^2\,.
\end{equation}
Albeit this definition assumes instant heating, the concept of \textit{heating efficiency} can be easily extended to non-instantaneous settings \cite{Rubio:2017gty}. The important point to keep in mind, however, is that the detailed description of the process is not needed in practice. In other words, since the heating stage in quintessential inflation requires almost compulsory an external sector with new parameters, the introduction of a single one compressing that information is a convenient choice. For example,
taking into account the value of the pivot scale $k_{\rm hc}=0.002 \, {\rm Mpc}^{-1}=1.27\times 10^{-32}$ eV used in the derivation of  the \textit{combined} 
Planck/BICEP2 results \cite{Planck:2015sxf}, the number of relativistic degrees of freedom at matter-radiation equality $g_{\rm mat}=3.36$ and the present value of the radiation temperature $T_0\simeq 2.73\, {\rm K}\simeq 2.35\times 10^{-4}$ eV, we can rewrite the required number of e-folds of inflation in terms of the heating efficiency \cite{Rubio:2017gty} 
\begin{equation}
  N\simeq 63.3 +\frac14\ln\left(\frac{r}{0.001}\right)-\frac14\ln\left(\frac{\Theta}{10^{-8}}\right)\,.
\end{equation}
Note that since comoving modes re-approach the Hubble scale faster during kination than during radiation or matter domination, the required number of $e$-folds needed to solve the flatness and horizon problems increases for smaller heating efficiencies. Assuming a fiducial value $H_{\rm kin}\sim 10^{11}$ GeV and taking into account that the heating efficiency per degree of freedom can vary between $\Theta\sim 10^{-19}$ and $\Theta \sim \mathcal{O}(1)$ depending on the specific heating mechanism \cite{Rubio:2017gty} (see below), the minimal number of $e$-folds in quintessential inflation scenarios is typically in the range $60 \lesssim N\lesssim 70$. Part of this window is, however, excluded by Big Bang Nucleosynthesis constraints. In particular,  the integrated bound in Eq.~\eqref{GWbounds} translates into a lower limit \cite{Rubio:2017gty}
\begin{equation}
    \Theta \gtrsim 10^{-16}\left(\frac{H_{\rm kin}}{10^{11} {\rm GeV}}\right)^2\,.
\end{equation}

Several heating mechanisms based on the exponential amplification of quantum fluctuations via some kind of instability have been put forward over the years \cite{Ford:1986sy,Spokoiny:1993kt,Chun:2009yu, Felder:1998vq,Felder:1999pv,Campos:2002yk,Dimopoulos:2017tud,Feng:2002nb,BuenoSanchez:2007jxm,Dimopoulos:2018wfg,Opferkuch:2019zbd,Bettoni:2021zhq,Dimopoulos:2019ogl,Karciauskas:2021fdu}. On general grounds, they can be classified according to the presence or absence of direct couplings to auxiliary matter sectors and to whether they produce particles via adiabaticity violation or tachyonic instability. To illustrate this explicitly, let us consider the following action for an exemplary and energetically-subdominant  scalar field $\chi$~
\begin{equation}\label{eq:exaamplechi}
    S=\int d^4x\sqrt{-g}\left[-\frac12 (\partial\chi)^2-\frac12 (M^2(\phi)+\xi R)\chi^2\right]\,.
\end{equation}
In this expression the inflaton field $\phi$ acts as a homogeneous background field, $M(\phi)$ is a field-dependent mass potentially including a bare mass contribution $m_\chi$, $\xi$ is a positive-definite non-minimal coupling and 
\begin{equation}\label{eq:Rw}
    R= 3H^2(1-3w)
\end{equation}
stands for the Ricci scalar in a FLRW background, written in terms the global equation of state $w$. Note that we have intentionally restricted ourselves to quadratic interactions in the scalar sector. This is indeed a good approximation for the first stages of kination, where the auxiliary field $\chi$ is close enough to the origin and particle production is most efficient. However, as we will see in Section \ref{subsec:spectator}, higher-order operators can play a crucial role at late times, especially in the presence of internal symmetries.

The evolution of the field $Y$ is more easily described in the conformal variables
\begin{equation}
\label{eq:Ycoord}
    \chi =\chi_*\frac{a_{\rm kin}}{a}Y\,,\quad\quad dz = a_{\rm kin}\chi_* d\tau\,,\quad\quad \vec y = a_{\rm kin}\chi_* \vec x\,,\qquad\vec \kappa = \frac{\vec k}{a_{\rm kin}\chi_*}\,,
\end{equation}
with $d\tau=dt/a(t)$ the conformal time and $\chi_*$ a potentially convenient normalization. Introducing also the Fourier transform
\begin{equation}
    Y(\vec y,z)=\int \frac{d^3 \kappa}{(2\pi)^{3/2}}Y_{\vec \kappa}(z)e^{i\vec\kappa\cdot\vec y}\,,
\end{equation}
the equation of motion following from the variation of Eq.~\eqref{eq:exaamplechi} with respect to $\chi$ can be written as
\begin{equation}
    Y_{\vec \kappa}''+\omega^2_\kappa (z) Y_{\vec \kappa} =0\,, 
\end{equation}
with 
\begin{equation}
\label{eq:freq_gen}
    \omega_\kappa^2(z) = \kappa^2+\frac{m_\chi^2+M^2_{\rm eff}(\phi)}{\chi_*^2}\left(\frac{a}{a_{\rm kin}}\right)^2 -3\mathcal H^2(3w-1)\left(\xi-\frac16\right)
\end{equation}
an effective frequency term, $\kappa=\vert \vec \kappa\vert$ and $\mathcal H=a'(z)/a(z) = a(z)H(z)$ the comoving Hubble rate. 

The quantization of the scalar field $\chi$ proceeds according to the standard procedure \cite{Birrell:1982ix,Mukhanov:2007zz}. The function $Y_{\vec \kappa}$ is promoted to a quantum operator 
\begin{equation}\label{eq:Yquant}
    \hat Y_{\vec \kappa} = \hat a_{\vec \kappa} f_\kappa(z) + \hat a^\dagger_{-\vec \kappa} f^*_\kappa(z) 
\end{equation}
satisfying the equal time commutation relations $[\hat Y_{\vec \kappa}(z),\hat \Pi_{\vec \kappa'}(z)] = i\delta(\vec \kappa+\vec \kappa')$, with $\hat \Pi_{\vec \kappa} = d\hat Y_{\vec \kappa}/dz$. Here, $a_{\vec \kappa}$ and $a^\dagger_{\vec \kappa'}$ are annihilation and creation operators displaying equal-time commutation relations $[a_{\vec \kappa},a^\dagger_{\vec \kappa'}]=\delta({\vec \kappa}-{\vec \kappa'})\,, [a_{\vec \kappa},a_{\vec \kappa'}]=[a^\dagger_{\vec \kappa},a^\dagger_{\vec \kappa'}]=0$ and $f_\kappa$ is a set of isotropic mode functions satisfying 
\begin{equation}\label{eq:feq}
    f_\kappa''+\omega^2_\kappa (z) f_\kappa =0\,, \hspace{20mm} 2\,{\rm Im}(f'_\kappa f^{*}_\kappa)=1\,.
\end{equation}
Note that the choice of the basis $\lbrace f_\kappa(z), f^*_\kappa(z) \rbrace$ is generically non unique, meaning that Eq.~\eqref{eq:Yquant}  should be understood as an implicit definition of the operators $\hat a$ and $\hat a^\dagger$. For constant positive square frequencies, a positive frequency mode  $f_\kappa\propto e^{-i\omega t}$ defines a unique vacuum state  $ \hat a_{\vec \kappa}\vert 0\rangle=0$. For non-constant non-positive square frequencies, the notion of vacuum become ambiguous. In particular, particle production can be achieved either because the adiabaticity condition $\vert \omega_\kappa'\vert/\omega_\kappa^2< 1$ is violated or because the frequency becomes tachyonic for a certain range of $\kappa$ modes, $\omega_\kappa^2 <0$ (i.e. $\omega_\kappa$ is imaginary). Given the complications and extensive phenomenology of the latter scenario, we will postpone its analysis to Section \ref{subsec:spectator}, focusing in what follows on the simplest case $\omega^2_\kappa (z)>0$. 

The amount of produced quanta for positive frequencies can be determined by comparing the instantaneous adiabatic vacua at two different times. Since the equation of motion \eqref{eq:feq} is second order, there must exist a linear \textit{Bogoliubov transformation} between two mode expansions $f_\kappa$ and $h_\kappa$ evaluated \textit{at the same time}, namely
\begin{equation}
    f_\kappa (z)= \alpha_{\kappa} \,h_\kappa(z) + \beta_{\kappa}\, h^*_\kappa(z)\,, 
\end{equation}
with the coefficients $\alpha_\beta$ and  $\beta_\kappa$ satisfying the normalization condition $\vert \alpha_\kappa \vert^2 -\vert \beta_\kappa\vert^2=1 $ following from the second expression in \eqref{eq:feq}. As shown for instance in Ref.~\cite{Mukhanov:2007zz}, the occupation number of a mode $\kappa$ is given by square of the second \textit{Bogoliubov coefficient} $\beta_\kappa$, 
\begin{equation}
 n_\kappa(z)=\left\vert\beta_\kappa(z)\right\vert^2  \,,
\end{equation}
whose expression in terms of the mode expansion functions can be obtained by solving the associated equation with proper ``in" and ``out" vacuum conditions.

Given the occupation numbers $n_\kappa(z)$ one can compute the total number of $Y$ particles and the associated energy density 
\begin{equation}
n(z)= \frac{1}{(2\pi)^3}\int d^3 \vec \kappa \,\, n_\kappa(z)  \,,    \hspace{15mm}
\rho(z)= \frac{1}{(2\pi)^3}\int d^3 \vec \kappa \,\, \omega_\kappa(z) n_\kappa(z)  \,.
\end{equation}
The resulting spectrum is not expected to be thermal. In fact, heating mechanisms tend to amplify IR fluctuations below a given characteristic scale, leading to a spectral shape initially peaked at low momenta. The subsequent evolution of this overoccupied system proceeds typically through a rather slow process of driven and free turbulence \cite{Micha:2002ey,Micha:2004bv,Bettoni:2021zhq},  for which thermal techniques are not generically applicable. Still, it is sometimes convenient to define an \textit{instantaneous temperature or energy scale}
\begin{equation}\label{eq:Tinst}
    T_{\rm r}(z)\equiv \left(\frac{30}{\pi^2 g_*(z)}\,\rho_{\rm r}(z)\right)^{1/4}\,,
\end{equation}
with  $g_*(z)$ the effective number of relativistic degrees of freedom at a given time $z$. The \textit{(re)heating temperature} $T_{r}(z_{\rm rad})$ signaling the onset of radiation domination is correspondingly defined by the condition $\rho_\phi(z_{\rm rad})=\rho_r(z_{rad})$. Assuming exact kination ($w=1$, $R=-6 H^2$, $\rho_\phi\sim a^{-6}$), relativistic decay products ($\rho_r\sim a^{-4}$) and entropy conservation, this quantity can be related to the \textit{maximum instantaneous temperature or energy scale} at the onset of kinetic domination, namely
\begin{equation} \label{Treh0}
    T_r(z_{\rm rad})
    =\left(\frac{g_{*,s}(z_{\rm kin})}{g_{*,s}(z_{\rm rad})}\right)^{1/3}\Theta^{1/2} \,T_r(z_{\rm kin})\,,
\end{equation}
with $g_{*,s}(z_{\rm kin})$ and $g_{*,s}(z_{\rm rad})$ the entropic degrees of freedom at the corresponding times. Accounting for all Standard Model degrees of freedom, $g_*(z_{\rm kin})=g_{*,s}(z_{\rm kin})=g_{*,s}(z_{\rm rad})=106.75$, we have 
\begin{eqnarray}
    T_r(z_{\rm rad})
        &\simeq& 2.7\times 10^8~\GeV\left(\frac{\Theta}{10^{-8}}\right)^\frac34\left(\frac{\Hkin}{10^{11}~\GeV}\right)^\frac12\,,   \label{Treheating} \\
T_r(z_{\rm kin})
        &\simeq& 2.7 \times 10^{12}~\GeV\left(\frac{\Theta}{10^{-8}}\right)^\frac14\left(\frac{\Hkin}{10^{11}~\GeV}\right)^\frac12 \lesssim 6  \times 10^{15}~{\rm GeV} \,\Theta^{1/4}\,, \label{Tkination}
\end{eqnarray}
with  the upper limit in the last expression following from \eqref{eq_rlimit} and the restriction \eqref{eq:max_inf_scale}. 

\paragraph{Specific cases:}
Having presented the generalities of preheating via violations of the adiabaticity condition $\vert \omega_\kappa'\vert/\omega_\kappa^2< 1$, let us discuss some specific mechanisms connected with the choices of functions and parameters in Eq.~\eqref{eq:freq_gen}. As before, we assume exact kinetic domination ($w=1$, $R=-6 H^2$).
\begin{itemize}
\item \textit{Gravitational heating}  \cite{Ford:1986sy,Spokoiny:1993kt,Chun:2009yu}

In this scenario, the production of particles is purely associated to the expansion of the Universe, being essentially triggered by any rapid change in the Ricci scalar $R$ \cite{Ford:1986sy,Damour:1995pd,Giovannini:1998bp}.
This mechanism  applies therefore to non-conformally coupled scalar fields only ($\xi\neq1/6$), since the evolution equations of gauge bosons and chiral fermions in a conformally flat geometry are invariant under Weyl rescalings. 

For a light scalar field ($m_\chi\ll H_{\rm kin}$) in kinetic domination, the effective frequency in Eq.~\eqref{eq:freq_gen} takes the form 
\begin{equation}\label{wgrav}
    \omega_\kappa^2 =
    \begin{cases}
    \kappa^2 + \mathcal H^2\,\quad &{\rm for}\quad \xi=0\,,\\
    \kappa^2 +(1-6\xi)\mathcal H^2\,\quad &{\rm for}\quad  \xi<1/6\,,
    \end{cases}
\end{equation}
with the condition in the last expression ensuring that $\omega_\kappa^2$ is positive definite for all $\kappa$, such that the particle production is only due to adiabaticity violation and the Bogoliubov formalism presented above can be straightforwardly applied. The case with $\xi>1/6$ is analyzed in detail in Section \ref{subsec:spectator}.

During a Hubble time in kinetic domination, the variation of the (relativistic) energy density per spectator field is of the order $\Delta \rho_r\sim \delta \times  (\Delta H)^4\sim (\Delta a)^{-12}$, with  $\delta$ an efficiency parameter potentially depending on the non-minimal coupling $\xi$ \cite{Spokoiny:1993kt,Ford:1986sy} and typically much smaller that one. In consequence, the production of particles is only efficient during the first stages of kination \cite{Rubio:2017gty}, leading to a relatively small heating efficiency \cite{Rubio:2017gty} 
\begin{equation}
\Theta=\frac{\delta \, {\cal N} }{1440\pi^2} \left(\frac{H_{\rm kin}}{M_P}\right)^2 =10^{-19}\,
\delta \, {\cal N}\left(\frac{H_{\rm kin}}{10^{11}{\rm GeV}}\right)^2\,,
\end{equation}
with ${\cal N}$ the number of light \textit{scalar} degrees of freedom.
Despite being quite natural, this gravitational heating mechanism suffers from several drawbacks. In particular, its low efficiency gives rise generically to a long-lasting period of kination, which makes difficult to satisfy the GW bound \eqref{GWbounds} unless a large number of species is introduced, ${\cal N} \gtrsim {\cal O} (10^{2}) \, \delta^{-1}$.
However, this solution can lead to the generation of sizeable isocurvature perturbations or secondary periods of inflation \cite{Felder:1999pv}. Another related possibility is to consider conformally-coupled particles with heavy masses, $m_\chi>H$ eventually decaying into lighter ones to recover the hBB era \cite{Haro:2018jtb,Haro:2018zdb,Salo:2021vdv}. 

\item \textit{Matter preheating} \cite{Felder:1998vq,Felder:1999pv,Campos:2002yk,Wetterich:2014gaa,Rubio:2017gty,Dimopoulos:2017tud,Bernal:2020bfj}

In this type of scenario the auxiliary field $\chi$ is directly coupled to the cosmon via the effective mass function $M^2(\phi)$. A simple choice ensuring decoupling at late times is given for instance by  \cite{Wetterich:2014gaa,Rubio:2017gty,Bernal:2020bfj} 
\begin{equation}\label{Mphiexample}
M^2(\phi)=\begin{cases} 
      g^2 \phi^2 & \quad\text{ for }\phi\leq 0\,, \\
      \tilde M^2 & \quad\text{ for }\phi >  0\,,
   \end{cases}
\end{equation}
with $g$ a dimensionless coupling constant and $\tilde M$ a constant mass parameter. This unconventional behaviour appears naturally in quintessential inflation models constructed on the basis of emergent scale symmetry in the vicinity of UV and IR fixed points~\cite{Wetterich:2014gaa,Rubio:2017gty}, cf.~Section \ref{sec:scalingframe}. Alternative choices smoothing the transition between negative and positive field values or modifying its \textit{timing} could be also considered without significantly changing the discussion below, see e.g. Ref.~\cite{Rubio:2017gty}. 

Given the form \eqref{Mphiexample}, and omitting the subdominant gravitational contribution in Eq.~\eqref{wgrav}, the effective frequency \eqref{eq:freq_gen} takes the form
\begin{equation}\label{wphiexample}
  \omega_\kappa^2 =\begin{cases} 
      \kappa^2 + g^2\left(\frac{a}{a_{\rm kin}}\right)^2\frac{\phi^2(z)}{\chi_*^2} & \quad\text{ for }\phi\leq 0\,, \\
      \kappa^2 + \left(\frac{a}{a_{\rm kin}}\right)^2\frac{\tilde M^2}{\chi_*^2} & \quad\text{ for }\phi >  0\,.
   \end{cases}
\end{equation}
For small $\kappa$ values and $\phi\leq 0$, the condition for the violation of the adiabaticity $\vert \omega_\kappa'\vert/\omega_\kappa^2< 1$  can be safely approximated by $ g|\phi'_0|\gtrsim \phi^2$, with $\vert \phi'_0\vert$ the cosmon velocity at zero crossing. For sufficiently large couplings, particle production takes place in a very narrow interval $\Delta \phi \sim (|\phi'_0|/ g)^{1/2}$ around $\phi=0$, being the process essentially instantaneous, $\Delta z \sim  \phi_*/|\phi'_0| \sim (g |\phi'_0|)^{-1/2}$, and independent of the particle spin. The momentum of the created particles follows directly from the uncertainty principle, $\Delta \kappa \sim (\Delta z)^{-1} \sim  (g |\phi'_0|)^{1/2}$. Provided it to significantly exceed the scale $\tilde M$ at positive cosmon values  $\phi>0$, ($\Delta \kappa  \gg \tilde M$), the produced particles are relativistic and can trigger the onset of radiation domination. As shown in Ref.~\cite{Bernal:2020bfj}, the heating efficiency in this particular example takes the form
\begin{equation}
\Theta\simeq 2 \times 10^{-8} \left(\frac{g}{0.02}\right)^2 \left(\frac{10^{11} \, {\rm GeV}}{H_{\rm kin}}\right)^2\left(\frac{\vert  \phi'_0\vert }{10^{-8} \mpl^2}\right)^2   \,. 
\end{equation} 

Alternative $Z_2$ symmetric-choices for the mass function $M^2(\phi)$ could be also considered \cite{Felder:1998vq,Felder:1999pv,Campos:2002yk,Bernal:2020bfj}.  In this case, the energy transfer happens through a combination of adiabaticity violation at zero crossing and a rapid enhancement of the effective particle masses as the cosmon field evolves towards large positive values. Since in this case the created particles become rapidly non-relativistic, their energy density redshifts as matter, $\rho_\chi\propto a^{-3}$, meaning that, in order to recover the hBB epoch, they must eventually decay into light degrees of freedom. This decay can be induced, for instance, by a direct Yukawa coupling $h\chi \bar\psi\psi$ between the spectator field $\chi$ and some fermionic species $\psi$. Since the mass of $\chi$ grows with $\phi$ after zero crossing, so it does the effective decay rate, $\Gamma \propto M(\phi)$, making this heating mechanism very efficient. 
\end{itemize}

\subsection{Hot big bang and dark energy} \label{subsec:hbb_DE}

Having discussed the production of matter fields coupled to the cosmon, let us consider now the evolution of the system during radiation and matter domination. The relevant equations can be conveniently written in terms of cosmological observables, namely
\cite{Wetterich:2003qb,Scherrer:2007pu,Chiba:2012cb} 
\begin{eqnarray}
&& \hspace{-10mm}\frac{w_\phi'}{1-w_\phi}=-3(1+w_\phi)+\lambda\sqrt{3(1+w_\phi)\Omega_\phi}\label{weq}\,,  \\
&& \hspace{-10mm}\,\Omega'_\phi=-\Omega_\phi (1-\Omega_\phi)\left(3(1+w_\phi)-n \right)\,,\label{Oeq}\\ 
&& \hspace{-10mm}\lambda'=-\sqrt{3(1+w_\phi)\Omega_\phi}(\Gamma-1)\lambda^2\,,\label{Leq}
\end{eqnarray}
with the primes denoting derivatives with respect to the number of $e$-folds and $n=4$ for radiation domination and $n=3$ for matter domination.  The parameters 
\be\label{SRpamDE}
\lambda\equiv -M_{\rm P}\frac{V_{,\phi}}{V}\,, \hspace{20mm} \Gamma\equiv\frac{VV_{,\phi \phi}}{V_{,\phi}^2}\,,
\ee
characterize, respectively, the local slope and curvature of the potential in the corresponding regime. Although, there is not a first principle guide to determine how the above system of equations will evolve during radiation and matter domination or whether it can eventually support an accelerated expansion era,  a dynamical analysis is enough to single out a few options. In fact, depending on the initial conditions and the specific shape of the cosmon potential, we can distinguish three cases \cite{Caldwell:2005tm}:
\begin{enumerate}
\item  \textit{Thawing models}: In this type of scenarios the Hubble friction freezes the cosmon field evolution during radiation and early matter domination, being its energy density completely dominated by the potential term. This corresponds to the fixed point $w_\phi=-1$ in Eq.~\eqref{weq}, which is, however, unstable for $\lambda \neq 0$. In particular, as soon as the Hubble rate becomes smaller than the mass of the scalar field, the cosmon will start rolling down the potential, forcing the effective equation-of-state parameter to increase. The growth of $w_\phi$ at late times can  be analytically computed in some specific limits. In particular, for approximately constant $\lambda$ and $\vert 1+w_\phi\vert\ll 1$, we have  \cite{Wetterich:1987fm,Copeland:1997et,Ferreira:1997hj,Scherrer:2007pu,Casas:2017wjh,Dutta:2008qn,Chiba:2009sj},
\begin{equation}\label{eqn:eos_omega_quintessence}
1+w_\phi= \frac{\lambda^2}{3} F(\Omega_{\phi})\,,  \hspace{20mm}
\Omega_{\phi}=\frac{1}{1+\Delta_{0}\,a^{-3}}\,,
\end{equation}
with
\begin{equation}
F(\Omega_{\phi})=\left[\frac{1}{\sqrt{\Omega_{\phi}}}-\Delta\tanh^{-1} \sqrt{\Omega_{\phi}}\right]^2
\end{equation}
a monotonically increasing function smoothly interpolating between $F(0)=0$ in  the deep radiation and matter dominated eras and $F(1)=1$ in the asymptotic dark energy dominated regime and 
\begin{equation}
\Delta  \equiv \frac{1-\Omega_{\phi}}{\Omega_{\phi}}\,,\hspace{20mm} 
\Delta_0  \equiv \frac{1-\Omega_{\phi,0}}{\Omega_{\phi,0}}\,,
\end{equation}
with the subscript $0$ marking quantities evaluated today. The present value of the dark-energy equation-of-state parameter follows directly from Eq.~\eqref{eqn:eos_omega_quintessence},
\begin{equation}
  \label{eqn:eos_omega_quintessence2}
\frac{1+w_{\phi,0}}{1+w_\phi(a)}= \frac{F(\Omega_{\phi,0})}{F(\Omega_{\phi})}.
\end{equation}
Note that these scenarios require a fine tuning of the local curvature and amplitude of the potential in order to ensure that the scalar field starts evolving at the correct time while reproducing the observed cosmological constant energy density \cite{Scherrer:2007pu}. 

\item \textit{Tracking freezing models}: In this case the cosmon field is initially in slow-roll motion and comes to a halt due the increasing flatness of the potential after a certain critical value. This limit corresponds to a fixed point
\begin{equation}\label{exprel}
\Omega_\phi=\frac{3(1+w_\phi)}{\lambda^2}\,, 
\end{equation}
with constant $w_\phi$. Since this configuration is stable, the scalar field trajectories will be dragged towards it for a large set of initial conditions, meaning that we are dealing with a tracker solution. If then $\Gamma>1$ \cite{Zlatev:1998tr,Steinhardt:1999nw}, for which  $\lambda$ approaches 0, the cosmon energy density will grow to dominate the background evolution.  Using the relation 
\begin{equation}
\frac{\Omega_{\phi}'}{\Omega_{\phi}}=-2\frac{\lambda'}{\lambda}    
\end{equation}
following from Eq.~\eqref{exprel} together with Eqs.~\eqref{Oeq} and \eqref{Leq} in the $\Omega_{\phi} \ll 1$ limit,\footnote{Improved results accounting for $\Omega_\phi$ as a perturbation to the zeroth-order solution can be found in Ref.~\cite{Chiba:2009gg}.}
the effective field equation of state along the tracker is given by 
\begin{equation}
w_\phi= -\frac{2(\Gamma-1)}{2\Gamma-1}\,.
\end{equation}
Note that, while the presence of the attractor alleviates the initial condition problem in this class of models, a certain degree of tuning is still required in order to recover the correct time at which the cosmon field starts to dominate the energy budget. 

\item \textit{Scaling freezing models}: In this scenario the cosmon field reaches as well a tracking solution but this time with  $\Omega_\phi ={\rm const}$.  A simple inspection of Eqs.~\eqref{weq} and \eqref{Oeq} reveals the existence of a stable fixed point at \cite{Wetterich:1987fm,Wetterich:1994bg,Ratra:1987rm,Copeland:1997et}
\begin{equation}\label{eq:tracking}
\Omega_\phi=\frac{n}{\lambda}\,,\hspace{20mm} w_\phi=\frac{n-3}{3}\,, 
\end{equation}
for constant $\lambda$.
These conditions are verified for $\Gamma = 1$ and correspond to an exponential potential
$V(\phi)=V_0 e^{-\lambda \phi/M_{P}}$, with
$V_0$ a constant. Note that in this scenario the cosmon equation-of-state parameter follows that of the background component during the radiation and matter dominated eras, opening up a new line of attack to the coincidence problem. The recovery of the late time acceleration of the Universe requires, however, the introduction of an exit mechanism from this scaling regime.  This can be triggered by several mechanisms such as double exponential potentials displaying two-late time attractors (one associated with the scaling solution and one where the cosmon field dominates the energy density) \cite{Barreiro:1999zs,Copeland:1997et} or growing neutrino masses. The first option does not provide, however, a satisfactory solution to the \textit{``why now"} problem since the amplitudes of the two exponential potentials must be properly chosen to ensure that the transition happens at the appropriate time. This is not the case of the growing neutrino mass scenario,  which we discuss in detail in Section \ref{subsec:GNQ}.   
\end{enumerate}
As pointed out in the Introduction, observations reveal that the dark energy equation-of-state parameter lies in a narrow strip around $w_\phi=-1$. For an extensive review of these constraints and their impact of the above scenarios, we refer the reader to Refs.~\cite{Chiba:2012cb,Wang:2011bi,Tsujikawa:2013fta} and references therein.

\section{Field relativity} \label{sec:fieldrel}

A canonical scalar field $\phi$ with a runaway potential is just one of the many ways of generating an early and a late accelerated expansion of the Universe using a single degree of freedom. A plethora of successful theories can be constructed for instance within a scalar-tensor framework
\begin{equation}\label{action0}
S[g_{\mu\nu}, \varphi,  f(\varphi), k(\varphi), V(\varphi)]=  \int  d^4 x \sqrt{- g}  \, \left[\frac{ f(\varphi)}{2}  R - \frac{ k(\varphi)}{2} \,  g^{\mu\nu }( \nabla_\mu \varphi) ( \nabla_\nu \varphi) - 
V(\varphi) \right]\,,
\end{equation}
with $ f(\varphi)$, $ k(\varphi)$ and $V(\varphi)$ arbitrary functions of a scalar field $\varphi$.  This parametrization leaves outside a large variety of ``safe" theories with second order derivatives of the scalar field, such as Hordenski, beyond Horndeski and Degenerate Higher-Order Scalar-Tensor (DHOST) theories \cite{Langlois:2018dxi}, but it is still able to accommodate several modified gravity scenarios such as $f(R)$ theories \cite{Sotiriou:2008rp} and the low-energy limits of string theories. On top of that, it can be easily extended to incorporate models where the interactions with the matter sector play an important role, such as the coupled quintessence settings discussed in Section \ref{subsec:CQ} or those in Refs.~\cite{Dimopoulos:2018eam,Dimopoulos:2019ogl,Karciauskas:2021fdu}.

By specifying the \textit{theory defining functions} in Eq.~\eqref{action0}, one can construct an \textit{equivalence} class of theories related to each other by a \textit{frame transformation} \cite{Flanagan:2004bz,Jarv:2014hma,Burns:2016ric}, defined this as the combination of a Weyl rescaling of the metric and a field reparametrization
\begin{equation}
\tilde g_{\mu\nu}=\Omega^{2}(\varphi)\, g_{\mu\nu} \,, \hspace{20mm} \left( \frac{d\tilde \varphi}{d\varphi}\right)^2 \equiv K (\varphi)\,.
\end{equation}
In particular, taking into account the transformation of the Ricci scalar under Weyl transformations,
\begin{equation}
\tilde R= \Omega^{-2}\,  R - 6 \, \Omega^{-3}\,  g^{\mu\nu} \nabla_\mu \nabla_\nu \Omega\,,
\end{equation}
it is possible to rewrite the action \eqref{action0} as
\begin{equation}
   \label{actionI}
 S[\tilde g_{\mu\nu}, \tilde \varphi, \tilde f( \tilde \varphi), \tilde k(\tilde \varphi), \widetilde V(\tilde \varphi)]=
 \int d^4 x \, \sqrt{ - \tilde g}\,  
 \left[     
  \frac{  \tilde f(   \tilde\varphi)}{2}  { \tilde R} 
   - \frac{ \tilde k(  \tilde\varphi)}{2} \,  \tilde g^{\mu\nu} 
   (\nabla_\mu  \tilde\varphi)  (\nabla_\nu  \tilde\varphi )
   -  {\widetilde V(   \tilde\varphi)} 
 \right]\, ,
\end{equation}
with 
\begin{eqnarray}
   \label{Imodelparamdef}
 \tilde f (   \tilde \varphi) &=& \Omega^{-2}\, f\, , \hspace{5mm}
 \tilde k(  \tilde \varphi) = \frac{\Omega^{-2}}{K}\, \Big( k
- 6\,  f\, \partial_{\varphi} \ln \Omega^2\:
+  6\,  f_{,\varphi}\, \partial_\varphi \ln \Omega_{,\varphi}  \Big)\, ,\hspace{5mm}
  \tilde V( \tilde \varphi) = \Omega^{-4}\, V\, .\nonumber
\end{eqnarray}
Note that even if the action $S$ is not invariant under Weyl rescalings and field reparametrizations, its functional form remains invariant under these combined transformations, 
\begin{equation}
S[\tilde g_{\mu\nu}, \tilde\varphi, \tilde f(\tilde\varphi), \tilde k(\tilde\varphi),\tilde V(\tilde\varphi)]=  S[g_{\mu\nu}, \varphi, f( \varphi), k( \varphi), V( \varphi)]\,.
\end{equation}
In this sense, frame transformations are somehow analog to coordinates transformations in classical mechanics. Different representations of a theory may give rise to strikingly different pictures (such as expanding Universes with constant particle masses or shrinking Universes with growing particles masses), but they describe, however, the same physical reality. 

In the so-called \textit{kinetial frame}, the gravitational part of the action takes the usual Einstein-Hilbert form, but the cosmon field displays a non-canonical normalization,
\begin{equation}
   \label{actionEgeneral}
 S=
 \int d^4 x \, \sqrt{ - g}\,  
 \left[     
  \frac{M_P^2}{2}  { R} 
   - \frac{k( \varphi)}{2} \, g^{\mu\nu} 
   (\nabla_\mu \varphi)  (\nabla_\nu \varphi )
   -  { V(  \varphi)} 
 \right]\,.
\end{equation}
Given the remaining freedom on the choice of the theory defining functions $k( \varphi)$ and $V(  \varphi)$, we can still distinguish some special cases:
\begin{enumerate}
\item \textit{Exponential basis}. This formulation makes use of a fixed exponential form for the potential
\begin{equation}
V(\varphi)=M^4 \exp\left(-\frac{\varphi}{M}\right)\,,    
\end{equation} 
with $M$ a constant, leaving the detailed model information for the \textit{kinetial} $k(\varphi)$ \cite{Wetterich:2013wza,Wetterich:2013jsa,Hebecker:2000zb,Wetterich:2019qzx}. This choice is particularly useful for model comparison since the relation between the scalar field value and the potential energy density is universal.  On top of that, all frame-covariant inflationary observables in Appendix \ref{app:frame-covariant} become functions of $k(\varphi)$ only \cite{Wetterich:2019qzx}. 
\item \textit{$\alpha$-attractor basis}. This formulation assumes a singular function \cite{Akrami:2017cir,Dimopoulos:2017zvq,Dimopoulos:2017tud,Garcia-Garcia:2018hlc},
\begin{equation}
k(\varphi)=\frac{1}{1-\frac{1}{6\alpha}
\frac{\varphi^2}{M^2_P}}\,,
\end{equation}
with $\vert \varphi\vert <\sqrt{6} M_P$. 
This choice can be motivated either by invoking conformal symmetry or extended supergravity scenarios, where the positive constant $\alpha$ describes the curvature of a hyperbolic K\"ahler manifold. For describing quintessential inflation, the potential in this setting is required to be asymmetric and non-singular, being the linear and exponential choices commonly used in the literature. Remarkably, the associated inflationary predictions are  independent of the specific choice of $V(\varphi)$, since this becomes exponentially stretched around the poles at $\vert \varphi\vert =\sqrt{6} M_P$ when moving to canonically normalized variables. On top of that, for moderate values of $\alpha$, and in contrast with the canonical expectation \eqref{eq:Hofa_kin}, the field excursion during kination is ${\cal O}(M_P)$, as required by swamplamp distance conjecture \cite{Palti:2019pca} (in case you take it seriously, of course). In spite of these positive aspects, $\alpha$-attractor models cannot solve the cosmological constant problem without invoking the inflationary landscape and anthropic considerations.
\item \textit{Flattening basis}. In this basis the theory defining functions are not singular, but satisfy the relations
\begin{equation}
k(\varphi)=\frac{1}{1+f(\varphi)} \,, \hspace{20mm}
V(\varphi)=\frac{M^4 f(\varphi)}{1+f(\varphi)}\,,
\end{equation}
with $M$ a constant with the dimension of mass and $f(\varphi)$ a dimensionless runaway function of the cosmon field. This peculiar structure appears naturally in Palatini scalar-tensor and modified gravity scenarios involving Jordan-frame potentials proportional to the square of the non-minimal coupling to gravity \cite{Kallosh:2013tua} or $R^2$ corrections \cite{Dimopoulos:2020pas}. Given $f(\varphi)$, this setting generates an inflationary plateau for $f(\varphi)\gg 1$, while reducing to the canonical form \eqref{actioncosmon} for $f(\varphi)\ll 1$. 
\end{enumerate}
Note that, while being rather general, the above bases should be understood just as convenient parametrizations for model building, not exhausting, however, all quintessential scenarios proposed in the literature. 

\section{Symmetry principles} \label{sec:scalingframe}

The naturalness of quintessential inflation is intimately related to that of the theory defining functions driving the early- and late-time acceleration of the Universe. In this regard, \textit{variable gravity} scenarios involving the realization of scale or dilatation symmetry constitute an interesting arena for model building \cite{Wetterich:2013jsa,Wetterich:2014gaa,Rubio:2017gty}. In this type of setting, inflation and dark energy take place in the vicinity of UV and IR fixed points extending in physical time to the infinite past and future. At these fixed points, any information about intrinsic mass scales is completely lost and scale invariance is manifestly realized even if broken in the underlying quantum field theory. The anomalous breaking of dilatations plays, however, a key role in the crossover among fixed points, which, depending on the model specifics, can take place in one or several stages and involve physical scales separated by many orders of magnitude, similarly to what happens in quantum chromodynamics, where the confinement scale is naturally much smaller than a given unification scale.

The relation of inflation and dark energy to scale invariance is more naturally discussed in terms of a quantum effective action including all fluctuation effects, since these are indeed needed for the emergence of fixed points. In the lack of a first principles' computation for the Standard Model non-minimally coupled to gravity, several authors have adopted the following parametrization for the graviscalar sector of the theory~\cite{Wetterich:2013jsa,Wetterich:2014gaa,Hossain:2014xha,Rubio:2017gty},
\be\label{actionJ}
S_{c}=  \int  d^4 x \sqrt{- g}  \, \left[
\frac{\varphi^2}{2}  R-\frac{1}{2}\left(B(\varphi/\mu)-6\right) g^{\mu\nu} \partial_\mu\varphi\partial_\nu\varphi -\mu^2\varphi^2\right]\,.
\ee
Here $\varphi$ stands for the cosmon field, $B(\varphi/\mu)$ is a positive dimensionless function of the dimensionless ratio $\varphi/\mu$ and $\mu$ is a mass parameter that can be associated with the dilatation anomaly. The absence of a quartic potential term in this expression implies the asymptotic vanishing of the observable cosmological constant when the theory is transformed to the Einstein frame \cite{Wetterich:1987fm}. This can be motivated by functional renormalization studies showing that the cosmon potential cannot increase as $\varphi^4$ when $\varphi\to \infty$, since this would lead to a singularity in the flow that tends to be avoided by strong gravity-induced renormalization effects \cite{Henz:2013oxa,Wetterich:2017ixo}. Formally, this corresponds to requiring conformal invariance ($B\to 0$) rather than scale symmetry ($B\to \textrm{constant}$)  at the infrared fixed point $\mu\to 0$ \cite{Luty:2012ww,Dymarsky:2013pqa}. In this conformal limit, the cosmon is no longer a propagating degree of freedom, as can be easily seen by transforming Eq.~\eqref{actionJ} to the Einstein frame.

The flow equation for $B(\varphi/\mu)$ encodes the relative change of the spontaneous mass scales proportional to $\varphi$ as compared to the intrinsic mass scales proportional to $\mu$. If all parameters with dimension of mass are either proportional to $\mu$ or $\varphi$, the flow of $\varphi$ at fixed $\mu$ can be related to the flow of $\mu$ at fixed $\varphi$, and viceversa,
\begin{equation}
\varphi\frac{\partial B}{\partial \varphi} \Big\vert_\mu  +\mu\frac{\partial B}{\partial \mu} \Big\vert_\varphi=0\,. 
\end{equation}
The UV fixed point $\mu\to \infty$ corresponds therefore to the $\varphi\to 0$ limit, while the IR fixed $\mu\to 0$ occurs for $\varphi\to \infty$. Due to the divergence of the mass term in Eq.~\eqref{actionJ} as $\mu\to \infty$, the existence of an UV fixed point requires the cosmon wave function renormalization to display an anomalous dimension. The minimal realization of this condition follows from a simple behaviour ~\cite{Wetterich:2014gaa,Rubio:2017gty},
\begin{equation}\label{simpleflow}
\mu\partial_\mu \ln  B =\sigma\,, \hspace{10mm}\longrightarrow \hspace{10mm}B=\left(\frac{m}{\varphi}\right)^\sigma   \,,
\end{equation}
with $\sigma>1$ and $m\equiv \mu\,  \exp\,(c_t)$ a crossover scale following from the dimensional transmutation of an integration constant $c_t$ determining the particular trajectory on the flow. For this specific choice, the effective action \eqref{actionJ} at the UV point is invariant under the simultaneous rescaling $ g_{\mu\nu}\,\to \,  \lambda^2 g_{\mu\nu}$ and $\varphi\, \to\,  \alpha^{-\frac{2}{2-\sigma}}\varphi$, with $\lambda$ a constant. 

Albeit generically subdominant for the cosmological solutions considered here, the graviscalar action \eqref{actionJ} can be complemented with higher-order curvature invariants such as Gauss-Bonnet or $R^2$ terms with coefficients depending on the dimensionless ratio $\varphi/\mu$. These are expected to play an important role at the UV fixed point \cite{Wetterich:2014zta}, where the coefficient of the Ricci scalar identically vanishes. On top of that, one should introduce a Standard Model sector describing the usual matter and radiation components. The main viability requirements are that i) the different renormalized dimensionless couplings (gauge, Yukawa and Higgs self-interactions) become sufficiently close to their IR fixed-point values before big bang 
nucleosynthesis and that ii) all mass parameters in the theory (including the confinement and Fermi scales) become proportional to $\varphi$ by that time \cite{Wetterich:2003qb}, such that the strict bounds on  fifth forces \cite{Adelberger:2009zz} and the temporal variation of fundamental constants \cite{Uzan:2010pm} are trivially satisfied ~\cite{Wetterich:2003qb,Shaposhnikov:2008xb,Garcia-Bellido:2011kqb,Ferreira:2016kxi,Burrage:2018dvt,Casas:2018fum}.  This restriction constitutes a key
difference with Brans-Dicke scenarios and similar scalar-tensor variants such as induced gravity theories \cite{Zee:1978wi}. It does not apply, however, to beyond the Standard Model sectors, where the crossover regime can take place even at the present cosmological epoch. 

\subsection{Ultraviolet regime}\label{subsec:UVregime}

\begin{figure}
	\begin{center}
\includegraphics[width=0.8\textwidth]{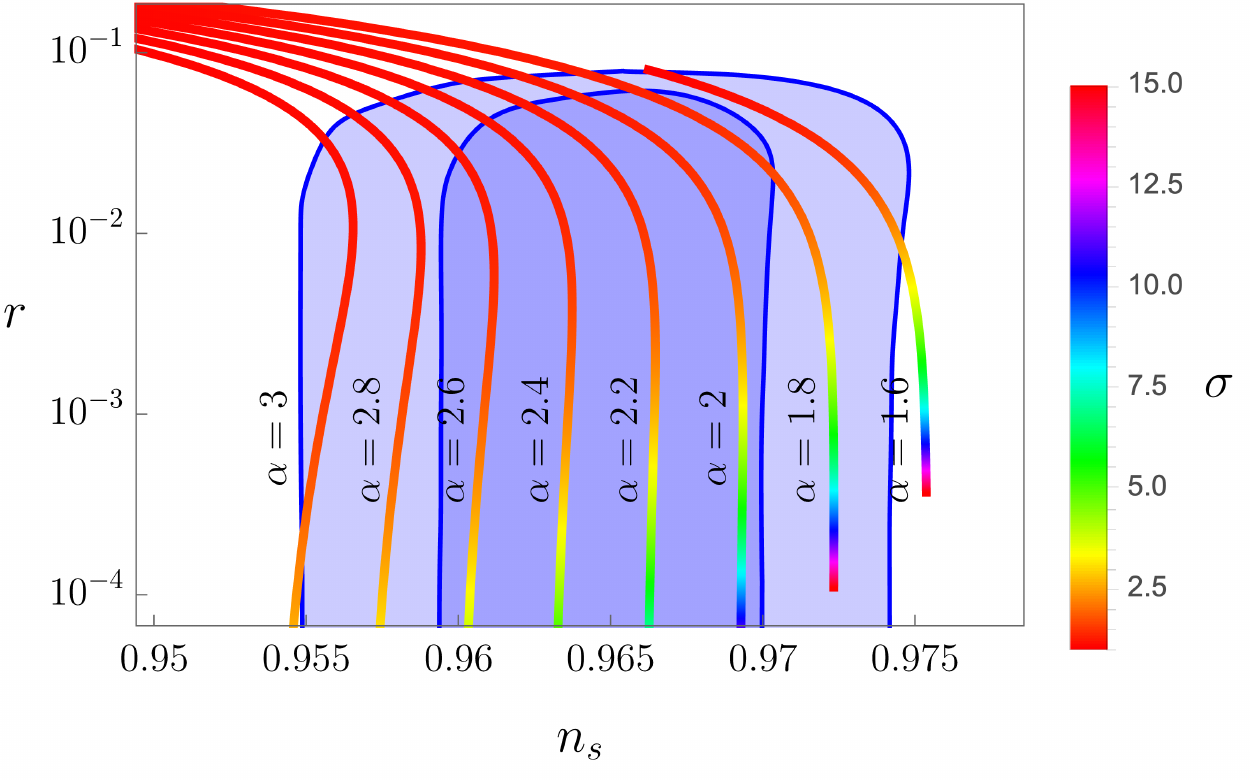}
		\caption{Comparison between the inflationary predictions \eqref{fig:nsrVG} for $N=60$ and the Planck/BICEP2 data at $1\sigma$ and $2\sigma$ C.L.}
		\label{fig:nsr}
	\end{center}
\end{figure}

The field equations for the graviscalar action \eqref{actionJ} admit an exact de Sitter solution able to support an accelerated expansion era at very early times \cite{Wetterich:2014gaa},
\begin{equation}
R_{\mu\nu}=\mu^2 g_{\mu\nu}\,, \hspace{20mm} \varphi=0\,.
\end{equation}
However, the anomalous violation of dilatations in the vicinity of this UV fixed point makes this situation unstable, forcing the cosmon field to evolve slowly with time while providing a graceful exit from the inflationary state. Once a specific form of $B$ is assumed, any subsequent analysis can be performed by using the “classical field equations” obtained by varying the quantum effective action \eqref{actionJ}. In particular, the inflationary observables following from this expression can be straightforwardly computed using the frame-covariant formalism in Appendix \ref{app:frame-covariant}, with  $f(\varphi)=\varphi^2$, $k(\varphi)\equiv B(\varphi/\mu)-6$ and $V(\varphi)=\mu^2\varphi^2$. The spectral tilt and the tensor-to-scalar ration can be then written in a rather suggestive way,
\begin{equation}\label{rq:nsr}
1-n_s= \frac{r}{8}\left(1+\frac12 \, \beta(\mu)\right)\,, \hspace{20mm} r = \frac{32}{B(\mu)} \,,
\end{equation}
with
\begin{equation}\label{eq:BNmaster}
\beta(\mu)\equiv \mu \, \partial_\mu \ln B=\frac12 \frac{\partial B}{\partial N} 
\end{equation}
the effective $\beta$ function for $B$, written in terms of the scale $\mu$ and the number of $e$-folds of inflation, 
\begin{equation}
N (\varphi)=\frac{1}{2} \int_\varphi^{\rm \varphi_{\rm end}} \frac{d\varphi'}{\varphi'} \, B(\varphi') \,,
\end{equation}
with $\varphi_\text{end}$ the value of $\varphi$ at the end of inflation. 
The emergent dilatation symmetry at the UV fixed point is responsible for both the flatness and amplitude of the primordial spectrum of density fluctuations. A explicit illustration of this can be found in Refs.~\cite{Wetterich:2014gaa,Rubio:2017gty}, where the fixed-point structure \eqref{simpleflow} was shown to generate a spectral amplitude ${\cal A}_s\propto \mu^2/m^2$,  leading naturally to tiny scalar fluctuations for moderate values of the integration constant $c_t$ previously defined. The tensor-to-scalar ratio following from this flow equation turns out to exceed, however, the current observational constraints \cite{BICEP:2021xfz}. Smaller values of this quantity can be easily obtained if the fixed point happens at some large but finite $B_*$ value \cite{Wetterich:2013wza} or if the behavior of $B$ in the small field regime is slightly modified. For instance, a suitable generalization
\begin{equation}\label{UVFP}
\mu\,\partial_\mu \ln  B =\sigma \,B^{1-\frac{1}{\alpha}}\,,\hspace{5mm}\longrightarrow \hspace{5mm} B(\bar N)=B_{\rm end} \frac{\bar N^\alpha }{c}\,, \hspace{10mm}
B(\varphi)=\left[ \frac{1-\alpha}{\alpha}\, \log \left(\frac{\tilde m}{\varphi
   }\right)^{\sigma}\right]^{\frac{\alpha}{1-\alpha}}\,, 
\end{equation}
with $\alpha, \, \sigma \geq 1$ translate into a spectral tilt and a tensor-to-scalar ratio 
\begin{equation}\label{fig:nsrVG}
 1-n_s=\frac{\alpha}{\bar N}+\frac{r}{8}  \,, \hspace{20mm} r =  \frac{16\,  c}{\bar N^{\alpha}}\,,
\end{equation}
with 
\begin{equation}\label{eq:BNex1}
\bar N=N+c^{1/\alpha} \,, \hspace{15mm} 
c\equiv B_{\rm end} \left(\frac{\alpha}{2\sigma}\right)^\alpha\,, \hspace{15mm} \tilde m\equiv \mu\,  \exp\,(\tilde c_t)\,,
\end{equation}
$B_{\rm end}\equiv B(N=0)=2$ the value of $B$ at the end of inflation, and $\tilde c_t$ an integration constant determining the particular trajectory on the flow. As shown in Fig.~\ref{fig:nsr}, values of these quantities compatible the latest Planck/BICEP2 data are obtained for moderate $\alpha$ and $\sigma$ values. 

\subsection{Crossover and infrared regimes}\label{subsec:IRregime}

Scale invariance is substantially violated when the coefficient of the kinetic term in Eq.~\eqref{actionJ} changes from positive to negative values. During this
crossover regime rapid variations of the dimensionless couplings in the theory are expected to occur as they transition from their UV to IR fixed-point values. Among other effects, this can play a major role on the onset of the standard hBB era. The minimal heating scenario involves the quantum effective potential for Higgs field $H$,
\begin{equation}
V_H=\frac12 \lambda(\varphi/\mu) \left(H^\dagger H-\alpha(\varphi/\mu)\varphi^2\right)^2\,,
\end{equation}
with the dimensionless couplings $\lambda(\varphi/\mu)$ and $\alpha(\varphi/\mu)$ strongly violating the adibaticity condition after the end of inflation and setting to their IR fixed-point values before big bang nucleosynthesis. When written in the Einstein-frame this behaviour leads to interactions of the runaway form \eqref{Mphiexample}, allowing to apply the techniques presented in Section \ref{subsec:std_heating}. Alternatively, one could consider a heating stage involving the production of new particles, which should eventually decay into the Standard Model content. 

After heating and entropy production, the cosmon field starts to approach the IR fixed point, with a characteristic time scale $\mu$. In the scaling frame \eqref{actionJ}, the Universe shrinks, while both temperature and particle masses increase with time and the size of atoms decrease \cite{Wetterich:2013aca}. For $\mu^{-1}=10^{10}\, \textrm{yr}= 1.2\times 10^{60} M_P^{-1}$, the current expectation value of the cosmon field equals the observed reduced Planck mass, $ \varphi(t_0) = 2.48 \times 10^{18}$ GeV, such that $\varphi/\mu= 1.2 \times 10^{60}$ \cite{Wetterich:2013jsa,Wetterich:2014gaa,Rubio:2017gty}. This picture should be understood as equivalent to the standard hBB picture of an expanding and cooling Universe, since the observable dimensionless ratios still coincide with those in the usual scenario. This can be seen explicitly by performing a Weyl redefinition of the metric bringing the action \eqref{actionJ} to a canonical form with constant Planck scale and fixed particle masses \cite{Wetterich:1987fm}. This equivalence can be extended to other relevant quantities for cosmology, such as the correlation functions of primordial density fluctuations \cite{Wetterich:2015ccd,Karam:2017zno}. 

Realistic scenarios involve, customarily, a constant or slowly varying form of $B$ following from instance from an IR flow equation with anomalous dimension $\sigma_{\rm IR}$ and small corrections proportional to $B$ \cite{Wetterich:2013jsa,Wetterich:1987fm,Wetterich:1994bg,Wetterich:2014gaa,Rubio:2017gty},
\begin{equation}
\mu\,\partial_\mu \ln B=\sigma_{\rm IR}+\kappa\, B\,.
\end{equation}
The dominant scaling violation at this time is associated with the potential term $\mu^2\varphi^2$ in Eq.~\eqref{actionJ}. In particular, the dimensionless ratio of this quantity to the fourth power of the effective Planck mass $\varphi$ decreases with time, reaching very small values at present times.  Note also that the enhanced conformal symmetry for $B=\mu=0$ implies that the $\beta$-function for $B$ must vanish at $B=0$. In this limit, $\sigma_{\rm IR}=0$.

Approximate tracking solutions in which the dark energy evolution follows the dominant radiation or matter components have been extensively studied in the literature \cite{Wetterich:1987fm,Wetterich:2013jsa,Wetterich:1994bg}. These scenarios include generically an additional beyond the Standard Model sector with a late crossover regime able to stop the tracking solution at the present cosmological era. A rather natural setup appears in seesaw \cite{Minkowski:1977sc,Yanagida:1979as,Gell-Mann:1979vob} and cascade \cite{Magg:1980ut,Lazarides:1980nt,Mohapatra:1980yp} scenarios in which the mass of heavy right-handed neutrinos or scalar triplets decreases with the cosmon field $\varphi$, leading to an increasing neutrino-to-electron mass ratio \cite{Wetterich:2007kr,Amendola:2007yx}. In the Einstein frame with fixed electron mass, this reduces to the growing neutrino mass mechanism described in Section \eqref{subsec:GNQ}.

\section{Phenomenology} \label{sec:pheno}

 One of the most crucial aspects in the era of precision cosmology is to gain knowledge on the fundamental mechanisms behind cosmological acceleration while identifying new signatures potentially detectable with present or near future observational campaigns. 
 
 In this second part of the review we will discuss the phenomenology associated with quintessential inflation scenarios and the interplay between the cosmon and other fields that might be present in the very early Universe. 
 
\subsection{Coupled quintessence}\label{subsec:CQ}

In this type of scenarios the cosmon field is allowed to exchange energy with a given matter component, such as baryons, radiation, cold dark matter or neutrinos. The interaction with baryons and photons is, however, strongly restricted by local gravity measurements \cite{Will:2005va,Elder:2019yyp,Brax:2018zfb} and the non-variation of fundamental constants \cite{Uzan:2002vq},\footnote{Ways of avoiding these constraints include symmetry principles as scale symmetry \cite{Garcia-Bellido:2011kqb,Ferreira:2016kxi,Burrage:2018dvt} and screening mechanisms  (see e.g. \cite{Clifton:2011jh,Ishak:2018his,Joyce:2014kja} and references therein).} leaving the coupling to dark matter \cite{Wetterich:1994bg,Amendola:1999er,Farrar:2003uw,Amendola:2003wa,Koivisto:2005nr,Amendola:2006dg,Boehmer:2008av,Baldi:2008ay,Baldi:2010vv,Gleyzes:2015pma} and/or neutrinos \cite{Fardon:2003eh,Brookfield:2005bz,Wetterich:2007kr,Amendola:2007yx, Mota:2008nj,Pettorino:2009vn,Wintergerst:2009fh,Ayaita:2011ay,Nunes:2011mw,Ayaita:2012xm,Ayaita:2014una,Fuhrer:2015xya,Casas:2016duf} as the main viable possibility. This coupling can be formulated either at the level of the equations of motion or at the level of the action. 

Heuristic and phenomenological modifications of the equations of motion have been extensively used in the literature, with typical interactions kernels involving local and non-local functions of the relevant energy densities \cite{Amendola:2006dg,delCampo:2006vv,Wei:2007ws,Caldera-Cabral:2008yyo,delCampo:2015vha,Chimento:2009hj}, their derivatives \cite{Verma:2013pya,SanchezG:2014snn,Shahalam:2015sja,Chimento:2009hj} and/or a given temporal scale, such as the Hubble parameter \cite{Amendola:2006dg,delCampo:2006vv,Wei:2007ws,Caldera-Cabral:2008yyo} or a constant decay rate \cite{Caldera-Cabral:2008yyo,Valiviita:2008iv}. These phenomenological approaches are mathematically simple but in many cases   physically ill-defined. On the one hand, they usually lack a covariant form for the energy-momentum transfer able to unambiguously describe realistic cosmologies involving perturbations. On the other hand, there is no clear reason for the effective cosmon-matter coupling to be dictated by global properties of the Universe, rather than by purely local ones.

The above problems are avoided in field theory formulations. In particular, cosmon-matter interactions appear naturally in variable and modified gravity scenarios when written in the Einstein frame \cite{Pettorino:2008ez}. To show this, one can consider an initial Jordan-frame action of the form \eqref{action0} complemented with an exemplary matter action for an uncoupled fermionic field $\psi$ with constant mass $\bar m_\psi$,~\footnote{The fermionic nature of the matter field is chosen purely for illustration purposes. For multifield generalizations and disformal couplings to matter, see, for instance, Refs.~\cite{vandeBruck:2016jgg,vandeBruck:2015ida,Koivisto:2012za}.}
\begin{equation}
S_\psi=\int d^4x \sqrt{-g}\left[
\frac{i}{2}\left(\bar{\psi}\gamma^\mu\nabla_\mu\psi
	-\bar{\psi}\overleftarrow{\nabla}_\mu\gamma^\mu\psi\right)-\bar m_\psi \bar{\psi}\psi\right]\,.
\end{equation}
Transforming both sectors to the Einstein frame and performing a Weyl rescaling of the spinor $\psi\to \Omega^{3/2}(\phi)\, \psi$ with 
\begin{equation}
\Omega^2=\frac{f(\phi)}{M_P^2}\,,\hspace{20mm} \left(\frac{d\phi}{d\varphi}\right)^2 = \frac32
\left(\frac{\partial \ln f}{\partial \varphi }\right)^2+\frac{k}{f}\,,
\end{equation}
we get
\begin{equation}\label{Eframeferimion}
    S = \int{\rm d}^4x\sqrt{-g}\left[\frac{M^2_{\rm P}}{2}R - \frac{1}{2}(\partial \phi)^2-V(\phi)+
\frac{i}{2}\left(\bar{\psi}\gamma^\mu\nabla_\mu\psi
	-\bar{\psi}\overleftarrow{\nabla}_\mu\gamma^\mu\psi\right)-m_\psi (\phi)\bar{\psi}\psi\right]\, \,.
\end{equation}
In this representation, the matter field $\psi$ interacts with the cosmon via the $\phi$-dependent mass term
\begin{equation}\label{masspsi}
m_\psi(\phi)\equiv \frac{\bar m_\psi \, M_P}{\sqrt{f(\phi)}} \,.
\end{equation}
The variation of the action \eqref{Eframeferimion} with respect to the Einstein-frame metric provides the usual Einstein’s equations.  The conservation of stress-energy in the presence of the aforementioned energy-momentum exchange can be then written as 
\begin{equation}\label{eq:continuity2}
\nabla_{\nu}T_{\phi}^{\mu\nu}= + \frac{\beta(\phi)}{M_P}T_{\psi}\partial^{\mu}\phi\,, \hspace{20mm}
\nabla_{\nu}T_{\psi}^{\mu\nu}= -\frac{\beta(\phi)}{M_P}T_{\psi}\partial^{\mu}\phi \,,
\end{equation}
with
\begin{equation}
\label{coup}
\beta(\phi)\equiv -M_P\frac{\partial \ln m_\psi(\phi)}{\partial \phi }\,,
\end{equation}
and $T_{\psi}=T_{\psi}^{\mu\nu}g_{\mu\nu}$ the trace of the energy momentum tensor for the $\psi$-field, which vanishes identically for relativistic particles. As anticipated, these expressions are fully covariant and determined by microphysics. The effective  coupling $\beta$ modifies equivalently the effective Klein-Gordon equation for the cosmon field 
\begin{equation}\label{eq:KGQNQ}
\nabla_{\mu}\nabla^{\mu}\phi=V'(\phi)+\beta(\phi) T_\psi\,, 
\end{equation}
and the evolution of neutrinos on a classical path \cite{Ayaita:2011ay}
\begin{equation}
	\frac{\dd u^\mu}{\dd \tau} + \Gamma^\mu_{\rho\sigma} u^\rho
	u^\sigma =\beta(\phi)\, u^\lambda \partial_\lambda \phi \, u^\mu+ \beta(\phi)\, \partial^\mu \phi\,,
	\label{eq:eomnu}
\end{equation}
with $u^\mu$ the four-velocity, $\Gamma_{\mu\nu}$ the Christoffel symbols associated to the metric $g_{\mu\nu}$ and $\tau$ the proper time. The right-hand side of Eq.~\eqref{eq:eomnu} plays the role of an additional fifth-force and consists of a velocity-dependent piece $\beta(\phi)\, u^\lambda \partial_\lambda \phi \, u^\mu$ ensuring momentum conservation for particles moving in a varying cosmon field and a velocity-independent piece $ \beta(\phi)\, \partial^\mu \phi$ becoming dominant for small velocities. 

In the Einstein frame, a given model within the paradigm is specified by a choice of the effective coupling $\beta(\phi)$. A simple possibility is to consider an effective coupling $\beta (\phi)=-g M_P/(m_0+g\phi)$ following from a renormalizable Yukawa interaction $m_\psi(\phi)\bar\psi \psi=m_0\bar \psi\psi +g \phi \bar \psi\psi$,  with $m_0$ a mass parameter and $g$ a dimensionless coupling. Another option is to consider a setup involving a constant $\beta$ coupling.  This describes dilatonic like interactions $ m_\psi(\phi)\bar\psi\psi=m_0 \exp\left(-\beta \phi/M_P\right)\bar\psi\psi$ like those appearing in induced and modified gravity scenarios \cite{Zee:1978wi,Wetterich:1987fk,Bernal:2020qyu}. 
At the phenomenological level, these choices have been applied in a variety of contexts, ranging from growing neutrino scenarios aiming to solve the coincidence problem \cite{Fardon:2003eh,Brookfield:2005bz,Wetterich:2007kr,Amendola:2007yx, Mota:2008nj,Pettorino:2009vn,Wintergerst:2009fh,Ayaita:2011ay,Nunes:2011mw,Ayaita:2012xm,Ayaita:2014una,Fuhrer:2015xya,Casas:2016duf} to strongly coupled cosmologies \cite{Amendola:1999er,Amendola:2001rc,Bonometto:2012qz,Bonometto:2013eva,Bonometto:2015mya,Maccio:2015iya,Bonometto:2017rdu,Bonometto:2017lhg} inducing the formation of primordial black holes \cite{Amendola:2017xhl,Bonometto:2018dmx,Flores:2020drq,Domenech:2021uyx} or similar compact objects \cite{Savastano:2019zpr} during radiation domination. In the following sections, we describe these applications in detail.

\subsubsection{Growing neutrino masses}\label{subsec:GNQ}

Every quintessential inflation model must predict the correct amount of dark energy, preferably without an excessive finetuning of the initial conditions. In this regard, tracking or attractor solutions as those presented in Section \ref{subsec:hbb_DE} are particularly interesting, albeit they require an additional exit mechanism able to initiate the current accelerated expansion of the Universe. Growing neutrino quintessence explains the transition from a scaling solution to a dark energy dominated era by identifying the fermion field $\psi$ in Eq.~\eqref{Eframeferimion} with the neutrino \cite{Fardon:2003eh,Brookfield:2005bz,Wetterich:2007kr,Amendola:2007yx, Mota:2008nj,Pettorino:2009vn,Wintergerst:2009fh,Ayaita:2011ay,Nunes:2011mw,Ayaita:2012xm,Ayaita:2014una,Fuhrer:2015xya,Casas:2016duf}. The special role of this particle can be motivated by a cross-over regime in a beyond the Standard Model sector, manifesting itself in the neutrino masses through a seesaw or cascade mechanism \cite{Wetterich:2014gaa}, cf. Section \ref{sec:scalingframe}.

In growing neutrino quintessence, the \textit{``why now"} problem becomes related to a physical trigger event: the recent change of the effective neutrino equation of state.
In particular, when neutrinos become non-relativistic at redshift $z \approx 5$ \cite{Mota:2008nj}, a negative choice of $\beta$ induces a fixed point in the Klein-Gordon equation \eqref{eq:KGQNQ} for the cosmon field, $ -\beta (\phi) \rho_{(\psi)}\simeq -V'(\phi)$, effectively stopping its evolution in the runaway potential $V(\phi)$ and pushing it away from the scaling solution. For neutrino masses in the eV (or sub-eV) range, this leads to an almost de Sitter equation of state \cite{Amendola:2007yx,Wetterich:2007kr}
\begin{equation}
1+w(t_0)=\frac{m_{\psi}(t_0)}{12\,{\rm eV}}\,,
\end{equation}
and an energy density roughly matching the observed value  $\rho_\phi(t_0)^{1/4}=2\cdot 10^{-3}$ eV, 
\begin{equation}
\rho_\phi^{1/4}(t_0)=1.27\left(\frac{\gamma_\psi\, m_\psi(t_0)}{{\rm eV}}\right)^{1/4} 10^{-3}\, {\rm eV}\,,
\end{equation}
with $\gamma_\psi=\tilde \gamma(t_0)$ a dimensionless ${\cal O}(1)$  parameter encoding the growth rate of the neutrino mass  \cite{Amendola:2007yx,Wetterich:2007kr}.

Under some circumstances, the long-range interactions mediated by the cosmon can give rise  to the formation of dense non-linear neutrino lumps ~\cite{Mota:2008nj,Wintergerst:2009fh,Pettorino:2010bv,Nunes:2011mw,Brouzakis:2010md,Baldi:2011es,Ayaita:2011ay,Ayaita:2012xm,Ayaita:2014una,Fuhrer:2015xya,Casas:2016duf}. The formation of these objects is conceptually similar to that of gravitational structures, being the main differences associated to the field dependent mass $m(\phi)$, which varies both in space and time. The impact of neutrino lumps on large scales depends crucially on the presence or absence of a variation of the effective coupling \eqref{coup}. In the constant coupling model, the instabilities are resolved by the formation of sizable stable lumps typically exceeding the size of large galaxy clusters. The associated backreaction effects lead to the suppression of the (averaged) energy-momentum trace in Eq.~\eqref{eq:KGQNQ}, weakening the deceleration of the cosmon field and making it difficult to obtain a realistic cosmology \cite{Fuhrer:2015xya}. This is due to two effects: i) the acceleration of neutrinos during collapse, which tends to make them relativistic close to the center of the lumps and ii ) the local negative value of the cosmon perturbation $\delta\phi$ within the lumps, which leads to neutrino masses $m_\psi(\bar\phi + \delta\phi)$ smaller than expected from the average field $\bar\phi$ \cite{Nunes:2011mw}. On the other hand, phenomenologically viable scenarios for variable $\beta$ and small neutrino masses seem a priori plausible \cite{Casas:2016duf}.

\subsubsection{Primordial structure formation}

The potential existence of an additional force stronger than gravity in coupled quintessence scenarios opens the possibility of inducing structure formation even in a radiation dominated epoch in which the standard gravitational collapse is inefficient. To this end, the fermion field $\psi$ must be identified with a beyond the Standard Model particle playing the role of dark matter and becoming non-relativistic prior to matter-radiation equality. As shown in Refs.~\cite{Wetterich:1994bg,Amendola:1999er,TocchiniValentini:2001ty,Amendola:2001rc,Bonometto:2012qz,Amendola:2017xhl}, for constant\footnote{See Refs.~\cite{Bonometto:2018dmx,Flores:2020drq,Domenech:2021uyx} for related settings with variable $\beta$.} $\beta\gg1$ and subleading fermion and cosmon components, this scenario admits a scaling solution during radiation domination in which the mass of the fermion decreases with time, $m_\psi\sim a^{-1}$, and
 \begin{equation}
\phi'=\frac{M_P}{\beta}\,,  \hspace{10mm}
\Omega_{\psi}=\frac{1}{3\beta^{2}}\,,\hspace{10mm}
\Omega_{\phi}=\frac{1}{6\beta^{2}}\,,\hspace{10mm}
  \Omega_{R}=1-\frac{1}{2\beta^{2}}\,,
\end{equation}
with $\Omega_{\psi},\Omega_{\phi},\Omega_{R}$ the energy density parameters of the $\phi$, $\psi$ and radiation fluids.  In the absence of shear or rotational components in the initial velocity field, the density contrast of the $\psi$ field satisfies approximately  \cite{Amendola:1999er,Amendola:2001rc,Bonometto:2012qz,Amendola:2017xhl}
\begin{equation}
\delta_\psi''-\delta_\psi'-(1+\delta_\psi)\delta_\psi-\frac{4}{3}\frac{\delta_\psi'^2}{(1+\delta_\psi)}=0 \,.
\end{equation}
At early times, this differential equation admits a linearized solution  \cite{Amendola:1999er,TocchiniValentini:2001ty,Amendola:2001rc,Bonometto:2012qz,Amendola:2017xhl}
\begin{equation}\label{growth}
\delta_\psi=\delta_{\psi, \rm in}\left(\frac{a}{a_{\rm in}}\right)^{p}\,,\hspace{20mm} p=\frac{1}{2}\left(1 + \sqrt{5}\right)\,,
\end{equation}
with $a_{\rm in}$ the scale factor at the onset of the scaling regime and $\delta_{\psi,\rm in}$ an initial value for the fluctuations which should be determined by requiring compatibility with CMB observations \cite{Amendola:2017xhl}.

Depending on the velocity distribution, the rapid growth \eqref{growth} can lead to the production of primordial black holes \cite{Amendola:2017xhl} or virialized halos  which could dominate the dark matter content of the Universe while passing the stringent microlensing constraints and cosmic microwave background energy injection bounds \cite{Savastano:2019zpr}. The formation process of primordial dark matter halos assumes the onset of a screening mechanism at the time of formation. The analysis of this stage is, however, complicated due to the highly non-linear and strong coupling regime of the problem at hand. In particular, most of the equations cannot be consistently linearized without leading to an unbounded growth of perturbations that becomes unphysical at late times. On top of that, the treatment implicitly neglects spatial density and temperature distributions and the potential existence of
smaller halos inside larger halos. Consequently, the stability of the lumps can be most probably addressed via $N$-body simulations only. If screening turns out to be inefficient, the long-range forces could potentially cause the emission of scalar waves by accelerated particles, dissipating energy from a halo and making it collapse into black holes \cite{Flores:2020drq}. 

\subsection{Hubble-induced phase transitions} \label{subsec:spectator}

In this section we present the phenomenology associated to the interplay between non-minimally coupled spectator fields and a period of kination. Before going into the details, it is worth to discuss the generality of this scenario. On the one hand, non-minimal couplings to gravity are to be expected from quantum field theory calculations on curved space-times \cite{Birrell:1982ix}. On the other, it is also natural to consider the spectator field to be massive and/or self interacting, including potentially higher-order operators as those generically expected in effective field theories not involving shift symmetry. Additionally, the scalar field could display additional global or local symmetries, being the former potentially broken by gravity \cite{Kallosh:1995hi}. As we shall see, the combination of a long lasting period of kination with the ingredients above leads unavoidably to the spontaneous breaking of the internal symmetry of the spectator field when the background transitions from inflation to kination. This \textit{Hubble-induced} spontaneous symmetry breaking (SSB) has been considered in the literature as a heating mechanism \cite{Figueroa:2016dsc,Nakama:2018gll,Dimopoulos:2018wfg,Opferkuch:2019zbd,Bettoni:2021zhq} but also in the context of gravitational waves production \cite{Bettoni:2018pbl}, Affleck-Dine baryogenesis \cite{Bettoni:2018utf} and dark matter \cite{Fairbairn:2018bsw,Laulumaa:2020pqi,Babichev:2020xeg}. 

\subsubsection{Spectator field dynamics} \label{subsec:spectdynamics}

To discuss the dynamics of Hubble-induced phase transitions, let us a consider a canonical quintessential scenario like the one in Section \ref{sec:standard_picture} supplemented with an exemplary action
\begin{equation}\label{eq:chiaction}
S_\chi=\int d^4x \sqrt{-g}\left[-\frac12 g^{\mu\nu}\partial_\mu\chi\partial_\nu\chi -\frac12 \left(m_\chi^2+\xi R\right)\chi^2-V(\chi)\right]
\end{equation}
for an energetically-subdominant and non-minimally coupled spectator field $\chi$ with bare mass $m_\chi$ and self-interaction potential
\begin{equation}\label{eq:pot_chi}
V(\chi)=\frac{\lambda}{2n}\frac{\chi^{2n}}{\Lambda^{2n-4}}\,,
\end{equation}
with $\lambda$ a dimensionless coupling constant and $\Lambda$ an UV suppression scale. For the sake of simplicity we have assumed the spectator field to be real and imposed a discrete $Z_2$ symmetry. These restrictions will not play, however, a key role in the following developments, being the phenomenology presented below easily extendable to more general settings and symmetry groups. 
 
From the point of view of the spectator field, the only relevant aspect of the cosmon evolution is the transition from inflation to kinetic domination, with the correspondent change in the equation-of-state parameter from $w=-1$ to $w=+1$. During inflation ($w=1$, $R=12H^2$) the effective square mass of the $\chi$ field is large and positive for sufficiently big non-minimal couplings, $\xi \gtrsim 1/12$, locking it at the origin potential and preventing the generation of sizeable isocurvature perturbations in potential conflict with CMB constraints \cite{Planck:2018jri}. However, as soon as the background evolution changes from inflation to kination ($w=1$, $R=-6H^2$), the origin of the potential becomes a local maximum for small bare masses $m^2_\chi<6\xi R$, forcing the spectator field to evolve towards the new time-dependent minima  appearing at large field values\footnote{Notice that this holds even if the potential is a sum of monomials with positive coefficients. The inclusion of negative coefficients in the power series could give rise, however, to the appearance of new local or global minima.}
\begin{equation}
\label{eq:chimin}
\vert\chi_{\rm min}\vert =H\left(\frac{6\xi -m^2_\chi/H^2}{\lambda}\right)^{\frac{1}{2(n-1)}}\left(\frac{\Lambda}{H}\right)^\frac{n-2}{n-1}\,.
\end{equation}
\begin{figure}
    \centering
    \includegraphics[scale=.7]{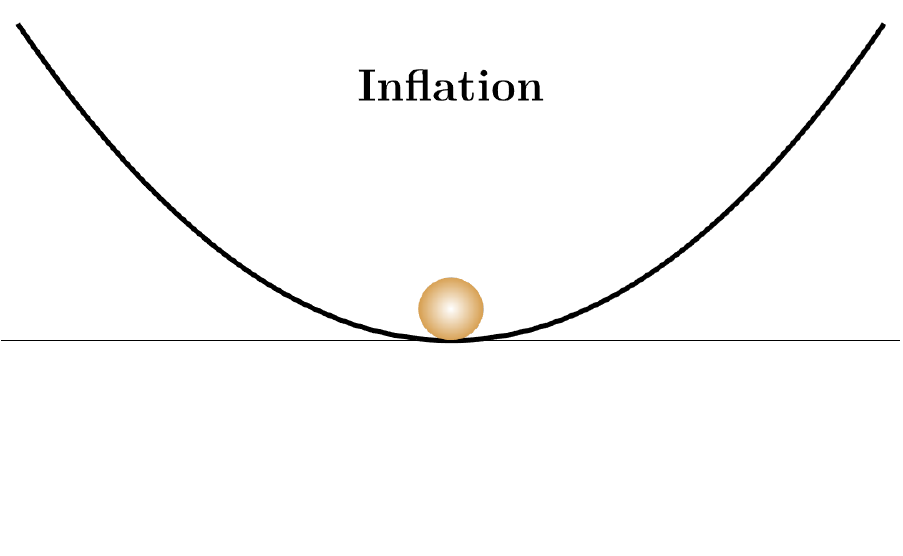}
    \includegraphics[scale=.7]{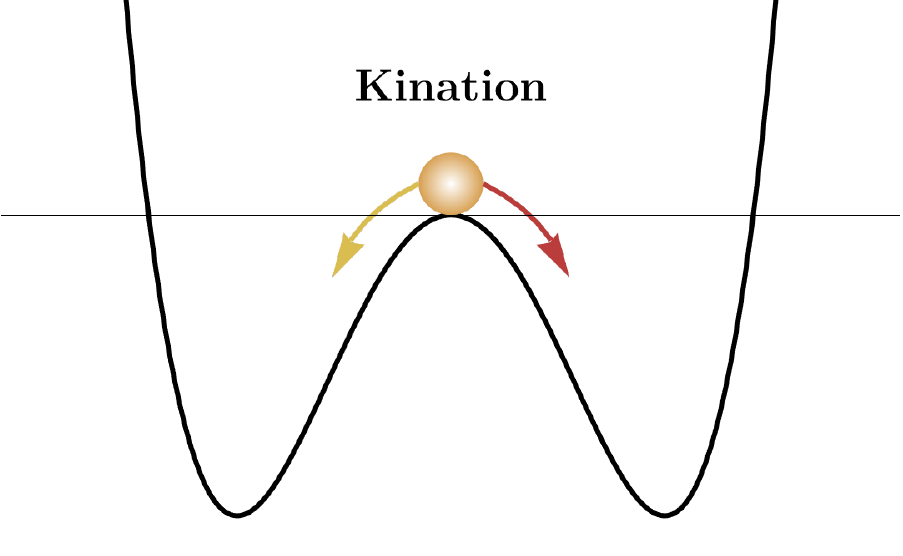}
    \includegraphics[scale=.7]{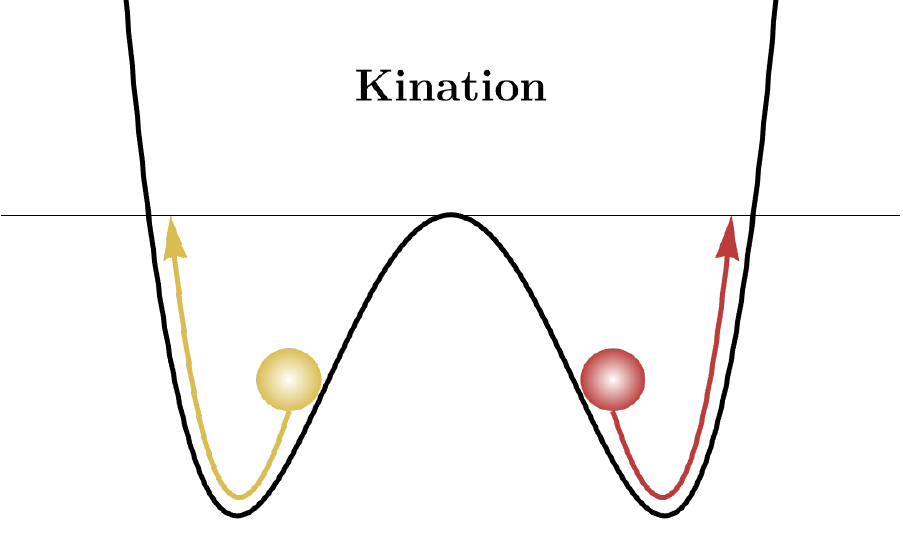}
    \includegraphics[scale=.7]{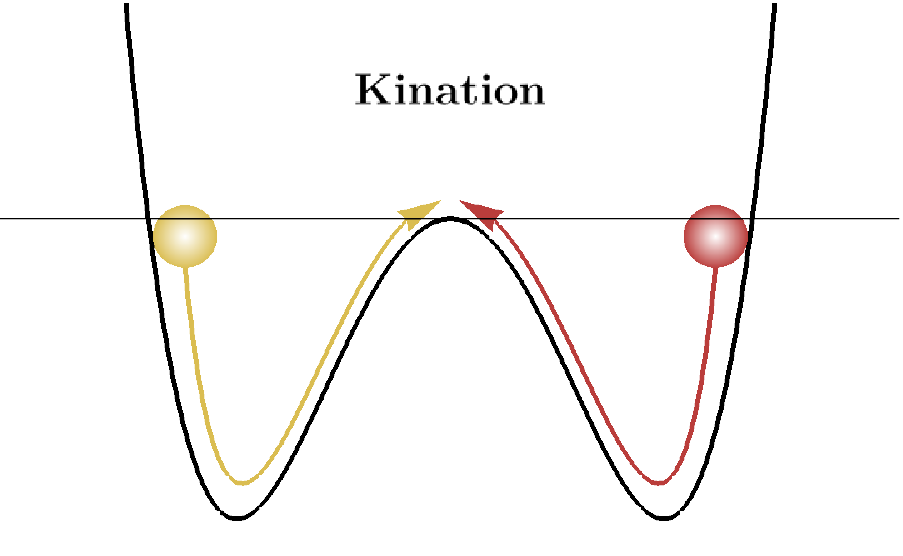}
    
    \caption{Schematic illustration of the evolution of the spectator field fluctuations during inflation and kination. During inflation the field distribution sits at the minimum of the effective potential. The internal symmetry of the field is, however, spontaneously broken at the onset of kination, forcing the field distribution to evolve towards two new minima appearing at large field values. Due to the time-dependence of the potential, the distribution is not trapped there, but rather oscillates with a bi-modal pattern around the origin.}
    \label{fig:HISB_Sketch}
\end{figure}
 The transition between the inflationary and kinetic dominated epochs depends on the details of the quintessential inflation potential.  If it occurs rapidly, the phase transition can be taken to be almost instantaneous. In other words, the spectator field is not moving away from the inflationary minimum while the potential is changing from the unbroken to the broken phase form. This is typically the situation realized in quintessential inflationary scenarios, where the cosmon potential at the end of inflation is steep. Another important aspect to keep in mind is that the background evolution is assumed to be homogeneous, and hence the transition occurs at the same time everywhere.~\footnote{Alternative pictures involving first-order phase transitions could appear in the presence of multiple higher-dimensional operators with positive and negative coefficients.}  Then, at the time of symmetry breaking the field has zero mean $\langle\chi\rangle=0$ and a certain variance $\langle \chi^2\rangle$ dictated by vacuum fluctuations. During the broken phase, the effective mass is negative and the system undergoes a stage of tachyonic or spinodal instability, which rapidly amplifies the quantum fluctuations while keeping their zero average value. In particular, the symmetry breaking process is local and any vacua or direction to it is equally probable. Then, the averaged value of the scalar field over scales much larger than the correlation length must identically vanish. This crucial observation implies that any description in terms of a homogeneous spectator field rolling down a potential is inaccurate to start with \cite{Felder:2000hj,Felder:2001kt,Bettoni:2019dcw,Bettoni:2021zhq}. This has two major implications.  First, the dynamics of the scalar field leads to the generation of \textit{temporary} topological defects. Second, the energy in gradients is non-negligible and the virialization of the system leads to an asymptotic equation-of-state parameter $w=1/3$, regardless of the specific self-interaction potential \eqref{eq:pot_chi}.

The tachyonic instability, which can be significantly long lasting since the new minima of the effective potential can be quite displaced from the origin and the kination period can last many $e$-folds, ends when the effective mass of the $\chi$ field turns positive. This can occur at the onset of radiation domination, where $R\simeq 0$, or be induced by the higher-order corrections or the bare mass parameter in Eq.~\eqref{eq:chiaction}, provided that these overcome the Ricci scalar contribution at a given time. At this point, the stability is \textit{fully} restored and the origin becomes again the true minimum of the potential. However, as we shall see below, there is still the possibility that the symmetry restoration is only \textit{effectively} realized. This peculiar characteristic of Hubble-induced phase transitions occurs when the energy lost by the spectator field in one oscillation is smaller than the flattening of the effective potential minimum, a quantity that can be easily inferred from Eq.~\eqref{eq:chimin}. In this case, the field distribution is able to pass back the origin of the potential, giving rise to an interacting bimodal distribution that oscillates around it.
The other possibility is that the field radiates enough energy as to get trapped in the new minima of the potential at a given time, tracking their subsequent evolution and giving rise to more standard topological defects with time-dependent tension and width \cite{Bettoni:2018pbl}. This interesting scenario could be realized, for instance, if the spectator field is allowed to decay into other species or if the effective friction experienced by it is somehow enhanced. 

\begin{figure}
    \centering
    \includegraphics[scale=0.6]{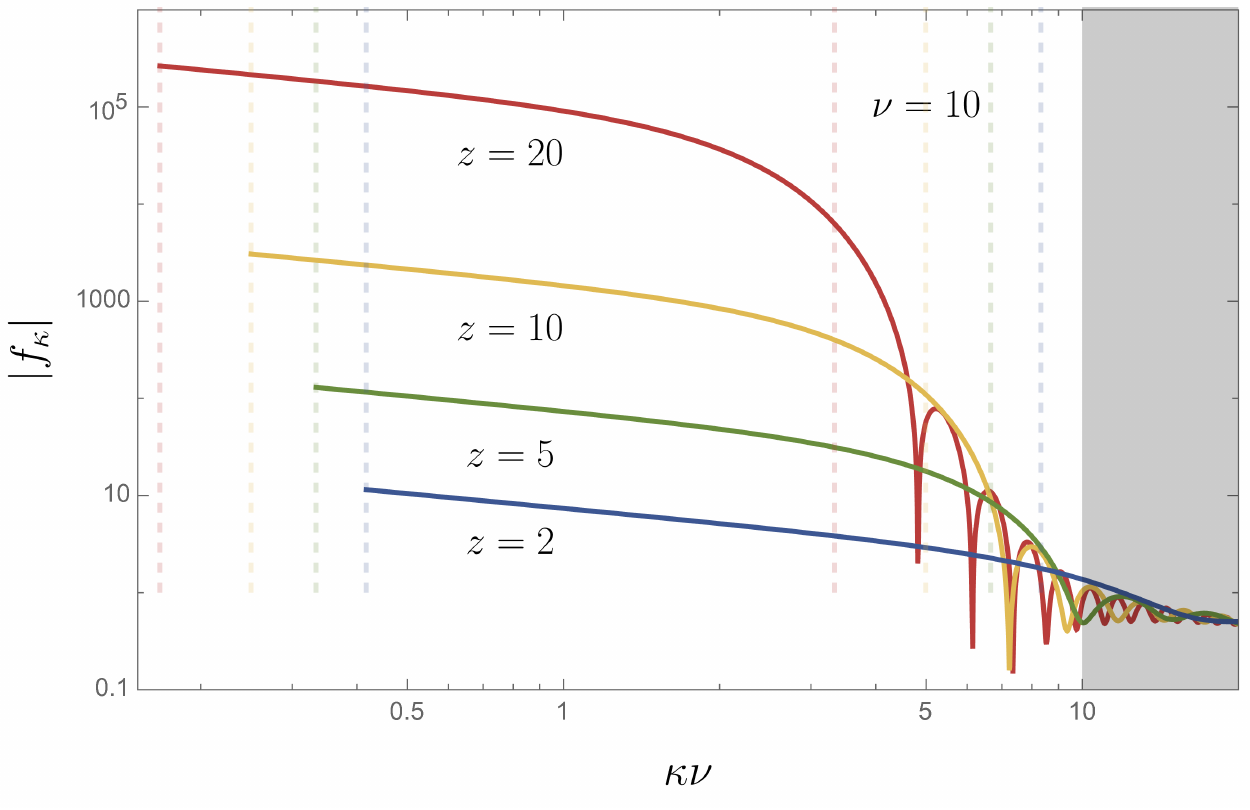}
    \includegraphics[scale=0.6]{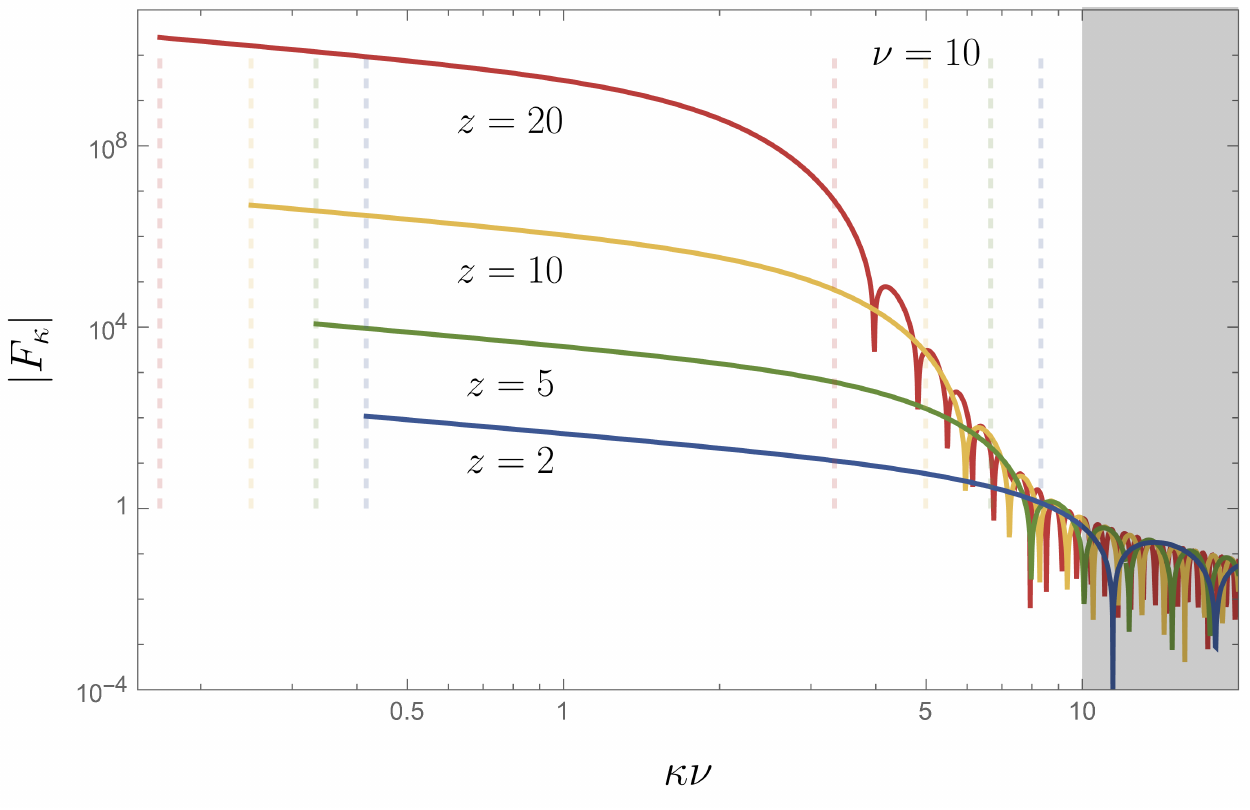}
    \caption{The functions $\vert f_\kappa\vert$ (left) and $\vert F_\kappa\vert$ (right) evaluated at different conformal times $z$. The shaded region corresponds to momenta outside the initial amplification band $\kappa>\kappa(z=0)$. The dashed vertical lines corresponds to the time dependent amplification band \eqref{eq:ampl_band} at the given times.}
    \label{fig:fkFk}
\end{figure}

To quantify the above qualitative description, we can rewrite the action \eqref{eq:chiaction} in terms of the coordinates and field variables in \eqref{eq:Ycoord}, getting
\begin{equation}
    S_\chi = \int d^3y\, dz\left[\frac12 (Y')^2-\frac12 \vert\nabla Y\vert^2 +\frac12 M^2(z)Y^2-\frac{\lambda_{\rm
    eff}}{2n}Y^{2n}\right]\,,
\end{equation}
where the prime denotes derivative with respect to the conformal time $z$, 
\begin{equation}
    M^2(z) = (4\nu^2-1)\mathcal H^2\,, \hspace{20mm}\lambda_{\rm eff} = \left(\frac{\chi_*}{\Lambda}\frac{a_{\rm kin}}{a}\right)^{2n-4}\,,
\end{equation}
and 
\begin{equation}
  \nu = \sqrt{3\xi/2}\,, \hspace{15mm}  \mathcal H(z) = \frac{1}{2(z+\nu)}\,,\hspace{15mm} \mathcal H' = -2\mathcal H^2\,,
\end{equation}
with the last relation holding during kination. At the onset of this expansion epoch, the spectator field is close to the origin and the higher-order terms in the potential can be safely neglected. This means that the early dynamics of the system can be well described in terms of a free scalar field with a time dependent mass $ M^2(z)$ and, therefore, exactly solved, as we discussed in Section \ref{sec:heating}. In fact, in terms of the dimensionless Fourier mode $\kappa$ in Eq.~\eqref{eq:Ycoord}, the problem reduces to find the solutions of the Schr\"odinger-like differential equation
\begin{equation}
    f''_\kappa + \omega^2_\kappa(z) f_\kappa = 0\,,\hspace{20mm} \omega_\kappa^2 = \kappa^2-M^2(z)\,,
\end{equation}
  for the mode function $f_\kappa$ with standard vacuum initial conditions $f_\kappa(0)=1/\sqrt{2\kappa}$ and $f'_\kappa(0) = -i\sqrt{\kappa/2}$. The detailed quantization of the system follows closely the treatment of  Refs.~\cite{Guth:1985ya,Polarski:1995jg,Kiefer:1998qe,Garcia-Bellido:2002fsq} and it is detailed for the present case in Ref.~\cite{Bettoni:2019dcw}.
The tachyonic instability induced by kinetic domination amplifies all subhorizon modes within the time-dependent window $\kappa_{\rm min}(z)$ and $\kappa_{\rm max}(z)$, where
\begin{equation}\label{eq:ampl_band}
    \kappa_{\rm min} = \mathcal{H}(z)\,,\hspace{20mm} \kappa_{\rm max} = (4\nu^2-1)^{1/2}\kappa_{\rm min}\,.
\end{equation}
The copious production of these modes translates eventually into a quantum-to-classical transition \cite{Polarski:1995jg,Lesgourgues:1996jc, Polarski:1995jg,Kiefer:1998qe,Kiefer:1998pb,Kiefer:1998jk}, meaning that the properties of the system can be well-approximated by those of a classical random field  \cite{Bettoni:2019dcw,Bettoni:2021zhq}. In particular, the above system can be solved in terms of Bessel's functions,
\begin{equation}
    f_\kappa(z) = \sqrt{z+\nu}\left[A_\kappa \mathcal J_\nu(\kappa(z+\ u))-B_\kappa\mathcal Y_\nu(\kappa(z+\nu))\right]\,,
\end{equation}
with $A_\kappa$ and $B_\kappa$ $\kappa$-dependent coefficients determined by the aforementioned initial conditions. As shown in Fig.~\ref{fig:fkFk}, the modes that belongs to the unstable region are exponentially amplified. Moreover, one can show that the Heisenberg uncertainty principle in Fourier space can be written as \cite{Polarski:1995jg,Lesgourgues:1996jc, Polarski:1995jg,Kiefer:1998qe,Kiefer:1998pb,Kiefer:1998jk,Bettoni:2019dcw,Bettoni:2021zhq}
\begin{equation}
\Delta Y_\kappa^2\,\Delta \Pi_\kappa^2 = \vert F_\kappa(z)\vert^2 + \frac14 \geq {\frac14}\, \Big\vert \langle [Y_\kappa(z),\ \Pi^\dagger_\kappa(z)]\rangle\Big\vert^2\,, 
\end{equation}
with 
\begin{equation}
F_\kappa(z) = {\rm Re}(f^*_\kappa f'_\kappa)\,,
\end{equation}
and the square brackets denoting the anticommutation of the corresponding variables. It is then clear that for $\vert F_\kappa\vert \gg 1$, the ordering of the operators becomes negligible, as expected for a classical system.

\begin{figure}[ht!]
    \centering
    \includegraphics[scale=0.165]{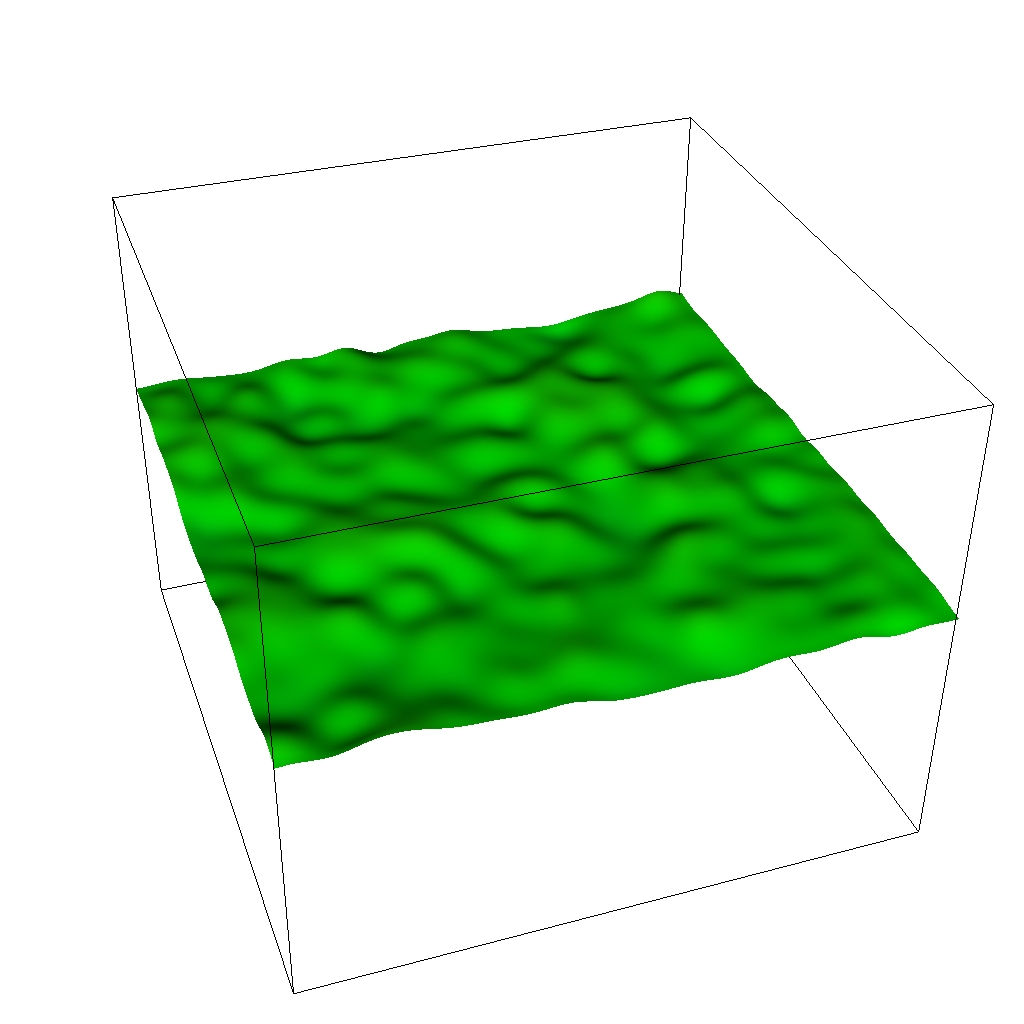}
    \includegraphics[scale=0.165]{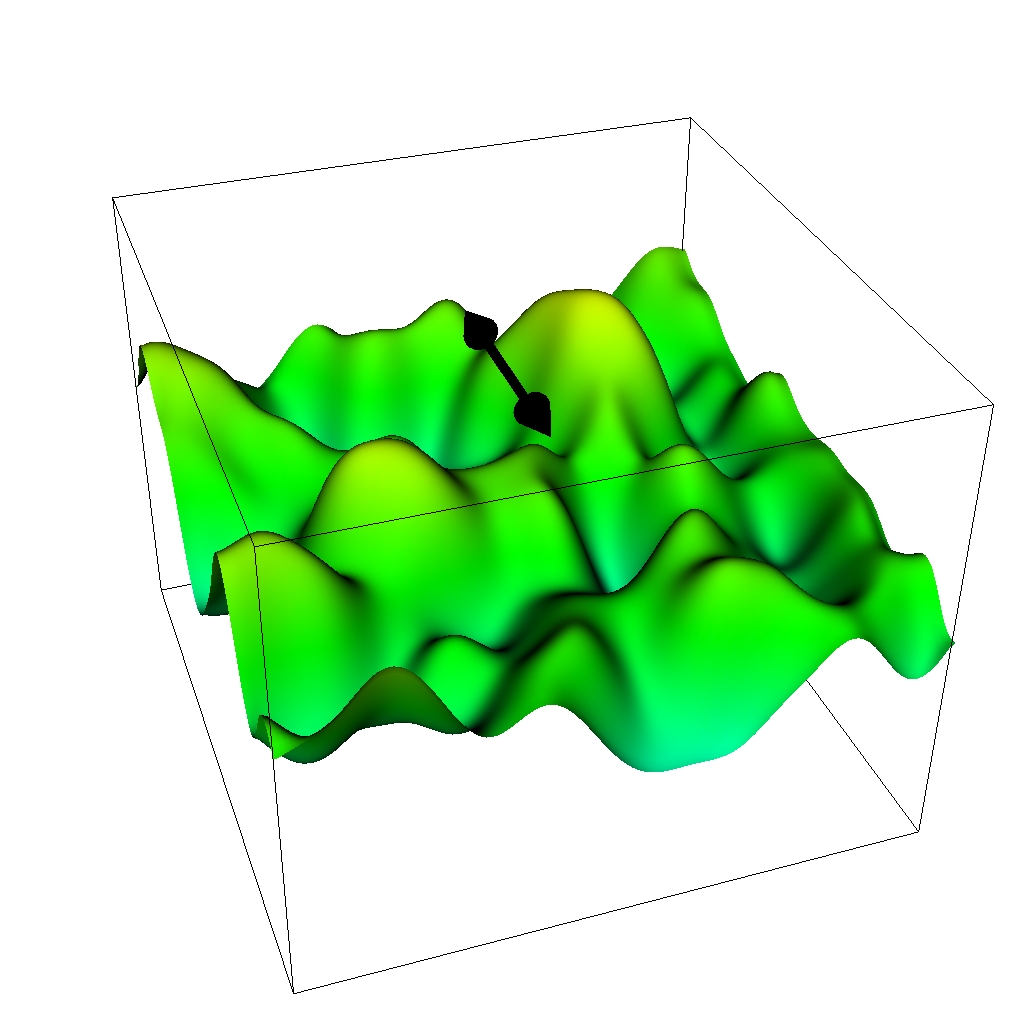}\\
    \includegraphics[scale=0.165]{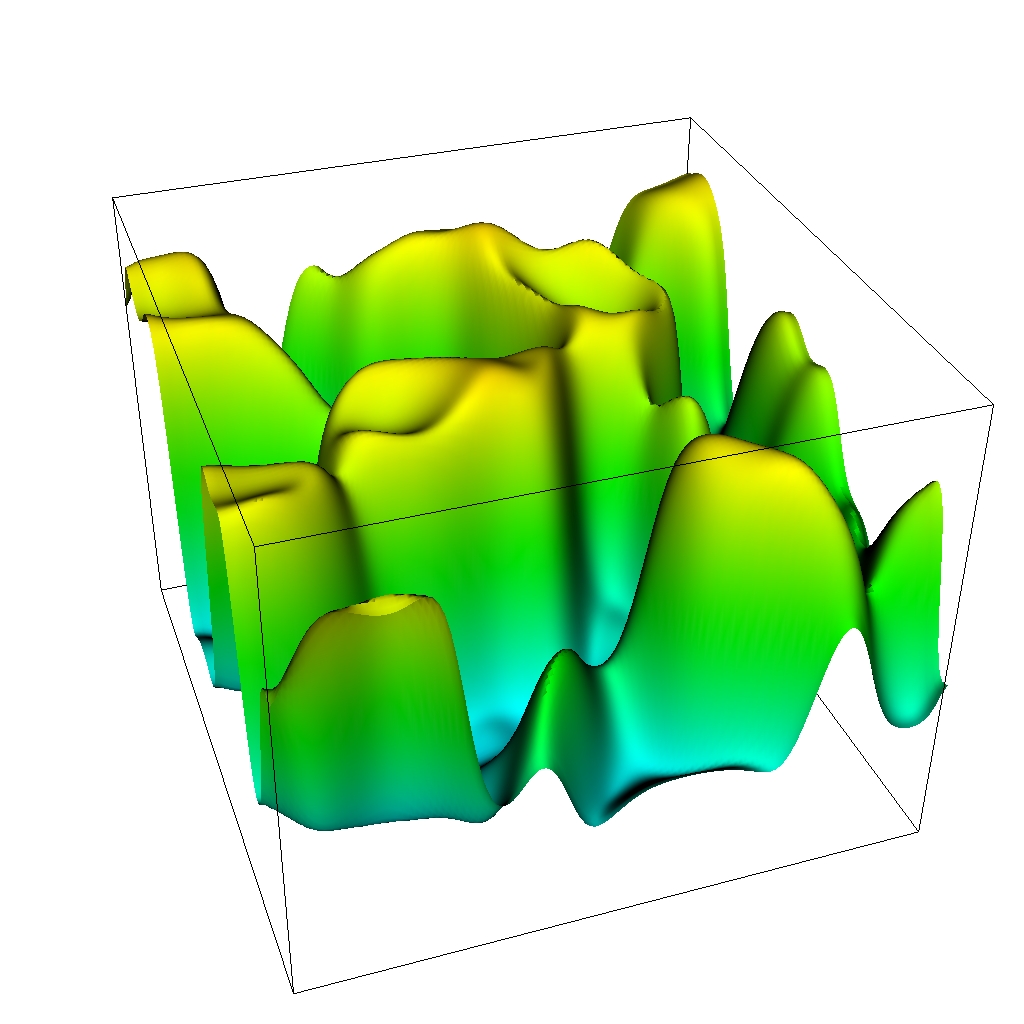}
    \includegraphics[scale=0.165]{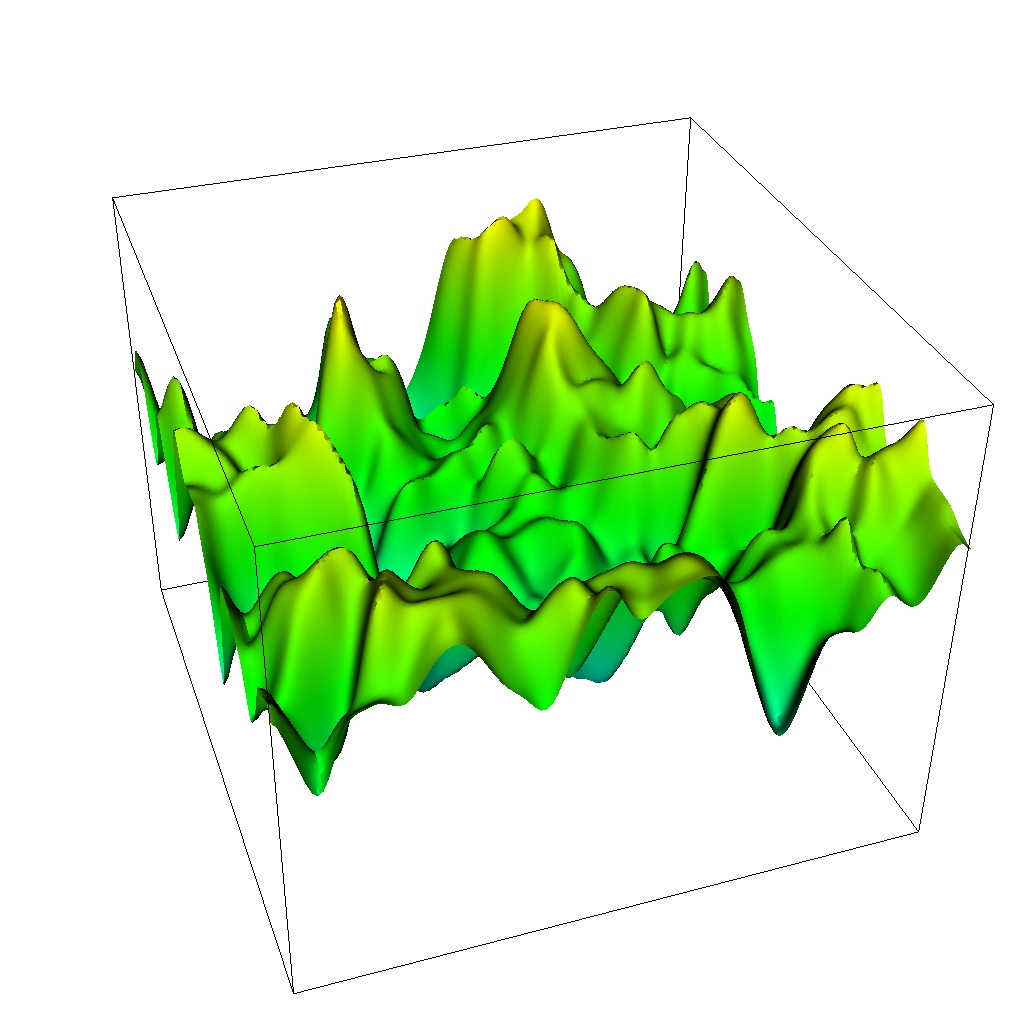}\\
    \includegraphics[scale=0.6]{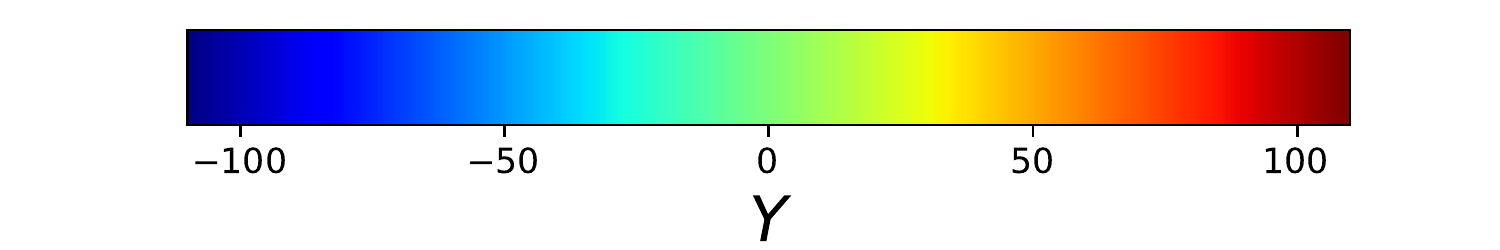}
    \caption{Spectator field growth for the benchmark point $\nu=10$, $\lambda=10^{-4}$. We display 2-dimensional sections of a fraction $V/8$ of the  3-dimensional simulation volume $V$ at times $z=5,\, 10,\, 15, \, 60$, with the $z$-axis indicating the corresponding field value at each point in this surface. The black segment in the second panel indicates the typical scale of the highest peaks. (Images from \cite{Bettoni:2021zhq})}
    \label{fig:snapshots}
\end{figure}

The above description ceases to be valid as soon as non-linearities kick in. At that moment, in order to properly account for the complete dynamics in the evolution equation
\begin{equation}
\label{eq:Yeq}
    Y'' -\nabla^2 Y-M^2(z)Y+\lambda_{\rm eff}Y^{2n-1}=0\,,
\end{equation}
one needs to resort to numerical simulations. Such simulations have been carried out in Ref.~\cite{Bettoni:2021zhq} for quartic  ($n=4$, $\lambda_{\rm eff}=\lambda$) and sextic potentials ($n=6$, $\lambda_{\rm eff}= \lambda(\chi_*/\Lambda\, a_{\rm kin}/a)^4$). Despite the complexity of the non-linear physics involved, it is still possible to picture the evolution during the broken phase in rather simple terms. As soon as the relevant scales can be treated classically, the field value moves \textit{locally} towards the new minima, where \textit{locally} means that the field spatial distribution will look homogeneous on scales smaller than the correlation length $\kappa_*=\sqrt{\nu+1}\mathcal H(z)$~\cite{Bettoni:2019dcw}.
This is exactly what is shown in the first three panels of Fig.~\ref{fig:snapshots}, where we display a few two-dimensional snapshots of the simulation ran in Ref.~\cite{Bettoni:2021zhq} for a quartic potential.\footnote{Movies covering the whole simulation are available at \href{https://www.youtube.com/playlist?list=PLl1K9-81ct6yHh6boyTiZAfNv_zlwux7-}{this URL}.} The first box shows the initial classicalized configuration, with the spectator field still very close to the vacuum initial conditions. The second to fourth boxes display the displacement of the spectator field from the origin, growing in amplitude and generating a rather inhomogeneous distribution. Since we are considering a $Z_2$ symmetric field, this implies the formation of regions of positive and negative field values separated by domain wall configurations. It is worth noticing that the characteristic scale of the inhomogeneity decreases with time. In particular, the profile of the inhomogeneities is smoother when the field rolls towards the broken phase minima, while it becomes more jagged in the subsequent evolution. This is due to the fact that after the field distribution reaches a minimum and bounces back, it is able to cross the origin towards the opposite minimum. Hence, the spectator field is effectively oscillating in a quartic potential, permitting energy exchange and making the oscillating defects fragment, while they diminish in amplitude and size. Despite this, the energy stored in the gradients of the spectator field is not decreasing with time and it represents a non-negligible fraction of the total field energy density for the total duration of the simulation. This confirms the idea that the Hubble-induced symmetry breaking is an inherently inhomogeneous process. 

\subsubsection{Applications}\label{subsec:applications}

In this last  part of the manuscript, we will focus on the expected phenomenology of the above scenario. In particular, we will describe the consequences of inhomogeneities and review some new aspects to be expected when changing the symmetry group or applying this mechanism to dark matter. 

\paragraph{Heating} \label{sec:heating}

\begin{figure}
    \centering
    \includegraphics[scale=0.47]{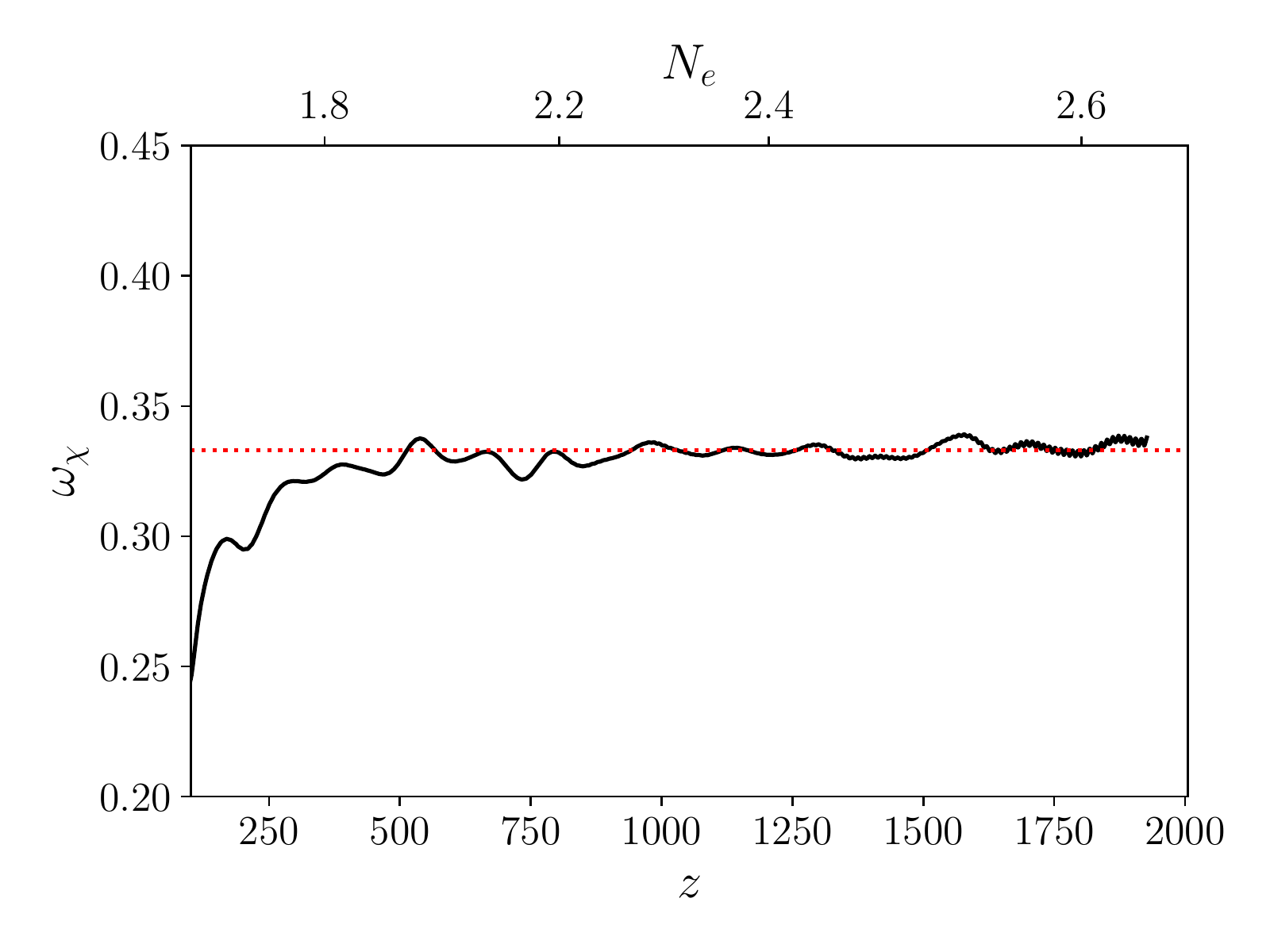}
    \includegraphics[scale=0.47]{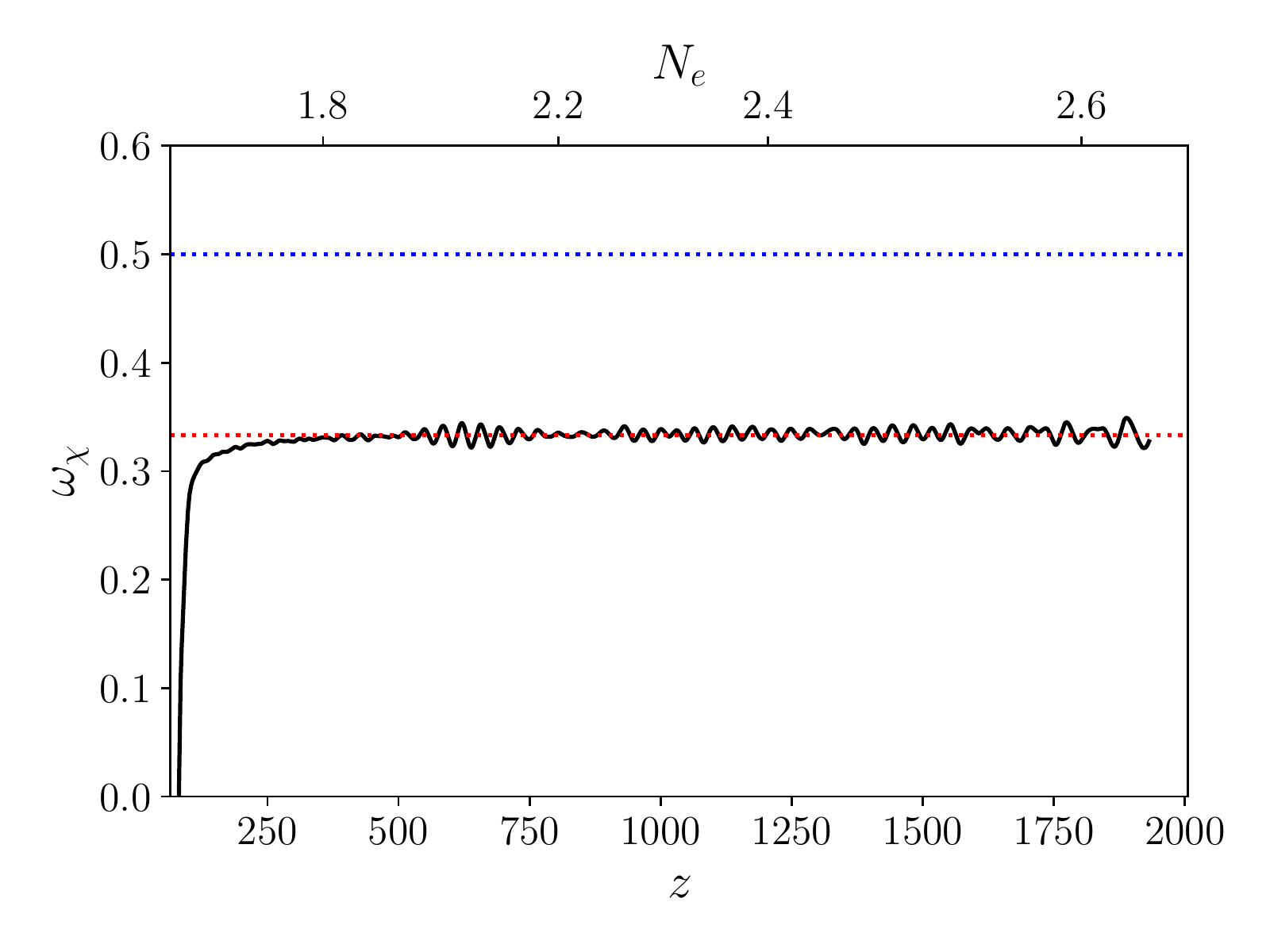}
    \caption{ (Left) Evolution of the volume and time averaged equation-of-state parameter \eqref{eq:wchi} for a quartic potential. (Right) Evolution of the same quantity for a sextic potential. The red horizontal line corresponds to the radiation value $1/3$ while the blue line corresponds to the value $1/2$. (Images credit \cite{Bettoni:2021zhq})}
    \label{fig:equation_of_state}
\end{figure}

The fast amplification of the the spectator field fluctuations triggered by the Hubble-induced SSB plays an important role when it comes to the heating of the Universe. This can be understood by realizing that the effect of the non-minimal coupling decreases with time, eventually becoming negligible in the determination of the spectator field dynamics. From there on, the field behaves almost as if it were minimally coupled to gravity. In particular, it will tend to \textit{virialize}, achieving an equation-of-state parameter asymptotically close to the radiation value $w=1/3$. In fact, for monomial potentials like the ones under consideration, the equations-of-state parameter can be written as \cite{Lozanov:2016hid,Lozanov:2017hjm}
\begin{equation}
\label{eq:wchi}
    w_\chi = \frac13 +\frac23 \frac{(n-2)}{(n+1) +\langle(\nabla\chi/a)^2\rangle/\langle V\rangle}\,,
\end{equation}
showing clearly that the field will behave as a radiation component provided that the gradients dominate over the potential and independently of the specific higher-order operators involved. For illustration purposes, we display in  Fig.~\ref{fig:equation_of_state} the time evolution of the equation-of-state parameter for quartic and sextic potentials.   It is worth stressing that if one were to consider the spectator field as a homogeneous condensate the equation of state would be that of radiation only for the first potential, taking the value $w=1/2$ for the sextic one. This crucial difference remarks how crucial inhomogeneities are in the context of SSB. 

\begin{figure}
    \centering
    \includegraphics[scale=0.5]{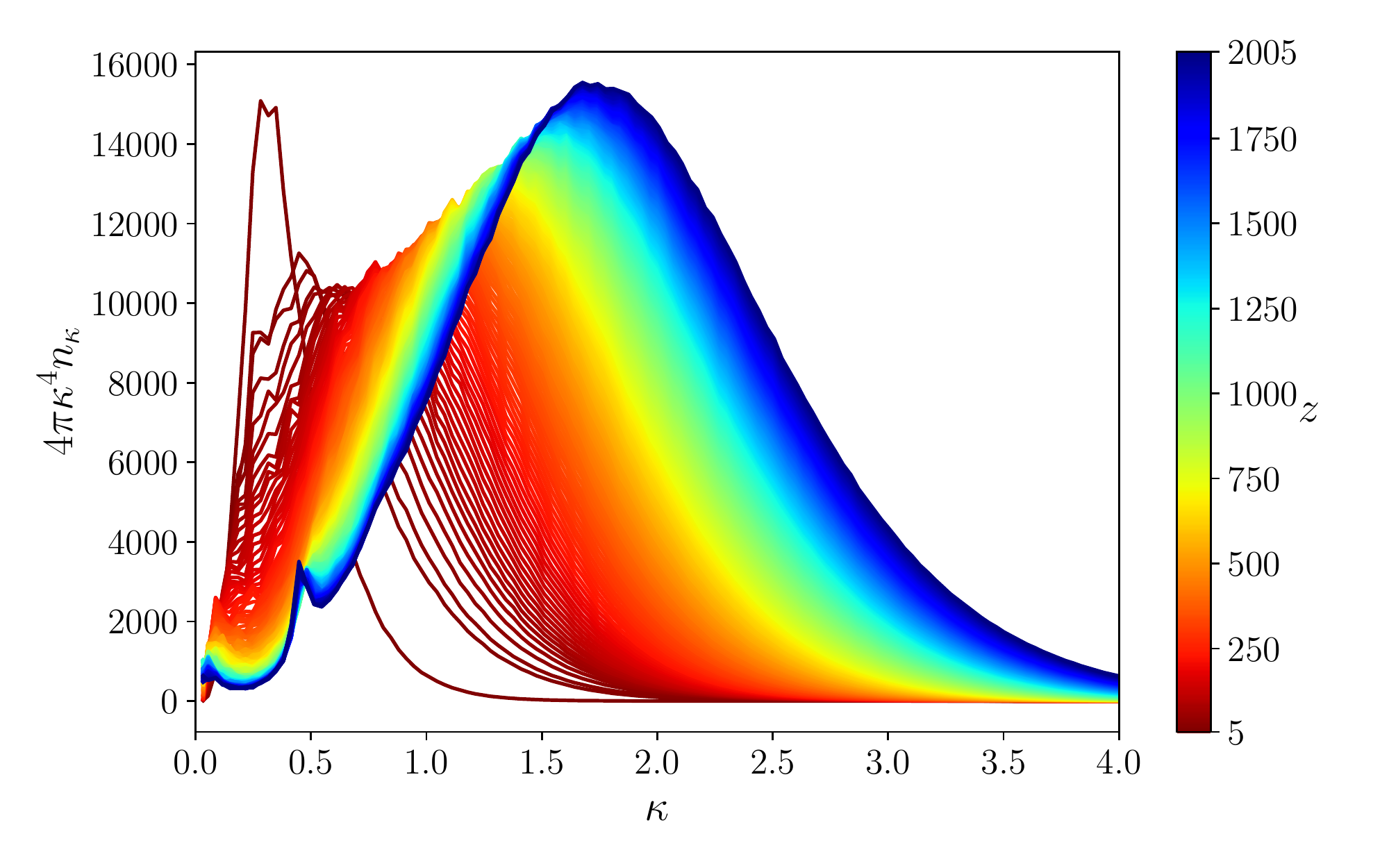}
    \caption{Time evolution of the occupation numbers for $\nu=10$ and $\lambda=10^{-4}$. Different colors correspond to different times, as indicated in the figure. (Image credit \cite{Bettoni:2021zhq})}
    \label{fig:spectrum}
\end{figure}

Another important aspect to keep in mind is that tachyonic instability only amplifies IR modes thus generating an initially far-from-thermal spectrum, as that displayed in Fig.~\ref{fig:spectrum}. However, the subsequent turbulent evolution of the system induces a slow shift of the spectral peak, which moves gradually towards the UV, leading to eventually a close-to-thermal distribution \cite{Micha:2002ey,Micha:2004bv}.  
When combined with the fast dilution of the background component discussed in Section \ref{sec:standard_picture}, this implies that the spectator field will eventually become dominant, starting an epoch of radiation domination and setting the conditions for the onset of the hBB era. Of course, much still remains to be said about this process. One might add, for instance, direct interactions between the spectator field and fermions, like in other heating mechanisms \cite{Felder:1998vq,Felder:1999pv,GarciaBellido:2008ab,Rubio:2015zia,Repond:2016sol,Fan:2021otj,Dufaux:2006ee}, or consider this radiation component as an inert, subdominant and extra radiation contribution (obviously satisfying all observational bounds). Also, if a significant bare mass parameter is allowed, the field could eventually oscillate around a quadratic potential, possibly behaving as dark matter (see below). 

\paragraph{Gravitational waves}
\label{sec:GW}

Another interesting phenomenological aspect related to the presence of defects and inhomogeneities is the associated emission of gravitational waves. Since the distribution of the spectator field after the Hubble-induced phase transition is highly inhomogeneous, the transverse traceless part of its energy momentum tensor will generically be non-vanishing, hence providing a source for GW emission.  The production of these tensor modes due K\"ahler-induced  potentials has been considered in supersymmetric extensions of the Standard Model \cite{Kamada:2015iga} and applied to the present Ricci-based scenario in Ref.~\cite{Bettoni:2018pbl} for a $U(1)$ symmetric spectator field. Although the generation of gravitational waves from topological defects is a very well-known phenomenon \cite{Dufaux:2010cf,Figueroa:2012kw,Giblin:2011yh}, the mechanism at hand possesses striking differences. First, the defects will have a typical size proportional to the Hubble rate $\mathcal H$, which makes them `fat', as in the context of axion dark matter \cite{Gorghetto:2018myk,Vaquero:2018tib}, and with a time-dependent tension and width, meaning that an exact scaling regime \cite{Bennett:1989ak,Perivolaropoulos:1992if} is never achieved. Second, the energy density in GW will be naturally amplified during kinetic domination. Although the precise relevance of these phenomena will ultimately depend on the parameters of the model, they represent  nonetheless a generic feature of it, potentially within the reach of forthcoming observational campaigns \cite{Gouttenoire:2019kij,Aggarwal:2020olq}, as shown explicitly in Ref.~\cite{Bettoni:2018pbl}. Interestingly, the obtained GW spectrum displays also a peaked structure, thus easing its discrimination from foreground signals.

\paragraph{Baryogenesis}
\label{sec:baryogenesis}

The observed asymmetry between matter and anti-matter in the Universe remains one of the biggest mysteries in particle physics. Even if the Standard Model could a priori produce a baryonic asymmetry during the electroweak phase transition \cite{Sakharov:1967dj}, the resulting value turns out too small as compared with observations. Traditionally, the amount of asymmetry is encoded in the baryon-to-entropy ratio 
\begin{equation}
    Y_B\equiv \frac{n_B}{s}=(0.861\pm0.008)\times 10^{-10}\,,
\end{equation}
where $n_B=n_b-n_{\bar b}$ being $n_{b(\bar b)}$ the number density of (anti)baryons. One of the most successful proposals to explain this small value of this quantity is the so-called Affleck-Dine mechanism \cite{Affleck:1984fy,Dine:1995kz}, which makes use of a gravity-induced SSB during inflation. Interestingly enough, Hubble-induced phase transitions are able to prompt a \textit{post-inflationary} Affleck-Dine asymmetry generation if the spectator field $\chi$ is charged under a weakly-broken $B-L$ symmetry \cite{Bettoni:2018utf}. In particular, the change of sign of the Ricci scalar at the onset of kinetic domination breaks  spontaneously the $U(1)_{B-L}$ symmetry, making the complex spectator field roll towards the new vacuum manifold of the potential with a certain phase that, at each space point, takes a random value. The additional explicit symmetry breaking requirement provides the charge non-conservation. In the polar representation, $\chi = \tfrac{1}{\sqrt{2}}\rho e^{i\theta}$, the associated symmetry-breaking operators adding to the potential\footnote{Notice that since here the symmetry is $U(1)$ one needs to replace in the action \eqref{eq:chiaction} the real scalar field by a complex one, i.e. $\chi^2\rightarrow \vert\chi\vert^2=\chi\chi^*$, being $\chi^*$ the complex conjugate of $\chi$.} \eqref{eq:pot_chi} can be written as
\begin{equation}
    \delta V = -\epsilon_0 \Lambda^{4-2n}\rho^{2n} F(\theta,\theta_0)\,,
    \end{equation}
with  $\epsilon_0$ and $\theta_0$ the amplitude and phase of a complex symmetry-breaking coefficient $\epsilon= \epsilon_0e^{i\theta_0}$. 
A net amount of particles over anti-particles can then be produced
\begin{equation}
    a^3n_{B-L}(t) = q \epsilon_0\Lambda^{4-2n}\int^t dt\, a^3(t)\rho^{2n} F'(\theta,\theta_0)\,,
\end{equation}
where the prime stands here for a derivative with respect to the phase $\theta$ and $n_{B-L}=q\rho^2\dot\theta$ is the $B-L$ current, being $q$ the associated charge. The quantity $n_{B-L}$ keeps growing until the symmetry is restored. In the case of a quartic operator, corresponding to $n=2$, the produced asymmetry is estimated to be  \cite{Bettoni:2018utf} 
\begin{equation}\label{eq:YBL}
    \frac{Y_{B-L}}{10\, {\rm GeV}} = \left(\frac{\Theta}{10^{-4}}\right)^{3/2}\frac{H_{\rm kin}}{10^7\, {\rm GeV}}\,,
\end{equation}
with $\Theta$ the heating efficiency defined in Eq.~\eqref{eq:heateff}.
As this point, it is important to emphasize a fundamental limitation of the procedure. The net baryonic excess \eqref{eq:YBL} is \textit{local}, and, therefore, the cosmological evolution will put in causal contact patches with different phase values, leading to a net zero charge on large scales! Albeit not usually emphasized, this problem is common to other post-inflationary mechanisms  for the generation of baryon asymmetry, as those considered, for instance, in Refs.~\cite{Sakstein:2017lfm,Sakstein:2017nns}. A possible way out is to consider a biased minimum, so that the potential is tilted and a unique global minimum exists, see e.g. \cite{Macpherson:1994wf,Coulson:1995nv,Avelino:2008qy}. Then, even if initially all spectator phases are equally probable, the dynamical evolution will generate a  flux towards the true minimum, thus producing a net \textit{global} asymmetry \cite{Bettoni:2018utf}. 

\paragraph{Dark matter}
\label{sec:darkmatter}

The Hubble-induced SSB has also been explored in the context of primordial dark matter production \cite{Markkanen:2015xuw,Fairbairn:2018bsw,Laulumaa:2020pqi,Babichev:2020xeg}. In particular, Ref.~\cite{Fairbairn:2018bsw} studied the super-Hubble creation of particles\footnote{Notice that this does not violate causality as long as it is sub-horizon, meaning by this that no modes smaller than the inflationary particle horizon are considered \cite{Bassett:1999mt}.} when the post-inflationary Ricci scalar can take negative values. This mechanism was also combined there with a post-inflationary period of kination. By solving the equation for the spectator field \eqref{eq:Yeq} at one loop precision, it is found a dark matter abundance
\begin{equation}
    \Omega_{\chi}h^2 = 0.078 \beta^{3/4}\left(\frac{m}{{\rm GeV}}\right)\left(\frac{H_{\rm rad}}{10^{13} {\rm GeV}}\right)^{3/2}\left(\frac{100}{g_{*,{\rm rad}}}\right)^{1/4}\,.
\end{equation}
Here $\beta$ is an $\mathcal O(1)$ parameter for  $\xi\lesssim 10$ representing the energy density of the produced $\chi$ quanta in units of $H_{\rm kin}$. $m$ is the dark matter bare mass parameter whose value is a priory free. However, upper bounds on the cross section for dark matter interactions obtained from galaxy clusters collisions impose the constraint 
\begin{equation}
    \frac{m}{{\rm GeV}} >0.027 \lambda^{2/3}\,.
\end{equation}
This bound is more constraining for larger values of $H_{\rm kin}$ and $\lambda$. However, even when these restrictions are combined with the requirement of avoiding dark matter overproduction, one still finds the very broad range of allowed dark matter masses keV $<m<10^4$ GeV \cite{Fairbairn:2018bsw}.

\section{Discussion and outlook} \label{sec:conclusions}

Inflation and dark energy are usually understood as two independent epochs in the expansion history of the Universe. In this review, we have described how these accelerated expansion eras can be unified into a \textit{quintessential inflation} framework in which a single degree of freedom, the cosmon, plays simultaneously the role of the inflaton and of the quintessence field. 
 
The naturalness of quintessential inflation is intimately related to its embedding in a fundamental framework able to produce a large hierarchy of scales without involving unacceptable finetunings. Natural realizations might appear, for instance, in variable gravity scenarios involving UV and IR fixed points where scale symmetry is manifestly realized. On top of that, this paradigm should provide a graceful solution to the \textit{``why now"} problem, potentially associated with some physical mechanism. In this respect, scaling solutions are able to provide a dark energy density comparable to the dominant matter content in the Universe for an extended period of time, reducing with it the tuning on the initial conditions. A suitable clock signalling the end of this scaling regime is naturally provided by neutrinos, which for sub-eV masses becomes non-relativistic around the same time in which dark energy should start to be dominant.  

Probably, the most distinctive phenomenology of quintessential inflation is associated to the unavoidable kinetic dominated regime. This unusual stage induces a relative amplification of all radiation and matter components in the Universe, including the products resulting from the heating stage and any stochastic background of gravitational waves of primordial origin. This can be used to put constraints on some of the theory defining parameters. For instance, the compatibility of the primordial tensor perturbations with BBN imposes bounds on the inflationary scale and the duration of kination, or equivalently, on the efficiency of the heating stage. Additional constraints on these quantities could follow from the spontaneous symmetry of internal symmetries of fields non-minimally coupled to gravity and the associated gravitational wave production. In this review we have discussed the Hubble-induced symmetry breaking of two specific symmetries, namely a global U(1) and a discrete Z$_2$. However, it is worth stressing that this mechanism applies to any symmetry group, provided the appearance of a time-dependent negative mass parameter in the effective potential leading to new minima with a non-trivial topological structure. Furthermore, the expansion rate may also affect a plethora of particle physics processes in the very early Universe, from the generation of the matter-antimatter asymmetry to the abundance of dark matter relics. 

\section*{Acknowledgments}

DB acknowledges support from Programa II: Contratos postdoctorales by Salamanca University and from Project PGC2018-096038-B-I00  funded by the Spanish ``Ministerio de Ciencia e Innovación" and FEDER ``A way of making Europe". JR thanks the Funda\c c\~ao para a Ci\^encia e Tecnologia (FCT), Portugal, for the financial support to the Center for Astrophysics and Gravitation-CENTRA, Instituto Superior T\'ecnico,  Universidade de Lisboa, through the Project No.~UIDB $/00099/2020$ and for the financial support through the CEECIND $/01091/2018$ grant. 

\appendix

\section{Frame-invariant formulation of inflation}\label{app:frame-covariant}
In this Appendix, we summarize the frame-invariant formulation of inflation in Refs.~\cite{Kaiser:1994vs,Burns:2016ric}, to which refer the reader for further details. 

The variation of the action \eqref{actionJ} with respect to the FLRW metric and the scalar field $\varphi$ provides respectively the Friedmann equations \cite{Kaiser:1994vs}, 
\begin{eqnarray}
H^2= \frac{fU}{3}\left(1-\frac{\epsilon_H}{3}+\frac{5\,\kappa_H}{3}  +\frac{\theta_H}{3} \right)^{-1} \,, 
\hspace{15mm} \epsilon_H +\kappa_H  - \theta_H = \frac{k \dot \varphi^2  }{2H^2f}   \,, \label{eq:Friedman} 
\end{eqnarray}
and the Klein-Gordon equation 
\begin{eqnarray}
{\ddot \varphi} + (3+\sigma_H)H{\dot \varphi}
+\frac{f^3}{E} U_{,\varphi}\ = 0\,, \label{eq:KG}
\end{eqnarray}
with
\begin{equation}
E\equiv k f  + \frac{3}{2} f ^2_{,\varphi}\,,  \hspace{20mm}  U\equiv\frac{V}{f^2}\,.
\end{equation} 
The quantities
\begin{equation}\label{eq:HubbleSR}
\epsilon_H\equiv - \frac{\dot H}{H^2}\,, \hspace{15mm} \kappa_H\equiv  \frac12 \frac{\dot f}{H f}\,,\hspace{15mm} \sigma_H\equiv \frac12\frac{\dot E}{H E}\,, \hspace{15mm} \theta_H\equiv \frac12 \frac{\ddot f }{H^2 f} \,,
\end{equation} 
in these expressions can be understood as a generalization of Hubble slow-roll parameters. Assuming these to be small during the inflationary stage, we can rewrite Eqs.~\eqref{eq:Friedman} and \eqref{eq:KG} as 
\begin{eqnarray}\label{eq:motionSR}
&& H^2\simeq     \frac{f U}{3} \,,\hspace{20mm}
   3 H{\dot \varphi} \simeq -
\frac{f^3}{E} U_{,\varphi}\,,
\end{eqnarray}
and compute the scalar and tensor power spectra
\begin{eqnarray}
P_s&\approx&  \frac{k^3}{24\pi^2 }\, \frac{U}{\epsilon_U + \kappa_U}\,, \hspace{15mm}
P_t \approx \frac{2 k^3 }{3\pi^2}\, U\,, 
\end{eqnarray}
 and the associated spectral tilt and tensor-to-scalar ratio, 
\begin{eqnarray}
1-n_s &=& 4\epsilon_U + 2(\sigma_U-\delta_U -\kappa_U)\,, \hspace{15mm} r =  16 (\epsilon_U +  \kappa_U)  \,,
\end{eqnarray}
with 
\begin{eqnarray}
 \epsilon_U\
&\equiv& \ \frac{1}{2}\, \frac{f  U_{,\varphi} (f U)_{,\varphi}}{EU^2}
\ , \hspace{10mm}  
\delta_U
\equiv
 \frac{1}{2}\, \frac{f  U_{,\varphi} (f U)_{,\varphi}}{EU^2}
 +\left(\frac{f^2 U_{,\varphi}}{ E U} \right)_{,\varphi}
 \,,  \\
\label{kappaU}
\kappa_U&\equiv& -\frac{f_{,\varphi}  }{2  }\, \frac{f  U_{,\varphi} }{EU  } 
\,, \hspace{10mm}
\sigma_U \equiv - \frac{1}{2}\,\frac{E_{,\varphi} }{E^2   }  \frac{f^2 U_{,\varphi} }{U}
\,,
\end{eqnarray}
the generalization of the potential slow-roll parameters. These expressions are intended to be evaluated at a number of $e$-folds $N=\int H\, dt$, computed out of Eqs.~\eqref{eq:motionSR},      
\begin{equation}\label{eq:Nefolds}
N (\varphi)\ =\ -\int_{\varphi}^{\varphi_\text{end}}  d\varphi' \frac{E (\varphi') }{f (\varphi')^2 }\,  \frac{ U (\varphi')  }{U (\varphi')_{ ,\varphi'}}
\,,
\end{equation}
with $\varphi_\text{end}$ the value of $\varphi$ at the end of inflation, determined by a frame-invariant generalization of the standard condition $\epsilon_U=1$,
\begin{equation}\label{eq:end_inf_inv}
 \max(\epsilon_U+\kappa_U, |\epsilon_U + \delta_U +4\kappa_U - \sigma_U | )= 1\,.   
\end{equation}
Note that, at the leading order in the slow-roll approximation, the inflationary observables are invariant under the separate action of conformal transformations and field reparametrizations even if the slow-roll parameters are themselves transformed, i.e.,
$
\widetilde n_s (\tilde \varphi)= n_s (\varphi)$, $ \widetilde r  (\tilde \varphi) = r(\varphi)$. Also, as written in Eq.~\eqref{eq:Nefolds}, the number of $e$-folds is invariant under frame transformations. Consequently, we have $
\widetilde n_s (\widetilde N)=n_s(N)$ and $
\widetilde r  (\widetilde N)= r(N)$. 

\section{A workout example of crossover regime}

For illustration purposes, we present here a specific flow equation smoothly interpolating between the fixed points considered in the main text and allowing for explicit analytical solutions,  
\begin{equation}\label{flowtotal}
 \mu \, \partial_\mu \ln B=\frac{\kappa\, \sigma B}{\sigma+\kappa \, B^{1/\alpha}}\,.
\end{equation}
Using Eq.~\eqref{eq:Nefolds} we get 
\begin{equation}\label{eq:BN}
B(\bar N)= \Big\lbrace\frac{\sigma}{\kappa
   }  {\cal W}\left[\frac{\kappa}{\sigma }
   \left(B_{\rm end}\,  e^{2 \kappa 
  \bar N}\right)^{p}\right]\Big \rbrace^{1/p } \,,\hspace{15mm}
\end{equation}
with 
\begin{equation}
\bar N=N+c^p \,, \hspace{15mm}  c\equiv B_{\rm end} \left(\frac{\alpha}{2\sigma}\right)^\frac1p\,.
\end{equation}
The above expression can be alternatively written as
\begin{equation}
B(\bar N)=  B_{\rm end} \Bigg\lbrace \left( \frac{1}{2\,\kappa \,p \, c^{p}} \right) {\cal W}\left[2\,\kappa\, p \, c^{p}
\,  e^{2\, \kappa\, p\, \bar N}\right]\Bigg\rbrace^{1/p} \,.
\end{equation}
The associated form $B(\varphi)$ can be trivially computed out of this expression through the relation 
\begin{equation}\label{eq:difBN}
\frac{\partial \ln \varphi}{\partial N}=-\frac{2}{B}\,,
\end{equation}
getting
\begin{equation}
\ln \left(\frac{\varphi}{ m\, e^{-\frac1\sigma \frac{1}{1-q}}}\right)^\sigma= \frac{\sigma}{\kappa} (B^{-1}) +\frac{(B^{-1})^{1-q}}{1-q}\,,
\end{equation}
where we have chosen the integration constant as $C=m\, e^{-\frac1\sigma \frac{\alpha}{\alpha-1}}$. This can be alternatively written as 
\begin{equation}\label{eq:flowsol}
\ln \left(\frac{\varphi}{ m }\right)^\sigma= \frac{\sigma}{\kappa}(B^{-1})+\ln_q B^{-1} \,,
\end{equation}
with 
\begin{equation}
\ln_q x = 
\begin{cases}
\ln x \hspace{15mm} {\rm if} \hspace{5mm} x>0 \,\, \& \,\, q=1 \,, \\
\frac{x^{1-q}-1}{1-q} \hspace{11mm} {\rm if} \hspace{5mm} x>0 \,\, \& \,\, q\neq 1\,, \\
\textrm{undefined}  \hspace{6mm} {\rm if} \hspace{5mm} x\leq 0 \,,
\end{cases}
\end{equation}
the so-called $q$-logarithm function used by Tsallis in non-extensive statistical mechanics \cite{da_Silva_2019}. This function is related to the $q$-exponential 
\begin{equation}
e_q^x = 
\begin{cases}
e^x \hspace{52mm} {\rm if} \hspace{5mm}  q=1 \,, \\
(1+(1-q)x)^{1/(1-q)} \hspace{22mm} {\rm if} \hspace{5mm}  q\neq 1\,,\,\,  \&\,\, 1+(1-q)x\geq0 \,,   \\
0  \hspace{54mm} {\rm if} \hspace{5mm} q\neq 1\,,\,\,  \&\,\, 1+(1-q)x<0 \,,
\end{cases}
\end{equation}
(such that $\exp_q (\ln_q x)=x$ and $\ln_q(\exp_q x)=x$) and to the Lambert-Tsallis function, defined as \cite{da_Silva_2019}
\begin{equation}
{\cal W}_q(x) e^{{\cal W}_q(x)}=x    \,, 
\end{equation}
and satisfying \cite{da_Silva_2019}
\begin{equation}
{\cal W}_q(x)=\ln_q x-_q\ln {\cal W}_q(x)\,,    
\end{equation}
with
\begin{equation}
a-_q b = \frac{a-b}{1+(1-q)b}\,.
\end{equation}
In general, Eq.~\eqref{eq:flowsol} cannot be inverted. However, some solutions can be found for specific values of $\alpha$, namely
\begin{eqnarray}
&& \alpha=1/2 \,,\hspace{5mm} B(\varphi)=\frac{1}{2} \left(1-\log \left(\frac{\varphi
   }{m}\right)^{\sigma }\right)+\sqrt{\frac{ \sigma }{\kappa }+\frac14\left(1-\log \left(\frac{\varphi
   }{m}\right)^{\sigma }\right)^2}\,,\\
&&\alpha=1 \,, \hspace{5mm} B(\varphi) =  \frac{\kappa }{\sigma}\, {\cal W}^{-1}\left[\frac{\sigma}{\kappa}\left(\frac{\varphi}{m}\right)^\sigma\right]\,,  \\
&&\alpha=2 \,, \hspace{5mm} B(\varphi)=\left( \frac{1+
   \sqrt{1+\frac{\sigma}{\kappa} (2+\ln \left(\frac{\varphi}{ m }\right)^\sigma) 
   }}{2+\ln \left(\frac{\varphi}{ m }\right)^\sigma}\right)^2\,.
\end{eqnarray}
The cosmological evolution following from the $\alpha=1$ case has been carefully analyzed in Ref.~\cite{Rubio:2017gty}, where we refer the interested reader for further details. 

\footnotesize
\small{
\bibliographystyle{JHEP.bst}
\bibliography{QIR}

\providecommand{\href}[2]{#2}\begingroup\raggedright\begin{thebibliography}{100}

\bibitem{Kamionkowski:2015yta}
M.~Kamionkowski and E.D.~Kovetz, \emph{{The Quest for B Modes from Inflationary
  Gravitational Waves}},
  \href{https://doi.org/10.1146/annurev-astro-081915-023433}{\emph{Ann. Rev.
  Astron. Astrophys.} {\bfseries 54} (2016) 227}
  [\href{https://arxiv.org/abs/1510.06042}{{\ttfamily 1510.06042}}].

\bibitem{SupernovaCosmologyProject:1997zqe}
{\scshape Supernova Cosmology Project} collaboration, \emph{{Discovery of a
  supernova explosion at half the age of the Universe and its cosmological
  implications}}, \href{https://doi.org/10.1038/34124}{\emph{Nature} {\bfseries
  391} (1998) 51} [\href{https://arxiv.org/abs/astro-ph/9712212}{{\ttfamily
  astro-ph/9712212}}].

\bibitem{SupernovaSearchTeam:1998fmf}
{\scshape Supernova Search Team} collaboration, \emph{{Observational evidence
  from supernovae for an accelerating universe and a cosmological constant}},
  \href{https://doi.org/10.1086/300499}{\emph{Astron. J.} {\bfseries 116}
  (1998) 1009} [\href{https://arxiv.org/abs/astro-ph/9805201}{{\ttfamily
  astro-ph/9805201}}].

\bibitem{Planck:2018jri}
{\scshape Planck} collaboration, \emph{{Planck 2018 results. X. Constraints on
  inflation}}, \href{https://doi.org/10.1051/0004-6361/201833887}{\emph{Astron.
  Astrophys.} {\bfseries 641} (2020) A10}
  [\href{https://arxiv.org/abs/1807.06211}{{\ttfamily 1807.06211}}].

\bibitem{BICEP:2021xfz}
{\scshape BICEP, Keck} collaboration, \emph{{Improved Constraints on Primordial
  Gravitational Waves using Planck, WMAP, and BICEP/Keck Observations through
  the 2018 Observing Season}},
  \href{https://doi.org/10.1103/PhysRevLett.127.151301}{\emph{Phys. Rev. Lett.}
  {\bfseries 127} (2021) 151301}
  [\href{https://arxiv.org/abs/2110.00483}{{\ttfamily 2110.00483}}].

\bibitem{2dFGRS:2001csf}
{\scshape 2dFGRS} collaboration, \emph{{The 2dF Galaxy Redshift Survey: The
  Power spectrum and the matter content of the Universe}},
  \href{https://doi.org/10.1046/j.1365-8711.2001.04827.x}{\emph{Mon. Not. Roy.
  Astron. Soc.} {\bfseries 327} (2001) 1297}
  [\href{https://arxiv.org/abs/astro-ph/0105252}{{\ttfamily
  astro-ph/0105252}}].

\bibitem{2dFGRS:2001ybp}
{\scshape 2dFGRS} collaboration, \emph{{Evidence for a non-zero lambda and a
  low matter density from a combined analysis of the 2dF Galaxy Redshift Survey
  and cosmic microwave background anisotropies}},
  \href{https://doi.org/10.1046/j.1365-8711.2002.05215.x}{\emph{Mon. Not. Roy.
  Astron. Soc.} {\bfseries 330} (2002) L29}
  [\href{https://arxiv.org/abs/astro-ph/0109152}{{\ttfamily
  astro-ph/0109152}}].

\bibitem{SDSS:2005xqv}
{\scshape SDSS} collaboration, \emph{{Detection of the Baryon Acoustic Peak in
  the Large-Scale Correlation Function of SDSS Luminous Red Galaxies}},
  \href{https://doi.org/10.1086/466512}{\emph{Astrophys. J.} {\bfseries 633}
  (2005) 560} [\href{https://arxiv.org/abs/astro-ph/0501171}{{\ttfamily
  astro-ph/0501171}}].

\bibitem{Peccei:1987mm}
R.D.~Peccei, J.~Sola and C.~Wetterich, \emph{{Adjusting the Cosmological
  Constant Dynamically: Cosmons and a New Force Weaker Than Gravity}},
  \href{https://doi.org/10.1016/0370-2693(87)91191-9}{\emph{Phys. Lett. B}
  {\bfseries 195} (1987) 183}.

\bibitem{WaliHossain:2014usl}
M.~Wali~Hossain, R.~Myrzakulov, M.~Sami and E.N.~Saridakis, \emph{{Unification
  of inflation and dark energy \`a la quintessential inflation}},
  \href{https://doi.org/10.1142/S0218271815300141}{\emph{Int. J. Mod. Phys. D}
  {\bfseries 24} (2015) 1530014}
  [\href{https://arxiv.org/abs/1410.6100}{{\ttfamily 1410.6100}}].

\bibitem{deHaro:2021swo}
J.~de~Haro and L.A.~Sal\'o, \emph{{Analytical and numerical review of
  Quintessential Inflation}},
  \href{https://arxiv.org/abs/2108.11144}{{\ttfamily 2108.11144}}.

\bibitem{Spokoiny:1993kt}
B.~Spokoiny, \emph{{Deflationary universe scenario}},
  \href{https://doi.org/10.1016/0370-2693(93)90155-B}{\emph{Phys. Lett.}
  {\bfseries B315} (1993) 40}
  [\href{https://arxiv.org/abs/gr-qc/9306008}{{\ttfamily gr-qc/9306008}}].

\bibitem{Peebles:1998qn}
P.J.E.~Peebles and A.~Vilenkin, \emph{{Quintessential inflation}},
  \href{https://doi.org/10.1103/PhysRevD.59.063505}{\emph{Phys. Rev.}
  {\bfseries D59} (1999) 063505}
  [\href{https://arxiv.org/abs/astro-ph/9810509}{{\ttfamily
  astro-ph/9810509}}].

\bibitem{Peloso:1999dm}
M.~Peloso and F.~Rosati, \emph{{On the construction of quintessential inflation
  models}}, \href{https://doi.org/10.1088/1126-6708/1999/12/026}{\emph{JHEP}
  {\bfseries 12} (1999) 026}
  [\href{https://arxiv.org/abs/hep-ph/9908271}{{\ttfamily hep-ph/9908271}}].

\bibitem{Dimopoulos:2001ix}
K.~Dimopoulos and J.W.F.~Valle, \emph{{Modeling quintessential inflation}},
  \href{https://doi.org/10.1016/S0927-6505(02)00115-9}{\emph{Astropart. Phys.}
  {\bfseries 18} (2002) 287}
  [\href{https://arxiv.org/abs/astro-ph/0111417}{{\ttfamily
  astro-ph/0111417}}].

\bibitem{Giovannini:2003jw}
M.~Giovannini, \emph{{Low scale quintessential inflation}},
  \href{https://doi.org/10.1103/PhysRevD.67.123512}{\emph{Phys. Rev. D}
  {\bfseries 67} (2003) 123512}
  [\href{https://arxiv.org/abs/hep-ph/0301264}{{\ttfamily hep-ph/0301264}}].

\bibitem{Brax:2005uf}
P.~Brax and J.~Martin, \emph{{Coupling quintessence to inflation in
  supergravity}}, \href{https://doi.org/10.1103/PhysRevD.71.063530}{\emph{Phys.
  Rev.} {\bfseries D71} (2005) 063530}
  [\href{https://arxiv.org/abs/astro-ph/0502069}{{\ttfamily
  astro-ph/0502069}}].

\bibitem{BuenoSanchez:2006fhh}
J.C.~Bueno~Sanchez and K.~Dimopoulos, \emph{{Trapped Quintessential
  Inflation}},
  \href{https://doi.org/10.1016/j.physletb.2006.09.045}{\emph{Phys. Lett. B}
  {\bfseries 642} (2006) 294}
  [\href{https://arxiv.org/abs/hep-th/0605258}{{\ttfamily hep-th/0605258}}].

\bibitem{Hossain:2014xha}
M.W.~Hossain, R.~Myrzakulov, M.~Sami and E.N.~Saridakis, \emph{{Variable
  gravity: A suitable framework for quintessential inflation}},
  \href{https://doi.org/10.1103/PhysRevD.90.023512}{\emph{Phys. Rev.}
  {\bfseries D90} (2014) 023512}
  [\href{https://arxiv.org/abs/1402.6661}{{\ttfamily 1402.6661}}].

\bibitem{Agarwal:2017wxo}
A.~Agarwal, R.~Myrzakulov, M.~Sami and N.K.~Singh, \emph{{Quintessential
  inflation in a thawing realization}},
  \href{https://doi.org/10.1016/j.physletb.2017.04.066}{\emph{Phys. Lett.}
  {\bfseries B770} (2017) 200}
  [\href{https://arxiv.org/abs/1708.00156}{{\ttfamily 1708.00156}}].

\bibitem{Ahmad:2017itq}
S.~Ahmad, R.~Myrzakulov and M.~Sami, \emph{{Relic gravitational waves from
  Quintessential Inflation}},
  \href{https://doi.org/10.1103/PhysRevD.96.063515}{\emph{Phys. Rev. D}
  {\bfseries 96} (2017) 063515}
  [\href{https://arxiv.org/abs/1705.02133}{{\ttfamily 1705.02133}}].

\bibitem{Geng:2017mic}
C.-Q.~Geng, C.-C.~Lee, M.~Sami, E.N.~Saridakis and A.A.~Starobinsky,
  \emph{{Observational constraints on successful model of quintessential
  Inflation}}, \href{https://doi.org/10.1088/1475-7516/2017/06/011}{\emph{JCAP}
  {\bfseries 1706} (2017) 011}
  [\href{https://arxiv.org/abs/1705.01329}{{\ttfamily 1705.01329}}].

\bibitem{Dimopoulos:2018eam}
K.~Dimopoulos and T.~Markkanen, \emph{{Dark energy as a remnant of inflation
  and electroweak symmetry breaking}},
  \href{https://doi.org/10.1007/JHEP01(2019)029}{\emph{JHEP} {\bfseries 01}
  (2019) 029} [\href{https://arxiv.org/abs/1807.04359}{{\ttfamily
  1807.04359}}].

\bibitem{Dimopoulos:2019ogl}
K.~Dimopoulos, M.~Kar\v{c}iauskas and C.~Owen, \emph{{Quintessential inflation
  with a trap and axionic dark matter}},
  \href{https://doi.org/10.1103/PhysRevD.100.083530}{\emph{Phys. Rev. D}
  {\bfseries 100} (2019) 083530}
  [\href{https://arxiv.org/abs/1907.04676}{{\ttfamily 1907.04676}}].

\bibitem{Benisty:2020xqm}
D.~Benisty and E.I.~Guendelman, \emph{{Lorentzian Quintessential Inflation}},
  \href{https://doi.org/10.1142/S021827182042002X}{\emph{Int. J. Mod. Phys. D}
  {\bfseries 29} (2020) 2042002}
  [\href{https://arxiv.org/abs/2004.00339}{{\ttfamily 2004.00339}}].

\bibitem{Benisty:2020qta}
D.~Benisty and E.I.~Guendelman, \emph{{Quintessential Inflation from Lorentzian
  Slow Roll}}, \href{https://doi.org/10.1140/epjc/s10052-020-8147-8}{\emph{Eur.
  Phys. J. C} {\bfseries 80} (2020) 577}
  [\href{https://arxiv.org/abs/2006.04129}{{\ttfamily 2006.04129}}].

\bibitem{Karciauskas:2021fdu}
M.~Kar\v{c}iauskas, S.~Rusak and A.~Saez, \emph{{Quintessential Inflation and
  the Non-Linear Effects of the Tachyonic Trapping Mechanism}},
  \href{https://arxiv.org/abs/2112.11536}{{\ttfamily 2112.11536}}.

\bibitem{Dimopoulos:2019gpz}
K.~Dimopoulos and L.~Donaldson-Wood, \emph{{Warm quintessential inflation}},
  \href{https://doi.org/10.1016/j.physletb.2019.07.017}{\emph{Phys. Lett. B}
  {\bfseries 796} (2019) 26}
  [\href{https://arxiv.org/abs/1906.09648}{{\ttfamily 1906.09648}}].

\bibitem{Feng:2002nb}
B.~Feng and M.-z.~Li, \emph{{Curvaton reheating in nonoscillatory inflationary
  models}}, \href{https://doi.org/10.1016/S0370-2693(03)00589-6}{\emph{Phys.
  Lett. B} {\bfseries 564} (2003) 169}
  [\href{https://arxiv.org/abs/hep-ph/0212213}{{\ttfamily hep-ph/0212213}}].

\bibitem{Kamali:2019xnt}
V.~Kamali, M.~Motaharfar and R.O.~Ramos, \emph{{Warm brane inflation with an
  exponential potential: a consistent realization away from the swampland}},
  \href{https://doi.org/10.1103/PhysRevD.101.023535}{\emph{Phys. Rev. D}
  {\bfseries 101} (2020) 023535}
  [\href{https://arxiv.org/abs/1910.06796}{{\ttfamily 1910.06796}}].

\bibitem{Guendelman:2014bva}
E.~Guendelman, R.~Herrera, P.~Labrana, E.~Nissimov and S.~Pacheva,
  \emph{{Emergent Cosmology, Inflation and Dark Energy}},
  \href{https://doi.org/10.1007/s10714-015-1852-1}{\emph{Gen. Rel. Grav.}
  {\bfseries 47} (2015) 10} [\href{https://arxiv.org/abs/1408.5344}{{\ttfamily
  1408.5344}}].

\bibitem{Guendelman:2015liz}
E.I.~Guendelman and R.~Herrera, \emph{{Curvaton reheating mechanism in a scale
  invariant two measures theory}},
  \href{https://doi.org/10.1007/s10714-015-1999-9}{\emph{Gen. Rel. Grav.}
  {\bfseries 48} (2016) 3} [\href{https://arxiv.org/abs/1511.08645}{{\ttfamily
  1511.08645}}].

\bibitem{Planck:2018vyg}
{\scshape Planck} collaboration, \emph{{Planck 2018 results. VI. Cosmological
  parameters}},
  \href{https://doi.org/10.1051/0004-6361/201833910}{\emph{Astron. Astrophys.}
  {\bfseries 641} (2020) A6}
  [\href{https://arxiv.org/abs/1807.06209}{{\ttfamily 1807.06209}}].

\bibitem{ParticleDataGroup:2020ssz}
{\scshape Particle Data Group} collaboration, \emph{{Review of Particle
  Physics}}, \href{https://doi.org/10.1093/ptep/ptaa104}{\emph{PTEP} {\bfseries
  2020} (2020) 083C01}.

\bibitem{Chiba:2012cb}
T.~Chiba, A.~De~Felice and S.~Tsujikawa, \emph{{Observational constraints on
  quintessence: thawing, tracker, and scaling models}},
  \href{https://doi.org/10.1103/PhysRevD.87.083505}{\emph{Phys. Rev.}
  {\bfseries D87} (2013) 083505}
  [\href{https://arxiv.org/abs/1210.3859}{{\ttfamily 1210.3859}}].

\bibitem{Geng:2015fla}
C.-Q.~Geng, M.W.~Hossain, R.~Myrzakulov, M.~Sami and E.N.~Saridakis,
  \emph{{Quintessential inflation with canonical and noncanonical scalar fields
  and Planck 2015 results}},
  \href{https://doi.org/10.1103/PhysRevD.92.023522}{\emph{Phys. Rev. D}
  {\bfseries 92} (2015) 023522}
  [\href{https://arxiv.org/abs/1502.03597}{{\ttfamily 1502.03597}}].

\bibitem{Durrive:2018quo}
J.-B.~Durrive, J.~Ooba, K.~Ichiki and N.~Sugiyama, \emph{{Updated observational
  constraints on quintessence dark energy models}},
  \href{https://doi.org/10.1103/PhysRevD.97.043503}{\emph{Phys. Rev. D}
  {\bfseries 97} (2018) 043503}
  [\href{https://arxiv.org/abs/1801.09446}{{\ttfamily 1801.09446}}].

\bibitem{Salopek:1990jq}
D.S.~Salopek and J.R.~Bond, \emph{{Nonlinear evolution of long wavelength
  metric fluctuations in inflationary models}},
  \href{https://doi.org/10.1103/PhysRevD.42.3936}{\emph{Phys. Rev. D}
  {\bfseries 42} (1990) 3936}.

\bibitem{Rubio:2017gty}
J.~Rubio and C.~Wetterich, \emph{{Emergent scale symmetry: Connecting inflation
  and dark energy}},
  \href{https://doi.org/10.1103/PhysRevD.96.063509}{\emph{Phys. Rev.}
  {\bfseries D96} (2017) 063509}
  [\href{https://arxiv.org/abs/1705.00552}{{\ttfamily 1705.00552}}].

\bibitem{Dimopoulos:2017zvq}
K.~Dimopoulos and C.~Owen, \emph{{Quintessential Inflation with
  $\alpha$-attractors}},
  \href{https://doi.org/10.1088/1475-7516/2017/06/027}{\emph{JCAP} {\bfseries
  1706} (2017) 027} [\href{https://arxiv.org/abs/1703.00305}{{\ttfamily
  1703.00305}}].

\bibitem{Akrami:2017cir}
Y.~Akrami, R.~Kallosh, A.~Linde and V.~Vardanyan, \emph{{Dark energy,
  $\alpha$-attractors, and large-scale structure surveys}},
  \href{https://doi.org/10.1088/1475-7516/2018/06/041}{\emph{JCAP} {\bfseries
  1806} (2018) 041} [\href{https://arxiv.org/abs/1712.09693}{{\ttfamily
  1712.09693}}].

\bibitem{Mukhanov:1981xt}
V.F.~Mukhanov and G.V.~Chibisov, \emph{{Quantum Fluctuations and a Nonsingular
  Universe}}, {\emph{JETP Lett.} {\bfseries 33} (1981) 532}.

\bibitem{Guth:1982ec}
A.H.~Guth and S.Y.~Pi, \emph{{Fluctuations in the New Inflationary Universe}},
  \href{https://doi.org/10.1103/PhysRevLett.49.1110}{\emph{Phys. Rev. Lett.}
  {\bfseries 49} (1982) 1110}.

\bibitem{Starobinsky:1982ee}
A.A.~Starobinsky, \emph{{Dynamics of Phase Transition in the New Inflationary
  Universe Scenario and Generation of Perturbations}},
  \href{https://doi.org/10.1016/0370-2693(82)90541-X}{\emph{Phys. Lett. B}
  {\bfseries 117} (1982) 175}.

\bibitem{Hawking:1982cz}
S.W.~Hawking, \emph{{The Development of Irregularities in a Single Bubble
  Inflationary Universe}},
  \href{https://doi.org/10.1016/0370-2693(82)90373-2}{\emph{Phys. Lett. B}
  {\bfseries 115} (1982) 295}.

\bibitem{Bardeen:1983qw}
J.M.~Bardeen, P.J.~Steinhardt and M.S.~Turner, \emph{{Spontaneous Creation of
  Almost Scale - Free Density Perturbations in an Inflationary Universe}},
  \href{https://doi.org/10.1103/PhysRevD.28.679}{\emph{Phys. Rev. D} {\bfseries
  28} (1983) 679}.

\bibitem{Lesgourgues:1996jc}
J.~Lesgourgues, D.~Polarski and A.A.~Starobinsky, \emph{{Quantum to classical
  transition of cosmological perturbations for nonvacuum initial states}},
  \href{https://doi.org/10.1016/S0550-3213(97)00224-1}{\emph{Nucl. Phys. B}
  {\bfseries 497} (1997) 479}
  [\href{https://arxiv.org/abs/gr-qc/9611019}{{\ttfamily gr-qc/9611019}}].

\bibitem{Polarski:1995jg}
D.~Polarski and A.A.~Starobinsky, \emph{{Semiclassicality and decoherence of
  cosmological perturbations}},
  \href{https://doi.org/10.1088/0264-9381/13/3/006}{\emph{Class. Quant. Grav.}
  {\bfseries 13} (1996) 377}
  [\href{https://arxiv.org/abs/gr-qc/9504030}{{\ttfamily gr-qc/9504030}}].

\bibitem{Kiefer:1998qe}
C.~Kiefer, D.~Polarski and A.A.~Starobinsky, \emph{{Quantum to classical
  transition for fluctuations in the early universe}},
  \href{https://doi.org/10.1142/S0218271898000292}{\emph{Int. J. Mod. Phys. D}
  {\bfseries 7} (1998) 455}
  [\href{https://arxiv.org/abs/gr-qc/9802003}{{\ttfamily gr-qc/9802003}}].

\bibitem{Kiefer:1998pb}
C.~Kiefer, J.~Lesgourgues, D.~Polarski and A.A.~Starobinsky, \emph{{The
  Coherence of primordial fluctuations produced during inflation}},
  \href{https://doi.org/10.1088/0264-9381/15/10/002}{\emph{Class. Quant. Grav.}
  {\bfseries 15} (1998) L67}
  [\href{https://arxiv.org/abs/gr-qc/9806066}{{\ttfamily gr-qc/9806066}}].

\bibitem{Kiefer:1998jk}
C.~Kiefer and D.~Polarski, \emph{{Emergence of classicality for primordial
  fluctuations: Concepts and analogies}},
  \href{https://doi.org/10.1002/andp.2090070302}{\emph{Annalen Phys.}
  {\bfseries 7} (1998) 137}
  [\href{https://arxiv.org/abs/gr-qc/9805014}{{\ttfamily gr-qc/9805014}}].

\bibitem{Joyce:1996cp}
M.~Joyce, \emph{{Electroweak Baryogenesis and the Expansion Rate of the
  Universe}}, \href{https://doi.org/10.1103/PhysRevD.55.1875}{\emph{Phys. Rev.
  D} {\bfseries 55} (1997) 1875}
  [\href{https://arxiv.org/abs/hep-ph/9606223}{{\ttfamily hep-ph/9606223}}].

\bibitem{Joyce:1997fc}
M.~Joyce and T.~Prokopec, \emph{{Turning around the sphaleron bound:
  Electroweak baryogenesis in an alternative postinflationary cosmology}},
  \href{https://doi.org/10.1103/PhysRevD.57.6022}{\emph{Phys. Rev. D}
  {\bfseries 57} (1998) 6022}
  [\href{https://arxiv.org/abs/hep-ph/9709320}{{\ttfamily hep-ph/9709320}}].

\bibitem{Bettoni:2018utf}
D.~Bettoni and J.~Rubio, \emph{{Quintessential Affleck-Dine baryogenesis with
  non-minimal couplings}},
  \href{https://doi.org/10.1016/j.physletb.2018.07.046}{\emph{Phys. Lett. B}
  {\bfseries 784} (2018) 122}
  [\href{https://arxiv.org/abs/1805.02669}{{\ttfamily 1805.02669}}].

\bibitem{Kamionkowski:1990ni}
M.~Kamionkowski and M.S.~Turner, \emph{{THERMAL RELICS: DO WE KNOW THEIR
  ABUNDANCES?}}, \href{https://doi.org/10.1103/PhysRevD.42.3310}{\emph{Phys.
  Rev. D} {\bfseries 42} (1990) 3310}.

\bibitem{Salati:2002md}
P.~Salati, \emph{{Quintessence and the relic density of neutralinos}},
  \href{https://doi.org/10.1016/j.physletb.2003.07.073}{\emph{Phys. Lett. B}
  {\bfseries 571} (2003) 121}
  [\href{https://arxiv.org/abs/astro-ph/0207396}{{\ttfamily
  astro-ph/0207396}}].

\bibitem{Profumo:2003hq}
S.~Profumo and P.~Ullio, \emph{{SUSY dark matter and quintessence}},
  \href{https://doi.org/10.1088/1475-7516/2003/11/006}{\emph{JCAP} {\bfseries
  11} (2003) 006} [\href{https://arxiv.org/abs/hep-ph/0309220}{{\ttfamily
  hep-ph/0309220}}].

\bibitem{Chung:2007vz}
D.J.H.~Chung, L.L.~Everett and K.T.~Matchev, \emph{{Inflationary cosmology
  connecting dark energy and dark matter}},
  \href{https://doi.org/10.1103/PhysRevD.76.103530}{\emph{Phys. Rev. D}
  {\bfseries 76} (2007) 103530}
  [\href{https://arxiv.org/abs/0704.3285}{{\ttfamily 0704.3285}}].

\bibitem{Visinelli:2009kt}
L.~Visinelli and P.~Gondolo, \emph{{Axion cold dark matter in non-standard
  cosmologies}}, \href{https://doi.org/10.1103/PhysRevD.81.063508}{\emph{Phys.
  Rev. D} {\bfseries 81} (2010) 063508}
  [\href{https://arxiv.org/abs/0912.0015}{{\ttfamily 0912.0015}}].

\bibitem{Redmond:2017tja}
K.~Redmond and A.L.~Erickcek, \emph{{New Constraints on Dark Matter Production
  during Kination}},
  \href{https://doi.org/10.1103/PhysRevD.96.043511}{\emph{Phys. Rev. D}
  {\bfseries 96} (2017) 043511}
  [\href{https://arxiv.org/abs/1704.01056}{{\ttfamily 1704.01056}}].

\bibitem{DEramo:2017gpl}
F.~D'Eramo, N.~Fernandez and S.~Profumo, \emph{{When the Universe Expands Too
  Fast: Relentless Dark Matter}},
  \href{https://doi.org/10.1088/1475-7516/2017/05/012}{\emph{JCAP} {\bfseries
  05} (2017) 012} [\href{https://arxiv.org/abs/1703.04793}{{\ttfamily
  1703.04793}}].

\bibitem{DEramo:2017ecx}
F.~D'Eramo, N.~Fernandez and S.~Profumo, \emph{{Dark Matter Freeze-in
  Production in Fast-Expanding Universes}},
  \href{https://doi.org/10.1088/1475-7516/2018/02/046}{\emph{JCAP} {\bfseries
  02} (2018) 046} [\href{https://arxiv.org/abs/1712.07453}{{\ttfamily
  1712.07453}}].

\bibitem{Visinelli:2017qga}
L.~Visinelli, \emph{{(Non-)thermal production of WIMPs during kination}},
  \href{https://doi.org/10.3390/sym10110546}{\emph{Symmetry} {\bfseries 10}
  (2018) 546} [\href{https://arxiv.org/abs/1710.11006}{{\ttfamily
  1710.11006}}].

\bibitem{Bernal:2020bfj}
N.~Bernal, J.~Rubio and H.~Veerm\"ae, \emph{{Boosting Ultraviolet Freeze-in in
  NO Models}}, \href{https://doi.org/10.1088/1475-7516/2020/06/047}{\emph{JCAP}
  {\bfseries 06} (2020) 047}
  [\href{https://arxiv.org/abs/2004.13706}{{\ttfamily 2004.13706}}].

\bibitem{Giovannini:1998bp}
M.~Giovannini, \emph{{Gravitational waves constraints on postinflationary
  phases stiffer than radiation}},
  \href{https://doi.org/10.1103/PhysRevD.58.083504}{\emph{Phys. Rev. D}
  {\bfseries 58} (1998) 083504}
  [\href{https://arxiv.org/abs/hep-ph/9806329}{{\ttfamily hep-ph/9806329}}].

\bibitem{Giovannini:1999bh}
M.~Giovannini, \emph{{Production and detection of relic gravitons in
  quintessential inflationary models}},
  \href{https://doi.org/10.1103/PhysRevD.60.123511}{\emph{Phys. Rev. D}
  {\bfseries 60} (1999) 123511}
  [\href{https://arxiv.org/abs/astro-ph/9903004}{{\ttfamily
  astro-ph/9903004}}].

\bibitem{Tashiro:2003qp}
H.~Tashiro, T.~Chiba and M.~Sasaki, \emph{{Reheating after quintessential
  inflation and gravitational waves}},
  \href{https://doi.org/10.1088/0264-9381/21/7/004}{\emph{Class. Quant. Grav.}
  {\bfseries 21} (2004) 1761}
  [\href{https://arxiv.org/abs/gr-qc/0307068}{{\ttfamily gr-qc/0307068}}].

\bibitem{Caprini:2018mtu}
C.~Caprini and D.G.~Figueroa, \emph{{Cosmological Backgrounds of Gravitational
  Waves}}, \href{https://doi.org/10.1088/1361-6382/aac608}{\emph{Class. Quant.
  Grav.} {\bfseries 35} (2018) 163001}
  [\href{https://arxiv.org/abs/1801.04268}{{\ttfamily 1801.04268}}].

\bibitem{Figueroa:2018twl}
D.G.~Figueroa and E.H.~Tanin, \emph{{Inconsistency of an inflationary sector
  coupled only to Einstein gravity}},
  \href{https://doi.org/10.1088/1475-7516/2019/10/050}{\emph{JCAP} {\bfseries
  10} (2019) 050} [\href{https://arxiv.org/abs/1811.04093}{{\ttfamily
  1811.04093}}].

\bibitem{Figueroa:2019paj}
D.G.~Figueroa and E.H.~Tanin, \emph{{Ability of LIGO and LISA to probe the
  equation of state of the early Universe}},
  \href{https://doi.org/10.1088/1475-7516/2019/08/011}{\emph{JCAP} {\bfseries
  08} (2019) 011} [\href{https://arxiv.org/abs/1905.11960}{{\ttfamily
  1905.11960}}].

\bibitem{Bernal:2019lpc}
N.~Bernal and F.~Hajkarim, \emph{{Primordial Gravitational Waves in Nonstandard
  Cosmologies}}, \href{https://doi.org/10.1103/PhysRevD.100.063502}{\emph{Phys.
  Rev. D} {\bfseries 100} (2019) 063502}
  [\href{https://arxiv.org/abs/1905.10410}{{\ttfamily 1905.10410}}].

\bibitem{Khlebnikov:1997di}
S.Y.~Khlebnikov and I.I.~Tkachev, \emph{{Relic gravitational waves produced
  after preheating}},
  \href{https://doi.org/10.1103/PhysRevD.56.653}{\emph{Phys. Rev. D} {\bfseries
  56} (1997) 653} [\href{https://arxiv.org/abs/hep-ph/9701423}{{\ttfamily
  hep-ph/9701423}}].

\bibitem{Easther:2006gt}
R.~Easther and E.A.~Lim, \emph{{Stochastic gravitational wave production after
  inflation}}, \href{https://doi.org/10.1088/1475-7516/2006/04/010}{\emph{JCAP}
  {\bfseries 04} (2006) 010}
  [\href{https://arxiv.org/abs/astro-ph/0601617}{{\ttfamily
  astro-ph/0601617}}].

\bibitem{Easther:2006vd}
R.~Easther, J.T.~Giblin, Jr. and E.A.~Lim, \emph{{Gravitational Wave Production
  At The End Of Inflation}},
  \href{https://doi.org/10.1103/PhysRevLett.99.221301}{\emph{Phys. Rev. Lett.}
  {\bfseries 99} (2007) 221301}
  [\href{https://arxiv.org/abs/astro-ph/0612294}{{\ttfamily
  astro-ph/0612294}}].

\bibitem{Dufaux:2007pt}
J.F.~Dufaux, A.~Bergman, G.N.~Felder, L.~Kofman and J.-P.~Uzan, \emph{{Theory
  and Numerics of Gravitational Waves from Preheating after Inflation}},
  \href{https://doi.org/10.1103/PhysRevD.76.123517}{\emph{Phys. Rev. D}
  {\bfseries 76} (2007) 123517}
  [\href{https://arxiv.org/abs/0707.0875}{{\ttfamily 0707.0875}}].

\bibitem{Garcia-Bellido:2007nns}
J.~Garcia-Bellido and D.G.~Figueroa, \emph{{A stochastic background of
  gravitational waves from hybrid preheating}},
  \href{https://doi.org/10.1103/PhysRevLett.98.061302}{\emph{Phys. Rev. Lett.}
  {\bfseries 98} (2007) 061302}
  [\href{https://arxiv.org/abs/astro-ph/0701014}{{\ttfamily
  astro-ph/0701014}}].

\bibitem{Garcia-Bellido:2007fiu}
J.~Garcia-Bellido, D.G.~Figueroa and A.~Sastre, \emph{{A Gravitational Wave
  Background from Reheating after Hybrid Inflation}},
  \href{https://doi.org/10.1103/PhysRevD.77.043517}{\emph{Phys. Rev. D}
  {\bfseries 77} (2008) 043517}
  [\href{https://arxiv.org/abs/0707.0839}{{\ttfamily 0707.0839}}].

\bibitem{Cui:2017ufi}
Y.~Cui, M.~Lewicki, D.E.~Morrissey and J.D.~Wells, \emph{{Cosmic Archaeology
  with Gravitational Waves from Cosmic Strings}},
  \href{https://doi.org/10.1103/PhysRevD.97.123505}{\emph{Phys. Rev. D}
  {\bfseries 97} (2018) 123505}
  [\href{https://arxiv.org/abs/1711.03104}{{\ttfamily 1711.03104}}].

\bibitem{Cui:2018rwi}
Y.~Cui, M.~Lewicki, D.E.~Morrissey and J.D.~Wells, \emph{{Probing the pre-BBN
  universe with gravitational waves from cosmic strings}},
  \href{https://doi.org/10.1007/JHEP01(2019)081}{\emph{JHEP} {\bfseries 01}
  (2019) 081} [\href{https://arxiv.org/abs/1808.08968}{{\ttfamily
  1808.08968}}].

\bibitem{Bettoni:2018pbl}
D.~Bettoni, G.~Dom\`enech and J.~Rubio, \emph{{Gravitational waves from global
  cosmic strings in quintessential inflation}},
  \href{https://doi.org/10.1088/1475-7516/2019/02/034}{\emph{JCAP} {\bfseries
  02} (2019) 034} [\href{https://arxiv.org/abs/1810.11117}{{\ttfamily
  1810.11117}}].

\bibitem{Chang:2019mza}
C.-F.~Chang and Y.~Cui, \emph{{Stochastic Gravitational Wave Background from
  Global Cosmic Strings}},
  \href{https://doi.org/10.1016/j.dark.2020.100604}{\emph{Phys. Dark Univ.}
  {\bfseries 29} (2020) 100604}
  [\href{https://arxiv.org/abs/1910.04781}{{\ttfamily 1910.04781}}].

\bibitem{Gouttenoire:2019rtn}
Y.~Gouttenoire, G.~Servant and P.~Simakachorn, \emph{{BSM with Cosmic Strings:
  Heavy, up to EeV mass, Unstable Particles}},
  \href{https://doi.org/10.1088/1475-7516/2020/07/016}{\emph{JCAP} {\bfseries
  07} (2020) 016} [\href{https://arxiv.org/abs/1912.03245}{{\ttfamily
  1912.03245}}].

\bibitem{Chang:2021afa}
C.-F.~Chang and Y.~Cui, \emph{{Gravitational Waves from Global Cosmic Strings
  and Cosmic Archaeology}},  \href{https://arxiv.org/abs/2106.09746}{{\ttfamily
  2106.09746}}.

\bibitem{Chung:2010cb}
D.J.H.~Chung and P.~Zhou, \emph{{Gravity Waves as a Probe of Hubble Expansion
  Rate During An Electroweak Scale Phase Transition}},
  \href{https://doi.org/10.1103/PhysRevD.82.024027}{\emph{Phys. Rev. D}
  {\bfseries 82} (2010) 024027}
  [\href{https://arxiv.org/abs/1003.2462}{{\ttfamily 1003.2462}}].

\bibitem{Allahverdi:2020bys}
R.~Allahverdi et~al., \emph{{The First Three Seconds: a Review of Possible
  Expansion Histories of the Early Universe}},
  \href{https://arxiv.org/abs/2006.16182}{{\ttfamily 2006.16182}}.

\bibitem{Maggiore:1999vm}
M.~Maggiore, \emph{{Gravitational wave experiments and early universe
  cosmology}}, \href{https://doi.org/10.1016/S0370-1573(99)00102-7}{\emph{Phys.
  Rept.} {\bfseries 331} (2000) 283}
  [\href{https://arxiv.org/abs/gr-qc/9909001}{{\ttfamily gr-qc/9909001}}].

\bibitem{Mangano:2005cc}
G.~Mangano, G.~Miele, S.~Pastor, T.~Pinto, O.~Pisanti and P.D.~Serpico,
  \emph{{Relic neutrino decoupling including flavor oscillations}},
  \href{https://doi.org/10.1016/j.nuclphysb.2005.09.041}{\emph{Nucl. Phys. B}
  {\bfseries 729} (2005) 221}
  [\href{https://arxiv.org/abs/hep-ph/0506164}{{\ttfamily hep-ph/0506164}}].

\bibitem{Cyburt:2015mya}
R.H.~Cyburt, B.D.~Fields, K.A.~Olive and T.-H.~Yeh, \emph{{Big Bang
  Nucleosynthesis: 2015}},
  \href{https://doi.org/10.1103/RevModPhys.88.015004}{\emph{Rev. Mod. Phys.}
  {\bfseries 88} (2016) 015004}
  [\href{https://arxiv.org/abs/1505.01076}{{\ttfamily 1505.01076}}].

\bibitem{Gouttenoire:2021jhk}
Y.~Gouttenoire, G.~Servant and P.~Simakachorn, \emph{{Kination cosmology from
  scalar fields and gravitational-wave signatures}},
  \href{https://arxiv.org/abs/2111.01150}{{\ttfamily 2111.01150}}.

\bibitem{Bassett:2005xm}
B.A.~Bassett, S.~Tsujikawa and D.~Wands, \emph{{Inflation dynamics and
  reheating}}, \href{https://doi.org/10.1103/RevModPhys.78.537}{\emph{Rev. Mod.
  Phys.} {\bfseries 78} (2006) 537}
  [\href{https://arxiv.org/abs/astro-ph/0507632}{{\ttfamily
  astro-ph/0507632}}].

\bibitem{Allahverdi:2010xz}
R.~Allahverdi, R.~Brandenberger, F.-Y.~Cyr-Racine and A.~Mazumdar,
  \emph{{Reheating in Inflationary Cosmology: Theory and Applications}},
  \href{https://doi.org/10.1146/annurev.nucl.012809.104511}{\emph{Ann. Rev.
  Nucl. Part. Sci.} {\bfseries 60} (2010) 27}
  [\href{https://arxiv.org/abs/1001.2600}{{\ttfamily 1001.2600}}].

\bibitem{Kofman:1997yn}
L.~Kofman, A.D.~Linde and A.A.~Starobinsky, \emph{{Towards the theory of
  reheating after inflation}},
  \href{https://doi.org/10.1103/PhysRevD.56.3258}{\emph{Phys. Rev. D}
  {\bfseries 56} (1997) 3258}
  [\href{https://arxiv.org/abs/hep-ph/9704452}{{\ttfamily hep-ph/9704452}}].

\bibitem{Ford:1986sy}
L.H.~Ford, \emph{{Gravitational Particle Creation and Inflation}},
  \href{https://doi.org/10.1103/PhysRevD.35.2955}{\emph{Phys. Rev. D}
  {\bfseries 35} (1987) 2955}.

\bibitem{Planck:2015sxf}
{\scshape Planck} collaboration, \emph{{Planck 2015 results. XX. Constraints on
  inflation}}, \href{https://doi.org/10.1051/0004-6361/201525898}{\emph{Astron.
  Astrophys.} {\bfseries 594} (2016) A20}
  [\href{https://arxiv.org/abs/1502.02114}{{\ttfamily 1502.02114}}].

\bibitem{Chun:2009yu}
E.J.~Chun, S.~Scopel and I.~Zaballa, \emph{{Gravitational reheating in
  quintessential inflation}},
  \href{https://doi.org/10.1088/1475-7516/2009/07/022}{\emph{JCAP} {\bfseries
  07} (2009) 022} [\href{https://arxiv.org/abs/0904.0675}{{\ttfamily
  0904.0675}}].

\bibitem{Felder:1998vq}
G.N.~Felder, L.~Kofman and A.D.~Linde, \emph{{Instant preheating}},
  \href{https://doi.org/10.1103/PhysRevD.59.123523}{\emph{Phys. Rev. D}
  {\bfseries 59} (1999) 123523}
  [\href{https://arxiv.org/abs/hep-ph/9812289}{{\ttfamily hep-ph/9812289}}].

\bibitem{Felder:1999pv}
G.N.~Felder, L.~Kofman and A.D.~Linde, \emph{{Inflation and preheating in NO
  models}}, \href{https://doi.org/10.1103/PhysRevD.60.103505}{\emph{Phys. Rev.
  D} {\bfseries 60} (1999) 103505}
  [\href{https://arxiv.org/abs/hep-ph/9903350}{{\ttfamily hep-ph/9903350}}].

\bibitem{Campos:2002yk}
A.H.~Campos, H.C.~Reis and R.~Rosenfeld, \emph{{Preheating in quintessential
  inflation}},
  \href{https://doi.org/10.1016/j.physletb.2003.09.064}{\emph{Phys. Lett. B}
  {\bfseries 575} (2003) 151}
  [\href{https://arxiv.org/abs/hep-ph/0210152}{{\ttfamily hep-ph/0210152}}].

\bibitem{Dimopoulos:2017tud}
K.~Dimopoulos, L.~Donaldson~Wood and C.~Owen, \emph{{Instant preheating in
  quintessential inflation with $\alpha$-attractors}},
  \href{https://doi.org/10.1103/PhysRevD.97.063525}{\emph{Phys. Rev. D}
  {\bfseries 97} (2018) 063525}
  [\href{https://arxiv.org/abs/1712.01760}{{\ttfamily 1712.01760}}].

\bibitem{BuenoSanchez:2007jxm}
J.C.~Bueno~Sanchez and K.~Dimopoulos, \emph{{Curvaton reheating allows TeV
  Hubble scale in NO inflation}},
  \href{https://doi.org/10.1088/1475-7516/2007/11/007}{\emph{JCAP} {\bfseries
  11} (2007) 007} [\href{https://arxiv.org/abs/0707.3967}{{\ttfamily
  0707.3967}}].

\bibitem{Dimopoulos:2018wfg}
K.~Dimopoulos and T.~Markkanen, \emph{{Non-minimal gravitational reheating
  during kination}},
  \href{https://doi.org/10.1088/1475-7516/2018/06/021}{\emph{JCAP} {\bfseries
  06} (2018) 021} [\href{https://arxiv.org/abs/1803.07399}{{\ttfamily
  1803.07399}}].

\bibitem{Opferkuch:2019zbd}
T.~Opferkuch, P.~Schwaller and B.A.~Stefanek, \emph{{Ricci Reheating}},
  \href{https://doi.org/10.1088/1475-7516/2019/07/016}{\emph{JCAP} {\bfseries
  07} (2019) 016} [\href{https://arxiv.org/abs/1905.06823}{{\ttfamily
  1905.06823}}].

\bibitem{Bettoni:2021zhq}
D.~Bettoni, A.~Lopez-Eiguren and J.~Rubio, \emph{{Hubble-induced phase
  transitions on the lattice with applications to Ricci reheating}},
  \href{https://arxiv.org/abs/2107.09671}{{\ttfamily 2107.09671}}.

\bibitem{Birrell:1982ix}
N.D.~Birrell and P.C.W.~Davies, \emph{{Quantum Fields in Curved Space}},
  Cambridge Monographs on Mathematical Physics, Cambridge Univ. Press,
  Cambridge, UK (2, 1984),
  \href{https://doi.org/10.1017/CBO9780511622632}{10.1017/CBO9780511622632}.

\bibitem{Mukhanov:2007zz}
V.~Mukhanov and S.~Winitzki, \emph{{Introduction to quantum effects in
  gravity}}, Cambridge University Press (6, 2007).

\bibitem{Micha:2002ey}
R.~Micha and I.I.~Tkachev, \emph{{Relativistic turbulence: A Long way from
  preheating to equilibrium}},
  \href{https://doi.org/10.1103/PhysRevLett.90.121301}{\emph{Phys. Rev. Lett.}
  {\bfseries 90} (2003) 121301}
  [\href{https://arxiv.org/abs/hep-ph/0210202}{{\ttfamily hep-ph/0210202}}].

\bibitem{Micha:2004bv}
R.~Micha and I.I.~Tkachev, \emph{{Turbulent thermalization}},
  \href{https://doi.org/10.1103/PhysRevD.70.043538}{\emph{Phys. Rev. D}
  {\bfseries 70} (2004) 043538}
  [\href{https://arxiv.org/abs/hep-ph/0403101}{{\ttfamily hep-ph/0403101}}].

\bibitem{Damour:1995pd}
T.~Damour and A.~Vilenkin, \emph{{String theory and inflation}},
  \href{https://doi.org/10.1103/PhysRevD.53.2981}{\emph{Phys. Rev. D}
  {\bfseries 53} (1996) 2981}
  [\href{https://arxiv.org/abs/hep-th/9503149}{{\ttfamily hep-th/9503149}}].

\bibitem{Haro:2018jtb}
J.~Haro, \emph{{Different reheating mechanisms in quintessence inflation}},
  \href{https://doi.org/10.1103/PhysRevD.99.043510}{\emph{Phys. Rev. D}
  {\bfseries 99} (2019) 043510}
  [\href{https://arxiv.org/abs/1807.07367}{{\ttfamily 1807.07367}}].

\bibitem{Haro:2018zdb}
J.~Haro, W.~Yang and S.~Pan, \emph{{Reheating in quintessential inflation via
  gravitational production of heavy massive particles: A detailed analysis}},
  \href{https://doi.org/10.1088/1475-7516/2019/01/023}{\emph{JCAP} {\bfseries
  01} (2019) 023} [\href{https://arxiv.org/abs/1811.07371}{{\ttfamily
  1811.07371}}].

\bibitem{Salo:2021vdv}
L.A.~Sal\'o and J.~de~Haro, \emph{{Gravitational particle production of
  superheavy massive particles in Quintessential Inflation: A numerical
  analysis}},  \href{https://arxiv.org/abs/2108.10795}{{\ttfamily 2108.10795}}.

\bibitem{Wetterich:2014gaa}
C.~Wetterich, \emph{{Inflation, quintessence, and the origin of mass}},
  \href{https://doi.org/10.1016/j.nuclphysb.2015.05.019}{\emph{Nucl. Phys. B}
  {\bfseries 897} (2015) 111}
  [\href{https://arxiv.org/abs/1408.0156}{{\ttfamily 1408.0156}}].

\bibitem{Wetterich:2003qb}
C.~Wetterich, \emph{{Cosmology with varying scales and couplings}},  in
  \emph{{Proceedings, 5th Internationa Conference on Strong and Electroweak
  Matter (SEWM 2002): Heidelberg, Germany, October 2-5, 2002}}, pp.~230--249,
  2003, \href{https://doi.org/10.1142/9789812704498_0022}{DOI}
  [\href{https://arxiv.org/abs/hep-ph/0302116}{{\ttfamily hep-ph/0302116}}].

\bibitem{Scherrer:2007pu}
R.J.~Scherrer and A.A.~Sen, \emph{{Thawing quintessence with a nearly flat
  potential}}, \href{https://doi.org/10.1103/PhysRevD.77.083515}{\emph{Phys.
  Rev.} {\bfseries D77} (2008) 083515}
  [\href{https://arxiv.org/abs/0712.3450}{{\ttfamily 0712.3450}}].

\bibitem{Caldwell:2005tm}
R.R.~Caldwell and E.V.~Linder, \emph{{The Limits of quintessence}},
  \href{https://doi.org/10.1103/PhysRevLett.95.141301}{\emph{Phys. Rev. Lett.}
  {\bfseries 95} (2005) 141301}
  [\href{https://arxiv.org/abs/astro-ph/0505494}{{\ttfamily
  astro-ph/0505494}}].

\bibitem{Wetterich:1987fm}
C.~Wetterich, \emph{{Cosmology and the Fate of Dilatation Symmetry}},
  \href{https://doi.org/10.1016/0550-3213(88)90193-9}{\emph{Nucl. Phys.}
  {\bfseries B302} (1988) 668}
  [\href{https://arxiv.org/abs/1711.03844}{{\ttfamily 1711.03844}}].

\bibitem{Copeland:1997et}
E.J.~Copeland, A.R.~Liddle and D.~Wands, \emph{{Exponential potentials and
  cosmological scaling solutions}},
  \href{https://doi.org/10.1103/PhysRevD.57.4686}{\emph{Phys. Rev.} {\bfseries
  D57} (1998) 4686} [\href{https://arxiv.org/abs/gr-qc/9711068}{{\ttfamily
  gr-qc/9711068}}].

\bibitem{Ferreira:1997hj}
P.G.~Ferreira and M.~Joyce, \emph{{Cosmology with a primordial scaling field}},
  \href{https://doi.org/10.1103/PhysRevD.58.023503}{\emph{Phys. Rev.}
  {\bfseries D58} (1998) 023503}
  [\href{https://arxiv.org/abs/astro-ph/9711102}{{\ttfamily
  astro-ph/9711102}}].

\bibitem{Casas:2017wjh}
S.~Casas, M.~Pauly and J.~Rubio, \emph{{Higgs-dilaton cosmology: An
  inflation–dark-energy connection and forecasts for future galaxy surveys}},
  \href{https://doi.org/10.1103/PhysRevD.97.043520}{\emph{Phys. Rev.}
  {\bfseries D97} (2018) 043520}
  [\href{https://arxiv.org/abs/1712.04956}{{\ttfamily 1712.04956}}].

\bibitem{Dutta:2008qn}
S.~Dutta and R.J.~Scherrer, \emph{{Hilltop Quintessence}},
  \href{https://doi.org/10.1103/PhysRevD.78.123525}{\emph{Phys. Rev. D}
  {\bfseries 78} (2008) 123525}
  [\href{https://arxiv.org/abs/0809.4441}{{\ttfamily 0809.4441}}].

\bibitem{Chiba:2009sj}
T.~Chiba, \emph{{Slow-Roll Thawing Quintessence}},
  \href{https://doi.org/10.1103/PhysRevD.80.109902}{\emph{Phys. Rev. D}
  {\bfseries 79} (2009) 083517}
  [\href{https://arxiv.org/abs/0902.4037}{{\ttfamily 0902.4037}}].

\bibitem{Zlatev:1998tr}
I.~Zlatev, L.-M.~Wang and P.J.~Steinhardt, \emph{{Quintessence, cosmic
  coincidence, and the cosmological constant}},
  \href{https://doi.org/10.1103/PhysRevLett.82.896}{\emph{Phys. Rev. Lett.}
  {\bfseries 82} (1999) 896}
  [\href{https://arxiv.org/abs/astro-ph/9807002}{{\ttfamily
  astro-ph/9807002}}].

\bibitem{Steinhardt:1999nw}
P.J.~Steinhardt, L.-M.~Wang and I.~Zlatev, \emph{{Cosmological tracking
  solutions}}, \href{https://doi.org/10.1103/PhysRevD.59.123504}{\emph{Phys.
  Rev. D} {\bfseries 59} (1999) 123504}
  [\href{https://arxiv.org/abs/astro-ph/9812313}{{\ttfamily
  astro-ph/9812313}}].

\bibitem{Chiba:2009gg}
T.~Chiba, \emph{{The Equation of State of Tracker Fields}},
  \href{https://doi.org/10.1103/PhysRevD.81.023515}{\emph{Phys. Rev. D}
  {\bfseries 81} (2010) 023515}
  [\href{https://arxiv.org/abs/0909.4365}{{\ttfamily 0909.4365}}].

\bibitem{Wetterich:1994bg}
C.~Wetterich, \emph{{The Cosmon model for an asymptotically vanishing time
  dependent cosmological 'constant'}}, {\emph{Astron. Astrophys.} {\bfseries
  301} (1995) 321} [\href{https://arxiv.org/abs/hep-th/9408025}{{\ttfamily
  hep-th/9408025}}].

\bibitem{Ratra:1987rm}
B.~Ratra and P.J.E.~Peebles, \emph{{Cosmological Consequences of a Rolling
  Homogeneous Scalar Field}},
  \href{https://doi.org/10.1103/PhysRevD.37.3406}{\emph{Phys. Rev.} {\bfseries
  D37} (1988) 3406}.

\bibitem{Barreiro:1999zs}
T.~Barreiro, E.J.~Copeland and N.J.~Nunes, \emph{{Quintessence arising from
  exponential potentials}},
  \href{https://doi.org/10.1103/PhysRevD.61.127301}{\emph{Phys. Rev. D}
  {\bfseries 61} (2000) 127301}
  [\href{https://arxiv.org/abs/astro-ph/9910214}{{\ttfamily
  astro-ph/9910214}}].

\bibitem{Wang:2011bi}
P.-Y.~Wang, C.-W.~Chen and P.~Chen, \emph{{Confronting Tracker Field
  Quintessence with Data}},
  \href{https://doi.org/10.1088/1475-7516/2012/02/016}{\emph{JCAP} {\bfseries
  02} (2012) 016} [\href{https://arxiv.org/abs/1108.1424}{{\ttfamily
  1108.1424}}].

\bibitem{Tsujikawa:2013fta}
S.~Tsujikawa, \emph{{Quintessence: A Review}},
  \href{https://doi.org/10.1088/0264-9381/30/21/214003}{\emph{Class. Quant.
  Grav.} {\bfseries 30} (2013) 214003}
  [\href{https://arxiv.org/abs/1304.1961}{{\ttfamily 1304.1961}}].

\bibitem{Langlois:2018dxi}
D.~Langlois, \emph{{Dark energy and modified gravity in degenerate higher-order
  scalar\textendash{}tensor (DHOST) theories: A review}},
  \href{https://doi.org/10.1142/S0218271819420069}{\emph{Int. J. Mod. Phys. D}
  {\bfseries 28} (2019) 1942006}
  [\href{https://arxiv.org/abs/1811.06271}{{\ttfamily 1811.06271}}].

\bibitem{Sotiriou:2008rp}
T.P.~Sotiriou and V.~Faraoni, \emph{{f(R) Theories Of Gravity}},
  \href{https://doi.org/10.1103/RevModPhys.82.451}{\emph{Rev. Mod. Phys.}
  {\bfseries 82} (2010) 451} [\href{https://arxiv.org/abs/0805.1726}{{\ttfamily
  0805.1726}}].

\bibitem{Flanagan:2004bz}
E.E.~Flanagan, \emph{{The Conformal frame freedom in theories of gravitation}},
  \href{https://doi.org/10.1088/0264-9381/21/15/N02}{\emph{Class. Quant. Grav.}
  {\bfseries 21} (2004) 3817}
  [\href{https://arxiv.org/abs/gr-qc/0403063}{{\ttfamily gr-qc/0403063}}].

\bibitem{Jarv:2014hma}
L.~Järv, P.~Kuusk, M.~Saal and O.~Vilson, \emph{{Invariant quantities in the
  scalar-tensor theories of gravitation}},
  \href{https://doi.org/10.1103/PhysRevD.91.024041}{\emph{Phys.\ Rev.\ D}
  {\bfseries 91} (2015) 024041}
  [\href{https://arxiv.org/abs/1411.1947}{{\ttfamily 1411.1947}}].

\bibitem{Burns:2016ric}
D.~Burns, S.~Karamitsos and A.~Pilaftsis, \emph{{Frame-Covariant Formulation of
  Inflation in Scalar-Curvature Theories}},
  \href{https://doi.org/10.1016/j.nuclphysb.2016.04.036}{\emph{Nucl. Phys. B}
  {\bfseries 907} (2016) 785}
  [\href{https://arxiv.org/abs/1603.03730}{{\ttfamily 1603.03730}}].

\bibitem{Wetterich:2013wza}
C.~Wetterich, \emph{{Cosmon inflation}},
  \href{https://doi.org/10.1016/j.physletb.2013.08.023}{\emph{Phys. Lett. B}
  {\bfseries 726} (2013) 15} [\href{https://arxiv.org/abs/1303.4700}{{\ttfamily
  1303.4700}}].

\bibitem{Wetterich:2013jsa}
C.~Wetterich, \emph{{Variable gravity Universe}},
  \href{https://doi.org/10.1103/PhysRevD.89.024005}{\emph{Phys. Rev. D}
  {\bfseries 89} (2014) 024005}
  [\href{https://arxiv.org/abs/1308.1019}{{\ttfamily 1308.1019}}].

\bibitem{Hebecker:2000zb}
A.~Hebecker and C.~Wetterich, \emph{{Natural quintessence?}},
  \href{https://doi.org/10.1016/S0370-2693(00)01339-3}{\emph{Phys. Lett. B}
  {\bfseries 497} (2001) 281}
  [\href{https://arxiv.org/abs/hep-ph/0008205}{{\ttfamily hep-ph/0008205}}].

\bibitem{Wetterich:2019qzx}
C.~Wetterich, \emph{{Quantum scale symmetry}},
  \href{https://arxiv.org/abs/1901.04741}{{\ttfamily 1901.04741}}.

\bibitem{Garcia-Garcia:2018hlc}
C.~García-García, E.V.~Linder, P.~Ruíz-Lapuente and M.~Zumalacárregui,
  \emph{{Dark energy from $\alpha$-attractors: phenomenology and observational
  constraints}},
  \href{https://doi.org/10.1088/1475-7516/2018/08/022}{\emph{JCAP} {\bfseries
  1808} (2018) 022} [\href{https://arxiv.org/abs/1803.00661}{{\ttfamily
  1803.00661}}].

\bibitem{Palti:2019pca}
E.~Palti, \emph{{The Swampland: Introduction and Review}},
  \href{https://doi.org/10.1002/prop.201900037}{\emph{Fortsch. Phys.}
  {\bfseries 67} (2019) 1900037}
  [\href{https://arxiv.org/abs/1903.06239}{{\ttfamily 1903.06239}}].

\bibitem{Kallosh:2013tua}
R.~Kallosh, A.~Linde and D.~Roest, \emph{{Universal Attractor for Inflation at
  Strong Coupling}},
  \href{https://doi.org/10.1103/PhysRevLett.112.011303}{\emph{Phys. Rev. Lett.}
  {\bfseries 112} (2014) 011303}
  [\href{https://arxiv.org/abs/1310.3950}{{\ttfamily 1310.3950}}].

\bibitem{Dimopoulos:2020pas}
K.~Dimopoulos and S.~S\'anchez~L\'opez, \emph{{Quintessential inflation in
  Palatini $f(R)$ gravity}},
  \href{https://doi.org/10.1103/PhysRevD.103.043533}{\emph{Phys. Rev. D}
  {\bfseries 103} (2021) 043533}
  [\href{https://arxiv.org/abs/2012.06831}{{\ttfamily 2012.06831}}].

\bibitem{Henz:2013oxa}
T.~Henz, J.M.~Pawlowski, A.~Rodigast and C.~Wetterich, \emph{{Dilaton Quantum
  Gravity}}, \href{https://doi.org/10.1016/j.physletb.2013.10.015}{\emph{Phys.
  Lett. B} {\bfseries 727} (2013) 298}
  [\href{https://arxiv.org/abs/1304.7743}{{\ttfamily 1304.7743}}].

\bibitem{Wetterich:2017ixo}
C.~Wetterich, \emph{{Graviton fluctuations erase the cosmological constant}},
  \href{https://doi.org/10.1016/j.physletb.2017.08.002}{\emph{Phys. Lett. B}
  {\bfseries 773} (2017) 6} [\href{https://arxiv.org/abs/1704.08040}{{\ttfamily
  1704.08040}}].

\bibitem{Luty:2012ww}
M.A.~Luty, J.~Polchinski and R.~Rattazzi, \emph{{The $a$-theorem and the
  Asymptotics of 4D Quantum Field Theory}},
  \href{https://doi.org/10.1007/JHEP01(2013)152}{\emph{JHEP} {\bfseries 01}
  (2013) 152} [\href{https://arxiv.org/abs/1204.5221}{{\ttfamily 1204.5221}}].

\bibitem{Dymarsky:2013pqa}
A.~Dymarsky, Z.~Komargodski, A.~Schwimmer and S.~Theisen, \emph{{On Scale and
  Conformal Invariance in Four Dimensions}},
  \href{https://doi.org/10.1007/JHEP10(2015)171}{\emph{JHEP} {\bfseries 10}
  (2015) 171} [\href{https://arxiv.org/abs/1309.2921}{{\ttfamily 1309.2921}}].

\bibitem{Wetterich:2014zta}
C.~Wetterich, \emph{{Eternal Universe}},
  \href{https://doi.org/10.1103/PhysRevD.90.043520}{\emph{Phys. Rev. D}
  {\bfseries 90} (2014) 043520}
  [\href{https://arxiv.org/abs/1404.0535}{{\ttfamily 1404.0535}}].

\bibitem{Adelberger:2009zz}
E.G.~Adelberger, J.H.~Gundlach, B.R.~Heckel, S.~Hoedl and S.~Schlamminger,
  \emph{{Torsion balance experiments: A low-energy frontier of particle
  physics}}, \href{https://doi.org/10.1016/j.ppnp.2008.08.002}{\emph{Prog.
  Part. Nucl. Phys.} {\bfseries 62} (2009) 102}.

\bibitem{Uzan:2010pm}
J.-P.~Uzan, \emph{{Varying Constants, Gravitation and Cosmology}},
  \href{https://doi.org/10.12942/lrr-2011-2}{\emph{Living Rev. Rel.} {\bfseries
  14} (2011) 2} [\href{https://arxiv.org/abs/1009.5514}{{\ttfamily
  1009.5514}}].

\bibitem{Shaposhnikov:2008xb}
M.~Shaposhnikov and D.~Zenhausern, \emph{{Scale invariance, unimodular gravity
  and dark energy}},
  \href{https://doi.org/10.1016/j.physletb.2008.11.054}{\emph{Phys. Lett.}
  {\bfseries B671} (2009) 187}
  [\href{https://arxiv.org/abs/0809.3395}{{\ttfamily 0809.3395}}].

\bibitem{Garcia-Bellido:2011kqb}
J.~Garcia-Bellido, J.~Rubio, M.~Shaposhnikov and D.~Zenhausern,
  \emph{{Higgs-Dilaton Cosmology: From the Early to the Late Universe}},
  \href{https://doi.org/10.1103/PhysRevD.84.123504}{\emph{Phys. Rev. D}
  {\bfseries 84} (2011) 123504}
  [\href{https://arxiv.org/abs/1107.2163}{{\ttfamily 1107.2163}}].

\bibitem{Ferreira:2016kxi}
P.G.~Ferreira, C.T.~Hill and G.G.~Ross, \emph{{No fifth force in a scale
  invariant universe}},
  \href{https://doi.org/10.1103/PhysRevD.95.064038}{\emph{Phys. Rev.}
  {\bfseries D95} (2017) 064038}
  [\href{https://arxiv.org/abs/1612.03157}{{\ttfamily 1612.03157}}].

\bibitem{Burrage:2018dvt}
C.~Burrage, E.J.~Copeland, P.~Millington and M.~Spannowsky, \emph{{Fifth
  forces, Higgs portals and broken scale invariance}},
  \href{https://arxiv.org/abs/1804.07180}{{\ttfamily 1804.07180}}.

\bibitem{Casas:2018fum}
S.~Casas, G.K.~Karananas, M.~Pauly and J.~Rubio, \emph{{Scale-invariant
  alternatives to general relativity. III. The inflation-dark energy
  connection}}, \href{https://doi.org/10.1103/PhysRevD.99.063512}{\emph{Phys.
  Rev.} {\bfseries D99} (2019) 063512}
  [\href{https://arxiv.org/abs/1811.05984}{{\ttfamily 1811.05984}}].

\bibitem{Zee:1978wi}
A.~Zee, \emph{{A Broken Symmetric Theory of Gravity}},
  \href{https://doi.org/10.1103/PhysRevLett.42.417}{\emph{Phys. Rev. Lett.}
  {\bfseries 42} (1979) 417}.

\bibitem{Wetterich:2013aca}
C.~Wetterich, \emph{{Universe without expansion}},
  \href{https://doi.org/10.1016/j.dark.2013.10.002}{\emph{Phys. Dark Univ.}
  {\bfseries 2} (2013) 184} [\href{https://arxiv.org/abs/1303.6878}{{\ttfamily
  1303.6878}}].

\bibitem{Wetterich:2015ccd}
C.~Wetterich, \emph{{Primordial cosmic fluctuations for variable gravity}},
  \href{https://doi.org/10.1088/1475-7516/2016/05/041}{\emph{JCAP} {\bfseries
  05} (2016) 041} [\href{https://arxiv.org/abs/1511.03530}{{\ttfamily
  1511.03530}}].

\bibitem{Karam:2017zno}
A.~Karam, T.~Pappas and K.~Tamvakis, \emph{{Frame-dependence of higher-order
  inflationary observables in scalar-tensor theories}},
  \href{https://doi.org/10.1103/PhysRevD.96.064036}{\emph{Phys. Rev. D}
  {\bfseries 96} (2017) 064036}
  [\href{https://arxiv.org/abs/1707.00984}{{\ttfamily 1707.00984}}].

\bibitem{Minkowski:1977sc}
P.~Minkowski, \emph{{$\mu \to e\gamma$ at a Rate of One Out of $10^{9}$ Muon
  Decays?}}, \href{https://doi.org/10.1016/0370-2693(77)90435-X}{\emph{Phys.
  Lett. B} {\bfseries 67} (1977) 421}.

\bibitem{Yanagida:1979as}
T.~Yanagida, \emph{{Horizontal gauge symmetry and masses of neutrinos}},
  {\emph{Conf. Proc. C} {\bfseries 7902131} (1979) 95}.

\bibitem{Gell-Mann:1979vob}
M.~Gell-Mann, P.~Ramond and R.~Slansky, \emph{{Complex Spinors and Unified
  Theories}}, {\emph{Conf. Proc. C} {\bfseries 790927} (1979) 315}
  [\href{https://arxiv.org/abs/1306.4669}{{\ttfamily 1306.4669}}].

\bibitem{Magg:1980ut}
M.~Magg and C.~Wetterich, \emph{{Neutrino Mass Problem and Gauge Hierarchy}},
  \href{https://doi.org/10.1016/0370-2693(80)90825-4}{\emph{Phys. Lett. B}
  {\bfseries 94} (1980) 61}.

\bibitem{Lazarides:1980nt}
G.~Lazarides, Q.~Shafi and C.~Wetterich, \emph{{Proton Lifetime and Fermion
  Masses in an SO(10) Model}},
  \href{https://doi.org/10.1016/0550-3213(81)90354-0}{\emph{Nucl. Phys. B}
  {\bfseries 181} (1981) 287}.

\bibitem{Mohapatra:1980yp}
R.N.~Mohapatra and G.~Senjanovic, \emph{{Neutrino Masses and Mixings in Gauge
  Models with Spontaneous Parity Violation}},
  \href{https://doi.org/10.1103/PhysRevD.23.165}{\emph{Phys. Rev. D} {\bfseries
  23} (1981) 165}.

\bibitem{Wetterich:2007kr}
C.~Wetterich, \emph{{Growing neutrinos and cosmological selection}},
  \href{https://doi.org/10.1016/j.physletb.2007.08.060}{\emph{Phys. Lett. B}
  {\bfseries 655} (2007) 201}
  [\href{https://arxiv.org/abs/0706.4427}{{\ttfamily 0706.4427}}].

\bibitem{Amendola:2007yx}
L.~Amendola, M.~Baldi and C.~Wetterich, \emph{{Quintessence cosmologies with a
  growing matter component}},
  \href{https://doi.org/10.1103/PhysRevD.78.023015}{\emph{Phys. Rev. D}
  {\bfseries 78} (2008) 023015}
  [\href{https://arxiv.org/abs/0706.3064}{{\ttfamily 0706.3064}}].

\bibitem{Will:2005va}
C.M.~Will, \emph{{The Confrontation between general relativity and
  experiment}}, \href{https://doi.org/10.12942/lrr-2006-3}{\emph{Living Rev.
  Rel.} {\bfseries 9} (2006) 3}
  [\href{https://arxiv.org/abs/gr-qc/0510072}{{\ttfamily gr-qc/0510072}}].

\bibitem{Elder:2019yyp}
B.~Elder, V.~Vardanyan, Y.~Akrami, P.~Brax, A.-C.~Davis and R.S.~Decca,
  \emph{{Classical symmetron force in Casimir experiments}},
  \href{https://doi.org/10.1103/PhysRevD.101.064065}{\emph{Phys. Rev. D}
  {\bfseries 101} (2020) 064065}
  [\href{https://arxiv.org/abs/1912.10015}{{\ttfamily 1912.10015}}].

\bibitem{Brax:2018zfb}
P.~Brax, A.-C.~Davis, B.~Elder and L.K.~Wong, \emph{{Constraining screened
  fifth forces with the electron magnetic moment}},
  \href{https://doi.org/10.1103/PhysRevD.97.084050}{\emph{Phys. Rev. D}
  {\bfseries 97} (2018) 084050}
  [\href{https://arxiv.org/abs/1802.05545}{{\ttfamily 1802.05545}}].

\bibitem{Uzan:2002vq}
J.-P.~Uzan, \emph{{The Fundamental Constants and Their Variation: Observational
  Status and Theoretical Motivations}},
  \href{https://doi.org/10.1103/RevModPhys.75.403}{\emph{Rev. Mod. Phys.}
  {\bfseries 75} (2003) 403}
  [\href{https://arxiv.org/abs/hep-ph/0205340}{{\ttfamily hep-ph/0205340}}].

\bibitem{Clifton:2011jh}
T.~Clifton, P.G.~Ferreira, A.~Padilla and C.~Skordis, \emph{{Modified Gravity
  and Cosmology}},
  \href{https://doi.org/10.1016/j.physrep.2012.01.001}{\emph{Phys. Rept.}
  {\bfseries 513} (2012) 1} [\href{https://arxiv.org/abs/1106.2476}{{\ttfamily
  1106.2476}}].

\bibitem{Ishak:2018his}
M.~Ishak, \emph{{Testing General Relativity in Cosmology}},
  \href{https://doi.org/10.1007/s41114-018-0017-4}{\emph{Living Rev. Rel.}
  {\bfseries 22} (2019) 1} [\href{https://arxiv.org/abs/1806.10122}{{\ttfamily
  1806.10122}}].

\bibitem{Joyce:2014kja}
A.~Joyce, B.~Jain, J.~Khoury and M.~Trodden, \emph{{Beyond the Cosmological
  Standard Model}},
  \href{https://doi.org/10.1016/j.physrep.2014.12.002}{\emph{Phys. Rept.}
  {\bfseries 568} (2015) 1} [\href{https://arxiv.org/abs/1407.0059}{{\ttfamily
  1407.0059}}].

\bibitem{Amendola:1999er}
L.~Amendola, \emph{{Coupled quintessence}},
  \href{https://doi.org/10.1103/PhysRevD.62.043511}{\emph{Phys. Rev. D}
  {\bfseries 62} (2000) 043511}
  [\href{https://arxiv.org/abs/astro-ph/9908023}{{\ttfamily
  astro-ph/9908023}}].

\bibitem{Farrar:2003uw}
G.R.~Farrar and P.J.E.~Peebles, \emph{{Interacting dark matter and dark
  energy}}, \href{https://doi.org/10.1086/381728}{\emph{Astrophys. J.}
  {\bfseries 604} (2004) 1}
  [\href{https://arxiv.org/abs/astro-ph/0307316}{{\ttfamily
  astro-ph/0307316}}].

\bibitem{Amendola:2003wa}
L.~Amendola, \emph{{Linear and non-linear perturbations in dark energy
  models}}, \href{https://doi.org/10.1103/PhysRevD.69.103524}{\emph{Phys. Rev.
  D} {\bfseries 69} (2004) 103524}
  [\href{https://arxiv.org/abs/astro-ph/0311175}{{\ttfamily
  astro-ph/0311175}}].

\bibitem{Koivisto:2005nr}
T.~Koivisto, \emph{{Growth of perturbations in dark matter coupled with
  quintessence}}, \href{https://doi.org/10.1103/PhysRevD.72.043516}{\emph{Phys.
  Rev. D} {\bfseries 72} (2005) 043516}
  [\href{https://arxiv.org/abs/astro-ph/0504571}{{\ttfamily
  astro-ph/0504571}}].

\bibitem{Amendola:2006dg}
L.~Amendola, G.~Camargo~Campos and R.~Rosenfeld, \emph{{Consequences of dark
  matter-dark energy interaction on cosmological parameters derived from SNIa
  data}}, \href{https://doi.org/10.1103/PhysRevD.75.083506}{\emph{Phys. Rev. D}
  {\bfseries 75} (2007) 083506}
  [\href{https://arxiv.org/abs/astro-ph/0610806}{{\ttfamily
  astro-ph/0610806}}].

\bibitem{Boehmer:2008av}
C.G.~Boehmer, G.~Caldera-Cabral, R.~Lazkoz and R.~Maartens, \emph{{Dynamics of
  dark energy with a coupling to dark matter}},
  \href{https://doi.org/10.1103/PhysRevD.78.023505}{\emph{Phys. Rev. D}
  {\bfseries 78} (2008) 023505}
  [\href{https://arxiv.org/abs/0801.1565}{{\ttfamily 0801.1565}}].

\bibitem{Baldi:2008ay}
M.~Baldi, V.~Pettorino, G.~Robbers and V.~Springel, \emph{{Hydrodynamical
  N-body simulations of coupled dark energy cosmologies}},
  \href{https://doi.org/10.1111/j.1365-2966.2009.15987.x}{\emph{Mon. Not. Roy.
  Astron. Soc.} {\bfseries 403} (2010) 1684}
  [\href{https://arxiv.org/abs/0812.3901}{{\ttfamily 0812.3901}}].

\bibitem{Baldi:2010vv}
M.~Baldi, \emph{{Time dependent couplings in the dark sector: from background
  evolution to nonlinear structure formation}},
  \href{https://doi.org/10.1111/j.1365-2966.2010.17758.x}{\emph{Mon. Not. Roy.
  Astron. Soc.} {\bfseries 411} (2011) 1077}
  [\href{https://arxiv.org/abs/1005.2188}{{\ttfamily 1005.2188}}].

\bibitem{Gleyzes:2015pma}
J.~Gleyzes, D.~Langlois, M.~Mancarella and F.~Vernizzi, \emph{{Effective Theory
  of Interacting Dark Energy}},
  \href{https://doi.org/10.1088/1475-7516/2015/08/054}{\emph{JCAP} {\bfseries
  08} (2015) 054} [\href{https://arxiv.org/abs/1504.05481}{{\ttfamily
  1504.05481}}].

\bibitem{Fardon:2003eh}
R.~Fardon, A.E.~Nelson and N.~Weiner, \emph{{Dark energy from mass varying
  neutrinos}}, \href{https://doi.org/10.1088/1475-7516/2004/10/005}{\emph{JCAP}
  {\bfseries 10} (2004) 005}
  [\href{https://arxiv.org/abs/astro-ph/0309800}{{\ttfamily
  astro-ph/0309800}}].

\bibitem{Brookfield:2005bz}
A.W.~Brookfield, C.~van~de Bruck, D.F.~Mota and D.~Tocchini-Valentini,
  \emph{{Cosmology of mass-varying neutrinos driven by quintessence: theory and
  observations}}, \href{https://doi.org/10.1103/PhysRevD.73.083515}{\emph{Phys.
  Rev. D} {\bfseries 73} (2006) 083515}
  [\href{https://arxiv.org/abs/astro-ph/0512367}{{\ttfamily
  astro-ph/0512367}}].

\bibitem{Mota:2008nj}
D.F.~Mota, V.~Pettorino, G.~Robbers and C.~Wetterich, \emph{{Neutrino
  clustering in growing neutrino quintessence}},
  \href{https://doi.org/10.1016/j.physletb.2008.03.060}{\emph{Phys. Lett. B}
  {\bfseries 663} (2008) 160}
  [\href{https://arxiv.org/abs/0802.1515}{{\ttfamily 0802.1515}}].

\bibitem{Pettorino:2009vn}
V.~Pettorino, D.F.~Mota, G.~Robbers and C.~Wetterich, \emph{{Clustering in
  growing neutrino cosmologies}},
  \href{https://doi.org/10.1063/1.3131514}{\emph{AIP Conf. Proc.} {\bfseries
  1115} (2009) 291} [\href{https://arxiv.org/abs/0901.1239}{{\ttfamily
  0901.1239}}].

\bibitem{Wintergerst:2009fh}
N.~Wintergerst, V.~Pettorino, D.F.~Mota and C.~Wetterich, \emph{{Very large
  scale structures in growing neutrino quintessence}},
  \href{https://doi.org/10.1103/PhysRevD.81.063525}{\emph{Phys. Rev. D}
  {\bfseries 81} (2010) 063525}
  [\href{https://arxiv.org/abs/0910.4985}{{\ttfamily 0910.4985}}].

\bibitem{Ayaita:2011ay}
Y.~Ayaita, M.~Weber and C.~Wetterich, \emph{{Structure Formation and
  Backreaction in Growing Neutrino Quintessence}},
  \href{https://doi.org/10.1103/PhysRevD.85.123010}{\emph{Phys. Rev. D}
  {\bfseries 85} (2012) 123010}
  [\href{https://arxiv.org/abs/1112.4762}{{\ttfamily 1112.4762}}].

\bibitem{Nunes:2011mw}
N.J.~Nunes, L.~Schrempp and C.~Wetterich, \emph{{Mass freezing in growing
  neutrino quintessence}},
  \href{https://doi.org/10.1103/PhysRevD.83.083523}{\emph{Phys. Rev. D}
  {\bfseries 83} (2011) 083523}
  [\href{https://arxiv.org/abs/1102.1664}{{\ttfamily 1102.1664}}].

\bibitem{Ayaita:2012xm}
Y.~Ayaita, M.~Weber and C.~Wetterich, \emph{{Neutrino lump fluid in growing
  neutrino quintessence}},
  \href{https://doi.org/10.1103/PhysRevD.87.043519}{\emph{Phys. Rev. D}
  {\bfseries 87} (2013) 043519}
  [\href{https://arxiv.org/abs/1211.6589}{{\ttfamily 1211.6589}}].

\bibitem{Ayaita:2014una}
Y.~Ayaita, M.~Baldi, F.~F\"uhrer, E.~Puchwein and C.~Wetterich,
  \emph{{Nonlinear growing neutrino cosmology}},
  \href{https://doi.org/10.1103/PhysRevD.93.063511}{\emph{Phys. Rev. D}
  {\bfseries 93} (2016) 063511}
  [\href{https://arxiv.org/abs/1407.8414}{{\ttfamily 1407.8414}}].

\bibitem{Fuhrer:2015xya}
F.~F\"uhrer and C.~Wetterich, \emph{{Backreaction in Growing Neutrino
  Quintessence}}, \href{https://doi.org/10.1103/PhysRevD.91.123542}{\emph{Phys.
  Rev. D} {\bfseries 91} (2015) 123542}
  [\href{https://arxiv.org/abs/1503.07995}{{\ttfamily 1503.07995}}].

\bibitem{Casas:2016duf}
S.~Casas, V.~Pettorino and C.~Wetterich, \emph{{Dynamics of neutrino lumps in
  growing neutrino quintessence}},
  \href{https://doi.org/10.1103/PhysRevD.94.103518}{\emph{Phys. Rev. D}
  {\bfseries 94} (2016) 103518}
  [\href{https://arxiv.org/abs/1608.02358}{{\ttfamily 1608.02358}}].

\bibitem{delCampo:2006vv}
S.~del Campo, R.~Herrera, G.~Olivares and D.~Pavon, \emph{{Interacting models
  of soft coincidence}},
  \href{https://doi.org/10.1103/PhysRevD.74.023501}{\emph{Phys. Rev. D}
  {\bfseries 74} (2006) 023501}
  [\href{https://arxiv.org/abs/astro-ph/0606520}{{\ttfamily
  astro-ph/0606520}}].

\bibitem{Wei:2007ws}
H.~Wei and S.N.~Zhang, \emph{{Interacting Energy Components and Observational
  $H(z)$ Data}},
  \href{https://doi.org/10.1016/j.physletb.2007.08.056}{\emph{Phys. Lett. B}
  {\bfseries 654} (2007) 139}
  [\href{https://arxiv.org/abs/0704.3330}{{\ttfamily 0704.3330}}].

\bibitem{Caldera-Cabral:2008yyo}
G.~Caldera-Cabral, R.~Maartens and L.A.~Urena-Lopez, \emph{{Dynamics of
  interacting dark energy}},
  \href{https://doi.org/10.1103/PhysRevD.79.063518}{\emph{Phys. Rev. D}
  {\bfseries 79} (2009) 063518}
  [\href{https://arxiv.org/abs/0812.1827}{{\ttfamily 0812.1827}}].

\bibitem{delCampo:2015vha}
S.~del Campo, R.~Herrera and D.~Pav\'on, \emph{{Interaction in the dark
  sector}}, \href{https://doi.org/10.1103/PhysRevD.91.123539}{\emph{Phys. Rev.
  D} {\bfseries 91} (2015) 123539}
  [\href{https://arxiv.org/abs/1507.00187}{{\ttfamily 1507.00187}}].

\bibitem{Chimento:2009hj}
L.P.~Chimento, \emph{{Linear and nonlinear interactions in the dark sector}},
  \href{https://doi.org/10.1103/PhysRevD.81.043525}{\emph{Phys. Rev. D}
  {\bfseries 81} (2010) 043525}
  [\href{https://arxiv.org/abs/0911.5687}{{\ttfamily 0911.5687}}].

\bibitem{Verma:2013pya}
M.M.~Verma and S.D.~Pathak, \emph{{The BICEP2 data and a single Higgs-like
  interacting scalar field}},
  \href{https://doi.org/10.1142/S0218271814500758}{\emph{Int. J. Mod. Phys. D}
  {\bfseries 23} (2014) 1450075}
  [\href{https://arxiv.org/abs/1312.1175}{{\ttfamily 1312.1175}}].

\bibitem{SanchezG:2014snn}
I.E.~Sanchez~G., \emph{{Dark matter interacts with variable vacuum energy}},
  \href{https://doi.org/10.1007/s10714-014-1769-0}{\emph{Gen. Rel. Grav.}
  {\bfseries 46} (2014) 1769}
  [\href{https://arxiv.org/abs/1405.1291}{{\ttfamily 1405.1291}}].

\bibitem{Shahalam:2015sja}
M.~Shahalam, S.D.~Pathak, M.M.~Verma, M.Y.~Khlopov and R.~Myrzakulov,
  \emph{{Dynamics of interacting quintessence}},
  \href{https://doi.org/10.1140/epjc/s10052-015-3608-1}{\emph{Eur. Phys. J. C}
  {\bfseries 75} (2015) 395}
  [\href{https://arxiv.org/abs/1503.08712}{{\ttfamily 1503.08712}}].

\bibitem{Valiviita:2008iv}
J.~Valiviita, E.~Majerotto and R.~Maartens, \emph{{Instability in interacting
  dark energy and dark matter fluids}},
  \href{https://doi.org/10.1088/1475-7516/2008/07/020}{\emph{JCAP} {\bfseries
  07} (2008) 020} [\href{https://arxiv.org/abs/0804.0232}{{\ttfamily
  0804.0232}}].

\bibitem{Pettorino:2008ez}
V.~Pettorino and C.~Baccigalupi, \emph{{Coupled and Extended Quintessence:
  theoretical differences and structure formation}},
  \href{https://doi.org/10.1103/PhysRevD.77.103003}{\emph{Phys. Rev. D}
  {\bfseries 77} (2008) 103003}
  [\href{https://arxiv.org/abs/0802.1086}{{\ttfamily 0802.1086}}].

\bibitem{vandeBruck:2016jgg}
C.~van~de Bruck, J.~Mifsud, J.P.~Mimoso and N.J.~Nunes, \emph{{Generalized dark
  energy interactions with multiple fluids}},
  \href{https://doi.org/10.1088/1475-7516/2016/11/031}{\emph{JCAP} {\bfseries
  11} (2016) 031} [\href{https://arxiv.org/abs/1605.03834}{{\ttfamily
  1605.03834}}].

\bibitem{vandeBruck:2015ida}
C.~van~de Bruck and J.~Morrice, \emph{{Disformal couplings and the dark sector
  of the universe}},
  \href{https://doi.org/10.1088/1475-7516/2015/04/036}{\emph{JCAP} {\bfseries
  04} (2015) 036} [\href{https://arxiv.org/abs/1501.03073}{{\ttfamily
  1501.03073}}].

\bibitem{Koivisto:2012za}
T.S.~Koivisto, D.F.~Mota and M.~Zumalacarregui, \emph{{Screening Modifications
  of Gravity through Disformally Coupled Fields}},
  \href{https://doi.org/10.1103/PhysRevLett.109.241102}{\emph{Phys. Rev. Lett.}
  {\bfseries 109} (2012) 241102}
  [\href{https://arxiv.org/abs/1205.3167}{{\ttfamily 1205.3167}}].

\bibitem{Wetterich:1987fk}
C.~Wetterich, \emph{{Cosmologies With Variable Newton's 'Constant'}},
  \href{https://doi.org/10.1016/0550-3213(88)90192-7}{\emph{Nucl. Phys.}
  {\bfseries B302} (1988) 645}.

\bibitem{Bernal:2020qyu}
N.~Bernal, J.~Rubio and H.~Veerm\"ae, \emph{{UV Freeze-in in Starobinsky
  Inflation}}, \href{https://doi.org/10.1088/1475-7516/2020/10/021}{\emph{JCAP}
  {\bfseries 10} (2020) 021}
  [\href{https://arxiv.org/abs/2006.02442}{{\ttfamily 2006.02442}}].

\bibitem{Amendola:2001rc}
L.~Amendola and D.~Tocchini-Valentini, \emph{{Baryon bias and structure
  formation in an accelerating universe}},
  \href{https://doi.org/10.1103/PhysRevD.66.043528}{\emph{Phys. Rev. D}
  {\bfseries 66} (2002) 043528}
  [\href{https://arxiv.org/abs/astro-ph/0111535}{{\ttfamily
  astro-ph/0111535}}].

\bibitem{Bonometto:2012qz}
S.A.~Bonometto, G.~Sassi and G.~La~Vacca, \emph{{Dark energy from dark
  radiation in strongly coupled cosmologies with no fine tuning}},
  \href{https://doi.org/10.1088/1475-7516/2012/08/015}{\emph{JCAP} {\bfseries
  08} (2012) 015} [\href{https://arxiv.org/abs/1206.2281}{{\ttfamily
  1206.2281}}].

\bibitem{Bonometto:2013eva}
S.A.~Bonometto and R.~Mainini, \emph{{Fluctuations in strongly coupled
  cosmologies}},
  \href{https://doi.org/10.1088/1475-7516/2014/03/038}{\emph{JCAP} {\bfseries
  03} (2014) 038} [\href{https://arxiv.org/abs/1311.6374}{{\ttfamily
  1311.6374}}].

\bibitem{Bonometto:2015mya}
S.A.~Bonometto, R.~Mainini and A.V.~Macci\`o, \emph{{Strongly coupled dark
  energy cosmologies: preserving $\Lambda$CDM success and easing low scale
  problems \textendash{} I. Linear theory revisited}},
  \href{https://doi.org/10.1093/mnras/stv1621}{\emph{Mon. Not. Roy. Astron.
  Soc.} {\bfseries 453} (2015) 1002}
  [\href{https://arxiv.org/abs/1503.07875}{{\ttfamily 1503.07875}}].

\bibitem{Maccio:2015iya}
A.V.~Macci\`o, R.~Mainini, C.~Penzo and S.A.~Bonometto, \emph{{Strongly coupled
  dark energy cosmologies: preserving \ensuremath{\Lambda}CDM success and
  easing low-scale problems \textendash{} II. Cosmological simulations}},
  \href{https://doi.org/10.1093/mnras/stv1680}{\emph{Mon. Not. Roy. Astron.
  Soc.} {\bfseries 453} (2015) 1371}
  [\href{https://arxiv.org/abs/1503.07867}{{\ttfamily 1503.07867}}].

\bibitem{Bonometto:2017rdu}
S.A.~Bonometto, M.~Mezzetti and R.~Mainini, \emph{{Strongly Coupled Dark Energy
  with Warm dark matter vs. LCDM}},
  \href{https://doi.org/10.1088/1475-7516/2017/10/011}{\emph{JCAP} {\bfseries
  10} (2017) 011} [\href{https://arxiv.org/abs/1703.05139}{{\ttfamily
  1703.05139}}].

\bibitem{Bonometto:2017lhg}
S.~Bonometto and R.~Mainini, \emph{{Coupled DM heating in SCDEW cosmologies}},
  \href{https://doi.org/10.3390/e19080398}{\emph{Entropy} {\bfseries 19} (2017)
  398} [\href{https://arxiv.org/abs/1707.06004}{{\ttfamily 1707.06004}}].

\bibitem{Amendola:2017xhl}
L.~Amendola, J.~Rubio and C.~Wetterich, \emph{{Primordial black holes from
  fifth forces}}, \href{https://doi.org/10.1103/PhysRevD.97.081302}{\emph{Phys.
  Rev. D} {\bfseries 97} (2018) 081302}
  [\href{https://arxiv.org/abs/1711.09915}{{\ttfamily 1711.09915}}].

\bibitem{Bonometto:2018dmx}
S.A.~Bonometto, R.~Mainini and M.~Mezzetti, \emph{{Strongly Coupled Dark Energy
  Cosmologies yielding large mass Primordial Black Holes}},
  \href{https://doi.org/10.1093/mnras/stz846}{\emph{Mon. Not. Roy. Astron.
  Soc.} {\bfseries 486} (2019) 2321}
  [\href{https://arxiv.org/abs/1807.11841}{{\ttfamily 1807.11841}}].

\bibitem{Flores:2020drq}
M.M.~Flores and A.~Kusenko, \emph{{Primordial Black Holes from Long-Range
  Scalar Forces and Scalar Radiative Cooling}},
  \href{https://doi.org/10.1103/PhysRevLett.126.041101}{\emph{Phys. Rev. Lett.}
  {\bfseries 126} (2021) 041101}
  [\href{https://arxiv.org/abs/2008.12456}{{\ttfamily 2008.12456}}].

\bibitem{Domenech:2021uyx}
G.~Dom\`enech and M.~Sasaki, \emph{{Cosmology of strongly interacting fermions
  in the early universe}},
  \href{https://doi.org/10.1088/1475-7516/2021/06/030}{\emph{JCAP} {\bfseries
  06} (2021) 030} [\href{https://arxiv.org/abs/2104.05271}{{\ttfamily
  2104.05271}}].

\bibitem{Savastano:2019zpr}
S.~Savastano, L.~Amendola, J.~Rubio and C.~Wetterich, \emph{{Primordial dark
  matter halos from fifth forces}},
  \href{https://doi.org/10.1103/PhysRevD.100.083518}{\emph{Phys. Rev. D}
  {\bfseries 100} (2019) 083518}
  [\href{https://arxiv.org/abs/1906.05300}{{\ttfamily 1906.05300}}].

\bibitem{Pettorino:2010bv}
V.~Pettorino, N.~Wintergerst, L.~Amendola and C.~Wetterich, \emph{{Neutrino
  lumps and the Cosmic Microwave Background}},
  \href{https://doi.org/10.1103/PhysRevD.82.123001}{\emph{Phys. Rev. D}
  {\bfseries 82} (2010) 123001}
  [\href{https://arxiv.org/abs/1009.2461}{{\ttfamily 1009.2461}}].

\bibitem{Brouzakis:2010md}
N.~Brouzakis, V.~Pettorino, N.~Tetradis and C.~Wetterich, \emph{{Nonlinear
  matter spectra in growing neutrino quintessence}},
  \href{https://doi.org/10.1088/1475-7516/2011/03/049}{\emph{JCAP} {\bfseries
  03} (2011) 049} [\href{https://arxiv.org/abs/1012.5255}{{\ttfamily
  1012.5255}}].

\bibitem{Baldi:2011es}
M.~Baldi, V.~Pettorino, L.~Amendola and C.~Wetterich, \emph{{Oscillating
  nonlinear large scale structure in growing neutrino quintessence}},
  \href{https://doi.org/10.1111/j.1365-2966.2011.19477.x}{\emph{Mon. Not. Roy.
  Astron. Soc.} {\bfseries 418} (2011) 214}
  [\href{https://arxiv.org/abs/1106.2161}{{\ttfamily 1106.2161}}].

\bibitem{TocchiniValentini:2001ty}
D.~Tocchini-Valentini and L.~Amendola, \emph{{Stationary dark energy with a
  baryon dominated era: Solving the coincidence problem with a linear
  coupling}}, \href{https://doi.org/10.1103/PhysRevD.65.063508}{\emph{Phys.
  Rev.} {\bfseries D65} (2002) 063508}
  [\href{https://arxiv.org/abs/astro-ph/0108143}{{\ttfamily
  astro-ph/0108143}}].

\bibitem{Kallosh:1995hi}
R.~Kallosh, A.D.~Linde, D.A.~Linde and L.~Susskind, \emph{{Gravity and global
  symmetries}}, \href{https://doi.org/10.1103/PhysRevD.52.912}{\emph{Phys. Rev.
  D} {\bfseries 52} (1995) 912}
  [\href{https://arxiv.org/abs/hep-th/9502069}{{\ttfamily hep-th/9502069}}].

\bibitem{Figueroa:2016dsc}
D.G.~Figueroa and C.T.~Byrnes, \emph{{The Standard Model Higgs as the origin of
  the hot Big Bang}},
  \href{https://doi.org/10.1016/j.physletb.2017.01.059}{\emph{Phys. Lett. B}
  {\bfseries 767} (2017) 272}
  [\href{https://arxiv.org/abs/1604.03905}{{\ttfamily 1604.03905}}].

\bibitem{Nakama:2018gll}
T.~Nakama and J.~Yokoyama, \emph{{Reheating through the Higgs amplified by
  spinodal instabilities and gravitational creation of gravitons}},
  \href{https://doi.org/10.1093/ptep/ptz014}{\emph{PTEP} {\bfseries 2019}
  (2019) 033E02} [\href{https://arxiv.org/abs/1803.07111}{{\ttfamily
  1803.07111}}].

\bibitem{Fairbairn:2018bsw}
M.~Fairbairn, K.~Kainulainen, T.~Markkanen and S.~Nurmi, \emph{{Despicable Dark
  Relics: generated by gravity with unconstrained masses}},
  \href{https://doi.org/10.1088/1475-7516/2019/04/005}{\emph{JCAP} {\bfseries
  04} (2019) 005} [\href{https://arxiv.org/abs/1808.08236}{{\ttfamily
  1808.08236}}].

\bibitem{Laulumaa:2020pqi}
L.~Laulumaa, T.~Markkanen and S.~Nurmi, \emph{{Primordial dark matter from
  curvature induced symmetry breaking}},
  \href{https://doi.org/10.1088/1475-7516/2020/08/002}{\emph{JCAP} {\bfseries
  08} (2020) 002} [\href{https://arxiv.org/abs/2005.04061}{{\ttfamily
  2005.04061}}].

\bibitem{Babichev:2020xeg}
E.~Babichev, D.~Gorbunov and S.~Ramazanov, \emph{{Gravitational misalignment
  mechanism of Dark Matter production}},
  \href{https://doi.org/10.1088/1475-7516/2020/08/047}{\emph{JCAP} {\bfseries
  08} (2020) 047} [\href{https://arxiv.org/abs/2004.03410}{{\ttfamily
  2004.03410}}].

\bibitem{Felder:2000hj}
G.N.~Felder, J.~Garcia-Bellido, P.B.~Greene, L.~Kofman, A.D.~Linde and
  I.~Tkachev, \emph{{Dynamics of symmetry breaking and tachyonic preheating}},
  \href{https://doi.org/10.1103/PhysRevLett.87.011601}{\emph{Phys. Rev. Lett.}
  {\bfseries 87} (2001) 011601}
  [\href{https://arxiv.org/abs/hep-ph/0012142}{{\ttfamily hep-ph/0012142}}].

\bibitem{Felder:2001kt}
G.N.~Felder, L.~Kofman and A.D.~Linde, \emph{{Tachyonic instability and
  dynamics of spontaneous symmetry breaking}},
  \href{https://doi.org/10.1103/PhysRevD.64.123517}{\emph{Phys. Rev. D}
  {\bfseries 64} (2001) 123517}
  [\href{https://arxiv.org/abs/hep-th/0106179}{{\ttfamily hep-th/0106179}}].

\bibitem{Bettoni:2019dcw}
D.~Bettoni and J.~Rubio, \emph{{Hubble-induced phase transitions: Walls are not
  forever}}, \href{https://doi.org/10.1088/1475-7516/2020/01/002}{\emph{JCAP}
  {\bfseries 01} (2020) 002}
  [\href{https://arxiv.org/abs/1911.03484}{{\ttfamily 1911.03484}}].

\bibitem{Guth:1985ya}
A.H.~Guth and S.-Y.~Pi, \emph{{The Quantum Mechanics of the Scalar Field in the
  New Inflationary Universe}},
  \href{https://doi.org/10.1103/PhysRevD.32.1899}{\emph{Phys. Rev. D}
  {\bfseries 32} (1985) 1899}.

\bibitem{Garcia-Bellido:2002fsq}
J.~Garcia-Bellido, M.~Garcia~Perez and A.~Gonzalez-Arroyo, \emph{{Symmetry
  breaking and false vacuum decay after hybrid inflation}},
  \href{https://doi.org/10.1103/PhysRevD.67.103501}{\emph{Phys. Rev. D}
  {\bfseries 67} (2003) 103501}
  [\href{https://arxiv.org/abs/hep-ph/0208228}{{\ttfamily hep-ph/0208228}}].

\bibitem{Lozanov:2016hid}
K.D.~Lozanov and M.A.~Amin, \emph{{Equation of State and Duration to Radiation
  Domination after Inflation}},
  \href{https://doi.org/10.1103/PhysRevLett.119.061301}{\emph{Phys. Rev. Lett.}
  {\bfseries 119} (2017) 061301}
  [\href{https://arxiv.org/abs/1608.01213}{{\ttfamily 1608.01213}}].

\bibitem{Lozanov:2017hjm}
K.D.~Lozanov and M.A.~Amin, \emph{{Self-resonance after inflation: oscillons,
  transients and radiation domination}},
  \href{https://doi.org/10.1103/PhysRevD.97.023533}{\emph{Phys. Rev. D}
  {\bfseries 97} (2018) 023533}
  [\href{https://arxiv.org/abs/1710.06851}{{\ttfamily 1710.06851}}].

\bibitem{GarciaBellido:2008ab}
J.~Garcia-Bellido, D.G.~Figueroa and J.~Rubio, \emph{{Preheating in the
  Standard Model with the Higgs-Inflaton coupled to gravity}},
  \href{https://doi.org/10.1103/PhysRevD.79.063531}{\emph{Phys. Rev.}
  {\bfseries D79} (2009) 063531}
  [\href{https://arxiv.org/abs/0812.4624}{{\ttfamily 0812.4624}}].

\bibitem{Rubio:2015zia}
J.~Rubio, \emph{{Higgs inflation and vacuum stability}},
  \href{https://doi.org/10.1088/1742-6596/631/1/012032}{\emph{J. Phys. Conf.
  Ser.} {\bfseries 631} (2015) 012032}
  [\href{https://arxiv.org/abs/1502.07952}{{\ttfamily 1502.07952}}].

\bibitem{Repond:2016sol}
J.~Repond and J.~Rubio, \emph{{Combined Preheating on the lattice with
  applications to Higgs inflation}},
  \href{https://doi.org/10.1088/1475-7516/2016/07/043}{\emph{JCAP} {\bfseries
  1607} (2016) 043} [\href{https://arxiv.org/abs/1604.08238}{{\ttfamily
  1604.08238}}].

\bibitem{Fan:2021otj}
J.~Fan, K.D.~Lozanov and Q.~Lu, \emph{{Spillway Preheating}},
  \href{https://doi.org/10.1007/JHEP05(2021)069}{\emph{JHEP} {\bfseries 05}
  (2021) 069} [\href{https://arxiv.org/abs/2101.11008}{{\ttfamily
  2101.11008}}].

\bibitem{Dufaux:2006ee}
J.F.~Dufaux, G.N.~Felder, L.~Kofman, M.~Peloso and D.~Podolsky,
  \emph{{Preheating with trilinear interactions: Tachyonic resonance}},
  \href{https://doi.org/10.1088/1475-7516/2006/07/006}{\emph{JCAP} {\bfseries
  07} (2006) 006} [\href{https://arxiv.org/abs/hep-ph/0602144}{{\ttfamily
  hep-ph/0602144}}].

\bibitem{Kamada:2015iga}
A.~Kamada and M.~Yamada, \emph{{Gravitational wave signals from short-lived
  topological defects in the MSSM}},
  \href{https://doi.org/10.1088/1475-7516/2015/10/021}{\emph{JCAP} {\bfseries
  10} (2015) 021} [\href{https://arxiv.org/abs/1505.01167}{{\ttfamily
  1505.01167}}].

\bibitem{Dufaux:2010cf}
J.-F.~Dufaux, D.G.~Figueroa and J.~Garcia-Bellido, \emph{{Gravitational Waves
  from Abelian Gauge Fields and Cosmic Strings at Preheating}},
  \href{https://doi.org/10.1103/PhysRevD.82.083518}{\emph{Phys. Rev. D}
  {\bfseries 82} (2010) 083518}
  [\href{https://arxiv.org/abs/1006.0217}{{\ttfamily 1006.0217}}].

\bibitem{Figueroa:2012kw}
D.G.~Figueroa, M.~Hindmarsh and J.~Urrestilla, \emph{{Exact Scale-Invariant
  Background of Gravitational Waves from Cosmic Defects}},
  \href{https://doi.org/10.1103/PhysRevLett.110.101302}{\emph{Phys. Rev. Lett.}
  {\bfseries 110} (2013) 101302}
  [\href{https://arxiv.org/abs/1212.5458}{{\ttfamily 1212.5458}}].

\bibitem{Giblin:2011yh}
J.T.~Giblin, Jr., L.R.~Price, X.~Siemens and B.~Vlcek, \emph{{Gravitational
  Waves from Global Second Order Phase Transitions}},
  \href{https://doi.org/10.1088/1475-7516/2012/11/006}{\emph{JCAP} {\bfseries
  11} (2012) 006} [\href{https://arxiv.org/abs/1111.4014}{{\ttfamily
  1111.4014}}].

\bibitem{Gorghetto:2018myk}
M.~Gorghetto, E.~Hardy and G.~Villadoro, \emph{{Axions from Strings: the
  Attractive Solution}},
  \href{https://doi.org/10.1007/JHEP07(2018)151}{\emph{JHEP} {\bfseries 07}
  (2018) 151} [\href{https://arxiv.org/abs/1806.04677}{{\ttfamily
  1806.04677}}].

\bibitem{Vaquero:2018tib}
A.~Vaquero, J.~Redondo and J.~Stadler, \emph{{Early seeds of axion
  miniclusters}},
  \href{https://doi.org/10.1088/1475-7516/2019/04/012}{\emph{JCAP} {\bfseries
  04} (2019) 012} [\href{https://arxiv.org/abs/1809.09241}{{\ttfamily
  1809.09241}}].

\bibitem{Bennett:1989ak}
D.P.~Bennett and F.R.~Bouchet, \emph{{Cosmic string evolution}},
  \href{https://doi.org/10.1103/PhysRevLett.63.2776}{\emph{Phys. Rev. Lett.}
  {\bfseries 63} (1989) 2776}.

\bibitem{Perivolaropoulos:1992if}
L.~Perivolaropoulos, \emph{{COBE versus cosmic strings: An Analytical model}},
  \href{https://doi.org/10.1016/0370-2693(93)91825-8}{\emph{Phys. Lett. B}
  {\bfseries 298} (1993) 305}
  [\href{https://arxiv.org/abs/hep-ph/9208247}{{\ttfamily hep-ph/9208247}}].

\bibitem{Gouttenoire:2019kij}
Y.~Gouttenoire, G.~Servant and P.~Simakachorn, \emph{{Beyond the Standard
  Models with Cosmic Strings}},
  \href{https://doi.org/10.1088/1475-7516/2020/07/032}{\emph{JCAP} {\bfseries
  07} (2020) 032} [\href{https://arxiv.org/abs/1912.02569}{{\ttfamily
  1912.02569}}].

\bibitem{Aggarwal:2020olq}
N.~Aggarwal et~al., \emph{{Challenges and Opportunities of Gravitational Wave
  Searches at MHz to GHz Frequencies}},
  \href{https://arxiv.org/abs/2011.12414}{{\ttfamily 2011.12414}}.

\bibitem{Sakharov:1967dj}
A.D.~Sakharov, \emph{{Violation of CP Invariance, C asymmetry, and baryon
  asymmetry of the universe}},
  \href{https://doi.org/10.1070/PU1991v034n05ABEH002497}{\emph{Pisma Zh. Eksp.
  Teor. Fiz.} {\bfseries 5} (1967) 32}.

\bibitem{Affleck:1984fy}
I.~Affleck and M.~Dine, \emph{{A New Mechanism for Baryogenesis}},
  \href{https://doi.org/10.1016/0550-3213(85)90021-5}{\emph{Nucl. Phys. B}
  {\bfseries 249} (1985) 361}.

\bibitem{Dine:1995kz}
M.~Dine, L.~Randall and S.D.~Thomas, \emph{{Baryogenesis from flat directions
  of the supersymmetric standard model}},
  \href{https://doi.org/10.1016/0550-3213(95)00538-2}{\emph{Nucl. Phys. B}
  {\bfseries 458} (1996) 291}
  [\href{https://arxiv.org/abs/hep-ph/9507453}{{\ttfamily hep-ph/9507453}}].

\bibitem{Sakstein:2017lfm}
J.~Sakstein and M.~Trodden, \emph{{Baryogenesis via Dark Matter-Induced
  Symmetry Breaking in the Early Universe}},
  \href{https://doi.org/10.1016/j.physletb.2017.09.059}{\emph{Phys. Lett. B}
  {\bfseries 774} (2017) 183}
  [\href{https://arxiv.org/abs/1703.10103}{{\ttfamily 1703.10103}}].

\bibitem{Sakstein:2017nns}
J.~Sakstein and A.R.~Solomon, \emph{{Baryogenesis in Lorentz-violating gravity
  theories}}, \href{https://doi.org/10.1016/j.physletb.2017.08.039}{\emph{Phys.
  Lett. B} {\bfseries 773} (2017) 186}
  [\href{https://arxiv.org/abs/1705.10695}{{\ttfamily 1705.10695}}].

\bibitem{Macpherson:1994wf}
A.L.~Macpherson and B.A.~Campbell, \emph{{Biased discrete symmetry breaking and
  Fermi balls}},
  \href{https://doi.org/10.1016/0370-2693(95)00080-5}{\emph{Phys. Lett. B}
  {\bfseries 347} (1995) 205}
  [\href{https://arxiv.org/abs/hep-ph/9408387}{{\ttfamily hep-ph/9408387}}].

\bibitem{Coulson:1995nv}
D.~Coulson, Z.~Lalak and B.A.~Ovrut, \emph{{Biased domain walls}},
  \href{https://doi.org/10.1103/PhysRevD.53.4237}{\emph{Phys. Rev. D}
  {\bfseries 53} (1996) 4237}.

\bibitem{Avelino:2008qy}
P.P.~Avelino, C.J.A.P.~Martins and L.~Sousa, \emph{{Dynamics of Biased Domain
  Walls and the Devaluation Mechanism}},
  \href{https://doi.org/10.1103/PhysRevD.78.043521}{\emph{Phys. Rev. D}
  {\bfseries 78} (2008) 043521}
  [\href{https://arxiv.org/abs/0805.4013}{{\ttfamily 0805.4013}}].

\bibitem{Markkanen:2015xuw}
T.~Markkanen and S.~Nurmi, \emph{{Dark matter from gravitational particle
  production at reheating}},
  \href{https://doi.org/10.1088/1475-7516/2017/02/008}{\emph{JCAP} {\bfseries
  02} (2017) 008} [\href{https://arxiv.org/abs/1512.07288}{{\ttfamily
  1512.07288}}].

\bibitem{Bassett:1999mt}
B.A.~Bassett, F.~Tamburini, D.I.~Kaiser and R.~Maartens, \emph{{Metric
  preheating and limitations of linearized gravity. 2.}},
  \href{https://doi.org/10.1016/S0550-3213(99)00495-2}{\emph{Nucl. Phys. B}
  {\bfseries 561} (1999) 188}
  [\href{https://arxiv.org/abs/hep-ph/9901319}{{\ttfamily hep-ph/9901319}}].

\bibitem{Kaiser:1994vs}
D.I.~Kaiser, \emph{{Primordial spectral indices from generalized Einstein
  theories}}, \href{https://doi.org/10.1103/PhysRevD.52.4295}{\emph{Phys.\
  Rev.\ D} {\bfseries 52} (1995) 4295}
  [\href{https://arxiv.org/abs/astro-ph/9408044}{{\ttfamily
  astro-ph/9408044}}].

\bibitem{da_Silva_2019}
G.~da~Silva and R.~Ramos, \emph{The lambert–tsallis wq function},
  \href{https://doi.org/10.1016/j.physa.2019.03.046}{\emph{Physica A:
  Statistical Mechanics and its Applications} {\bfseries 525} (2019)
  164–170}.

\end{thebibliography}\endgroup
}
\end{document}